\documentclass{aa}
\pdfoutput=1

\usepackage{color}
\usepackage{array}
\usepackage{lscape}
\usepackage{txfonts}
\usepackage{graphicx}
\usepackage{multicol}
\usepackage{blindtext}
\usepackage{dblfloatfix}
\usepackage{fixltx2e}

\newcommand{\AU}{\rm{\; AU}}

\newcommand{\K}{{\rm\; K}}
\newcommand{\G}{{\rm\; G}}
\newcommand{\g}{{\rm\; g}}

\newcommand{\m}{{\rm\; m}}
\newcommand{\mm}{{\rm\; mm}}
\newcommand{\cm}{{\rm\; cm}}

\newcommand{\km}{{\rm\; km}}
\newcommand{\ms}{{\rm\; m \; s^{-1}}}

\newcommand{\crit}{_{\rm crit}}

\newcommand{\Hp}{H_{\rm p}}
\newcommand{\tf}{t_{\rm f}}
\newcommand{\cs}{c_{\rm s}}
\newcommand{\vk}{v_{\rm k}}
\newcommand{\vr}{v_{\rm r}}
\newcommand{\vth}{v_{\rm th}}
\newcommand{\tauf}{\tau_{\rm f}}
\newcommand{\rhop}{\rho_{\rm p}}

\newcommand{\Omegak}{\Omega_{\rm k}}

\newcommand{\cminv}[1]{\rm{\; cm}^{-#1}}
\newcommand{\sinv}{\rm{\; s}^{-1}}

\begin{document}

   \title{How to form planetesimals from mm-sized chondrules \\ and chondrule aggregates}

   \author{Daniel Carrera \and Anders Johansen \and Melvyn B. Davies}
   
   \authorrunning{D. Carrera et al}
   
   \institute{Lund Observatory,
              Department of Astronomy and Theoretical Physics,
              Lund University, Box 43, 22100 Lund, Sweden \\
              \email{danielc@astro.lu.se, anders@astro.lu.se, mbd@astro.lu.se}
             }
  \abstract {
  The size distribution of asteroids and Kuiper belt objects in the solar system is difficult to reconcile with a bottom-up formation scenario due to the observed scarcity of objects smaller than $\sim$100 km in size. Instead, planetesimals appear to form top-down, with large $100-1000$ km bodies forming from the rapid gravitational collapse of dense clumps of small solid particles.
  In this paper we investigate the conditions under which solid particles can form dense clumps in a protoplanetary disk.
  We use a hydrodynamic code to model the interaction between solid particles and the gas inside a shearing box inside the disk, considering particle sizes from sub-millimeter-sized chondrules to meter-sized rocks.
  We find that particles down to millimeter sizes can form dense particle clouds through the run-away convergence of radial drift known as the streaming instability. We make a map of the range of conditions (strength of turbulence, particle mass-loading, disk mass, and distance to the star) which are prone to producing dense particle clumps. Finally, we estimate the distribution of collision speeds between mm-sized particles. We calculate the rate of sticking collisions and obtain a robust upper limit on the particle growth timescale of $\sim$$10^5$ years.
  This means that mm-sized chondrule aggregates can grow on a timescale much smaller than the disk accretion timescale ($\sim$$10^6 - 10^7$ years). Our results suggest a pathway from the mm-sized grains found in primitive meteorites to fully formed asteroids.
  We speculate that asteroids may form from a positive feedback loop in which coagualation leads to particle clumping driven by the streaming instability. This clumping, in turn reduces collision speeds and enhances coagulation. Future simulations should model coagulation and the streaming instability together to explore this feedback loop further.
  }

   \maketitle
%

\section{Introduction}

Planetesimals are super-kilometer-sized bodies that are the seeds of terrestrial planets as well as the cores of ice and gas giants \citep{Safronov_1972,Chiang_2010,Johansen_2014}. Hence, the production of planetesimals is an important step in planet formation. In the solar system, large asteroids and Kuiper belt objects are the left-over planetesimals that did not become incorporated into a planet. Similar planetesimal belts are known to exist around other stars \citep{Wyatt_2008}.

The size distribution of the solar system planetesimals (i.e. large asteroids and Kuiper belt objects) reveals a ``knee'' at around 100 km in size, with a relative scarcity of bodies smaller than $\sim$100 km. Some authors have argued that this knee cannot be reproduced by bottom-up coagulation scenarios, unless the minimal initial size of a planetesimal was also $\sim$100 km. Hence, it appears that the size of solids in the protoplanetary disk ``jumped'' from the sub-meter scale to $\sim$100 km without passing through the intermediate sizes, and that bodies smaller than 100 km formed later by collisional grinding \citep{Morbidelli_2009}. A similar bump is inferred for the trans-Neptunian population \citep{Fraser_2009,Sheppard_2010,Shankman_2013,Fraser_2014}. That said, \citet{Weidenschilling_2010} suggests an alternate scenario in which the ``knee'' is produced in a bottom-up process from an initial population of 0.1 km-sized bodies. In this scenario, the bump at $\sim$100 km is produced by the transition from dispersion-dominated runaway growth to a regime dominated by Keplerian shear.

There are also important theoretical difficulties in building planetesimals in a bottom-up process. The first hurdle is that small particles in the mm-cm size range do not easily stick to form larger objects. Indeed, the largest grains observed in a protoplanetary disk are mm-cm in size \citep{Testi_2014,van_der_Marel_2013}. To the extent that meter-sized objects do form, they quickly spiral into the central star due to friction with gas orbiting the star at sub-Keplerian speeds \citep{Weidenschilling_1977}. Axisymmetric pressure bumps may stop particle drift \citep{Johansen_2009}, but growth rates remain low \citep{Johansen_2008}. For example, bodies at 3 AU can grow to a maximum of 100 m in 1 Myr \citep{Windmark_2012}.

These constraints suggest a different picture of how planetesimals form. It begins with the growth of macroscopic particles by coagulation, followed by the accumulation of these particles by a hydrodynamical process like the streaming instability (see below) which increases the local concentration of solids,

\begin{equation}\label{eqn:Z}
   Z  \, = \, \frac{\Sigma_{\rm solid}}{\Sigma_{\rm total}}
      \, = \, \frac{\Sigma_{\rm solid}}{\Sigma_{\rm gas} + \Sigma_{\rm solid}},
\end{equation}
where $\Sigma_{\rm solid}$ and $\Sigma_{\rm gas}$ are the surface densities of the solid and gas components of the disk respectively. The solids also sediment toward the mid-plane of the disk, so that the volume density of solids vs gas at the midplane can be significantly higher than $Z$. If the density of solids reaches the Roche density,

\pagebreak

\begin{equation}\label{eqn:roche-density}
  \rho_{\rm R} = \frac{9 \Omega^2}{4 \pi \G}\,,
\end{equation}
where $\Omega$ is the Keplerian frequency, then the particle self-gravity overwhelms Keplerian shear, leading to gravitational collapse \citep{Goldreich_1973}. This allows a collection of small objects to coalesce directly into fully formed planetesimals, and explaining the large birth sizes.

The main obstacle to achieving high densities is disk turbulence, thought to be caused by the magneto-rotational instability \citep[MRI,][]{Balbus_1991}. But even inside an MRI-free ``dead zone'' \citep{Gammie_1996}, the Kelvin-Helmholtz instability can be very effective at producing turbulence that blocks particle sedimentation to the mid-plane before they can reach the Roche density \citep{Weidenschilling_1980,Youdin_2002,Lee_2010}. Particle densities high enough for a local gravitational collapse can be obtained by concentration in long-lived gaseous vortices \citep{Barge_1995,Meheut_2012}, or pressure bumps caused by the MRI \citep{Johansen_2009,Johansen_2011b}. These regions can quickly accumulate meter and decimeter sized bodies, and for meter-sized boulders, the concentrations can significantly reduce radial drift caused by friction with the surrounding gas \citep{Johansen_2006,Johansen_2007}.



The streaming instability is another powerful mechanism to concentrate particles in localized dense clumps and to prevent radial drift \citep{Youdin_2005,Johansen_2007,Johansen_2007b,Bai_2010b}. It is driven by the relative drift between the solid and gas components of the disk. The gas component of the disk experiences a radial pressure support that counters the gravitational force of the star. This leads to a speed difference $\Delta v$ between the gas and solid components of the disk. A useful way to understand the streaming instability is that solids experience a head wind from the gas, and the gas is, in turn, pushed forward by solids. A small overdensity of solids will have a stronger back-reaction on the gas, and hence, a lower radial drift than neighboring regions. The reduced radial drift leads to the further accumulation of solids as neighboring particles drift into the overdensity. In this way, the streaming instability bears some similarity to a traffic jam.

The behavior of solid particles embedded in a gaseous accretion disk is primarily determined by the Stokes number, $\tauf = \tf \, \Omegak$, which measures the particle friction time $\tf$ in terms of the Keplerian frequency $\Omegak$. Some authors have raised the concern that the streaming instability has mainly been studied for relatively large particles, with $\tauf \sim 0.1$, corresponding to dm-sized particles when applied to the asteroid belt \citep{Shi_2013}. Particles of this size are inconsistent with coagulation experiments which show that particles cannot grow beyond mm sizes \citep{Zsom_2010,Guttler_2010}. Therefore, the main goal of this paper is to study the streaming instability for particle sizes down to millimeter.

Chondrules make up most of the mass in primitive meteorites, with a typical size of $R \sim 0.3 \mm$, and large chondrules reaching $R \sim 1 \mm$ \citep[e.g.][]{jacquet_2014}. At this size range, chondrules couple with the small-scale turbulent eddies in the disk, and they concentrate in the high-pressure regions between the eddies \citep{Cuzzi_2001}. This alone does not lead to particle growth, because a typical collision between two chondrules in the disk will result in bouncing \citep{Guttler_2010}. However, a collision between a small $\mu$m-sized dust particle and a chondrule does result in sticking. \citet{Ormel_2008} have shown that chondrules acquire a porous dust layer around them, which absorbs some of the kinetic energy of collisions and allows the chondrules to stick more easily. Under typical disk conditions, most of the chondrule and dust mass is in the form of millimeter-sized aggregates. For low turbulence regions, the chondrule aggregates have typical sizes around $R \sim 4 \mm$, and maximum sizes as high as $R \sim 10 \cm$ \citep{Ormel_2008}. Our work establishes the connection between the streaming instability and chondrules and chondrule aggregates. While we test the limits of the streaming instability at both the small-particle and the large-particle limits, the bulk of our analysis focuses on the size range of chondrules and chondrule aggregates.

In addition to particle size, the streaming instability also depends on the magnitude of the radial pressure gradient \citep{Bai_2010c}, as well as the solid concentration \citep{Johansen_2009b,Bai_2010b}. In particular, \citet{Bai_2010b} have shown that the streaming instability is effective in producing particle clumps for particle size $\tauf \gtrsim 0.1$, and solid concentration $Z \gtrsim 0.02$.

A number of global processes increase the local solid concentration, including radial drift of particles from the outer disk into the inner disk \citep{Youdin_2002}, large scale pressure bumps or vortices \citep{Johansen_2009}, and lastly, there is a region near the ice line where ice-dust aggregates (``dirty snowballs'') break up on a timescale comparable to the infall timescale, leading to a significant accumulation of solids by as much as a factor of 6.7 \citep{Sirono_2011}. The second way to increase $Z$ is to remove gas from the disk, either through late-stage photoevaporation \citep[e.g.][]{Alexander_2006,Alexander_2006b} or through disk winds \citep{Bai_2013,Suzuki_2009}. Gas removal has the double effect of increasing  $\tauf$ and $Z$ simultaneously. Hence, gas removal may be a powerful way to trigger the streaming instability in the late stages of the disk.

This paper is structured as follows. In section \ref{sect:numerical} we present our numerical model, including the simulation setup and initial conditions. In section \ref{sect:results} we present our core result -- a map of the region in the $Z$ vs $\tauf$ phase space that is consistent with particle clumping (Fig.\ \ref{fig:final-results}). In section \ref{sect:ppd-models} we review the implications and offer 
new constraints on planetesimal formation models (Fig.\ \ref{fig:models}). In section \ref{sect:coagulation} we measure the collision speed for $\tauf = 0.003$ particles and find that particle growth occurs faster than the disk accretion timescale. In section \ref{sect:conclusions} we present our conclusions.


\section{Numerical model}\label{sect:numerical}

We use the Pencil Code \citep{Youdin_2007} to model a small vertical slice of the protoplanetary disk in which the solar system planets formed using a stratified model that includes vertical gravity. The canonical model for this disk is the minimum mass solar nebula \citep[MMSN,][]{Hayashi_1981}. In it, the surface density of the gas component of the disk follows the power law

\begin{equation}\label{eqn:hayashi}
	\Sigma = 1700 \g \cminv2 \, \left( \frac{r}{\AU} \right)^{-3/2}.
\end{equation}
In addition to the gas, the disk also contains solid particles that are initially $\mu$m in size and represent about 1\% of the disk mass. Over time, particle sizes can grow by coagulation, and the mass ratio between gas and solids may change if particles migrate through the disk, or if the gas becomes depleted.


\subsection{Aerodynamics of solid particles}

The gas component of the disk produces a drag force on the particles whenever the particles have a non-zero relative speed $v_{\rm rel}$ with respect to the gas. The form of the gas drag law depends on the size of the particle relative to the mean free path $\lambda$ of gas particles. Small particles, with size $R < 9\lambda/4$ experience Epstein drag, while larger particles experience Stokes drag. We will show in section \ref{sect:lambda} that the the smaller particles in our simulations lie in the Epstein regime. Particles in the Epstein regime experience the drag force

\begin{equation}\label{eqn:epstein}
    F_{\rm drag} \, = \, \frac{4\pi}{3} \, \rho \, R^2 \vth \, v_{\rm rel}
\end{equation}
where $\rho$ is the gas density, $R$ is the particle size, and $\vth$ is the mean thermal speed of gas molecules,

\[
    \vth \, = \, \sqrt{\frac{8 \, k \, T}{\pi \, \mu \, m_{\rm p}}}.
\]
Here $k$ is the Boltzmann constant, $T$ is the local temperature, $\mu$ is the mean molecular weight of the gas, and $m_{\rm p}$ is the proton mass \citep{Armitage_2010}. It is convenient to write $\vth$ in terms of the isothermal sound speed $\cs$,

\[
    \cs \, = \, \sqrt{\frac{k \, T}{\mu \, m_{\rm p}}}
    \;\;\;\Rightarrow\;\;\;
    \vth \, = \, \cs \sqrt{\frac{8}{\pi}}.
\]
The friction time

\[
    \tf \, = \, \frac{m \, v_{\rm rel}}{F_{\rm drag}}
\]
measures how long it takes for a particle with mass $m$ to have its relative speed changed by order unity as a result of gas drag. For a particle in the Epstein regime (Eq.\ \ref{eqn:epstein}) with material density $\rho_\bullet$, we can write the friction time as

\begin{equation}\label{eqn:friction-time}
    \tf \, = \, \frac{\rho_\bullet \, R}{\rho \, \vth}
        \, = \, \frac{\rho_\bullet \, R}{\rho \, \cs} \sqrt{\frac{\pi}{8}}\,,
\end{equation}
where $\rho$ is the gas density. We commonly express $\tf$ in terms of the Keplerian frequency $\Omegak$ to obtain the Stokes number

\begin{equation*}
    \tauf \, = \, \Omegak \, \tf
          \, = \, \Omegak \, \frac{\rho_\bullet \, R}{\rho \, \cs} \sqrt{\frac{\pi}{8}}.
\end{equation*}
The Stokes number is a scale-free measure of the stopping time. Since the disk scale height is $H = \cs / \Omegak$, the Stokes number can be written as

\begin{equation}\label{eqn:tauf}
    \tauf = \frac{\rho_\bullet \, R}{\rho \, H} \sqrt{\frac{\pi}{8}}.
\end{equation}


\subsection{Size of particles in the Epstein regime}

We are interested in the formation of planetesimals at the mid-plane of the disk, where the gas density is

\begin{equation}\label{eqn:rho}
    \rho = \frac{\Sigma}{H \, \sqrt{2\pi}} \,.
\end{equation}
With Eq.\ \ref{eqn:tauf}, this gives the particle Stokes number in terms of the local surface density

\begin{equation}\label{eqn:stokes-number}
    \tauf = \frac{\rho_\bullet \, R}{\Sigma} \frac{\pi}{2}.
\end{equation}
Combined with Eq.\ \ref{eqn:hayashi}, we obtain the relation between $R$ and $\tauf$ for the MMSN

\begin{equation}\label{eqn:particle-size}
    R \sim 78 \cm \; \tauf
      \left( \frac{r}{2.5 \, \AU} \right)^{-3/2}
      \left( \frac{\rho_\bullet}{3.5 \g \cminv3} \right)^{-1}.
\end{equation}
Note that $\rho_\bullet = 3.5 \g \cminv3$ is a typical density for a mm-sized chondrule \citep{Hughes_1980}, and $r = 2.5 \AU$ lies in the inner part of the present day asteroid belt. Note also that this equation is only valid within the Epstein regime. In section \ref{sect:lambda} we show that this corresponds to $\tauf \le 1$. Our two largest particle sizes, $\tauf = 3$ and 10, lie in the Stokes regime. These particles are a few meters in size.


\subsection{Pressure support}

The gas component in a protoplanetary disk experiences a radial pressure support that partially counters the gravitational force from the central star. As a result, the gas in the disk orbits at sub-Keplerian speed. The difference between Keplerian speed $\vk$ and and gas speed $u_{\rm \phi}$ can be written as

\begin{equation}
	\Delta v \equiv \vk - u_{\rm \phi} = \eta \; \vk,
\end{equation}
where the pressure gradient parameter $\eta$ is given by

\begin{equation}
	\eta = - \frac{1}{2} \left( \frac{\cs}{\vk} \right)^2
                       \frac{\partial \ln P}{\partial \ln r},
\end{equation}
where $P$ is the gas pressure \citep{Nakagawa_1986}. In the absence of gas, the solid component of the disk would orbit at the Keplerian speed $\vk$. In the presence of gas, the solids experience a ``headwind'' from the gas component, with a wind speed of $\Delta v$. This headwind is a source of drag that can lead to radial drift. For scale-free numerical simulations, it is helpful to normalize the headwind speed by the sound speed and write

\begin{equation}\label{eqn:Delta}
  \Delta \equiv \frac{\Delta v}{\cs}
       = \eta \; \frac{\vk}{\cs}
       = - \frac{1}{2} \left( \frac{\cs}{\vk} \right)
                 \frac{\partial \ln P}{\partial \ln r}.
\end{equation}
This $\Delta$ is a free parameter that measures the strength of the pressure support and the headwind. The pressure is $P = \rho \, \cs^2$. Equation \ref{eqn:rho} gives the gas density at the midplane, which combined with $\Omegak = \cs / H$ allows us to write 

\[
    P \, = \, \frac{\Sigma}{H \, \sqrt{2\pi}} \, \cs^2
      \, = \, \frac{\Sigma \, \Omegak \, \cs}{\sqrt{2\pi}}
      \, \propto \, r^{-3} \, \cs\,.
\]
The isothermal sound speed $\cs$ is determined primarily by the local temperature

\begin{equation}\label{eqn:cs}
    \cs \, = \, \sqrt{\frac{k \, T}{\mu \, m_{\rm p}}}
        \, = \, 60 \ms \left( \frac{\mu}{2.3} \right)^{-1/2}
                       \left( \frac{T}{\K} \right)^{1/2},
\end{equation}
where $k$ is the Boltzmann constant, $T$ is the temperature, $m_{\rm p}$ is the proton mass, and $\mu$ is the mean molecular weight. In the MMSN the disk is assumed to be optically thin, so that its temperature profile is $T = 280 \K \, (r/\AU)^{-1/2}$. However, \citet{Chiang_2010} proposed a colder disk, with $T = 120 \K \, (r/\AU)^{-3/7}$. For the moment, then, we will simply write the temperature profile as

\[
    T = T_0 \, \left( \frac{r}{\AU} \right)^{-\beta}
    \;\;\Rightarrow\;\;
    \cs = 60 \ms \sqrt{T_0} \left( \frac{r}{\AU} \right)^{-\beta/2}.
\]
Therefore, the pressure $P \propto r^{-3} \, \cs$ follows a power law with exponent

\[
    \frac{\partial \ln P}{\partial \ln r}
        \, = \, - \left(3 + \frac{\beta}{2}\right).
\]
Equation\ \ref{eqn:Delta} can now be written as

\begin{equation}
    \Delta 
        \, = \, 10^{-3} \sqrt{T_0} \left(3 + \frac{\beta}{2}\right)
          \left( \frac{r}{\AU} \right)^{(1-\beta)/2}.
\end{equation}
Using the parameters from the MMSN and the Chiang-Youdin nebula (CY) we get

\begin{eqnarray}
   \Delta_{\rm MMSN} \, &=& \, 0.068 \, \left( \frac{r}{2.5 \AU} \right)^{1/4}, \\
   \Delta_{\rm CY}   \, &=& \, 0.046 \, \left( \frac{r}{2.5 \AU} \right)^{2/7}.
\end{eqnarray}
Given the uncertainty in the disk model, we choose $\Delta = 0.05$ as a typical value for $r \sim 2.5 \AU$ (present day asteroid belt). This is also the value chosen by \cite{Bai_2010b}, so that using $\Delta = 0.05$ facilitates comparison with their work. We also ran simulations with $\Delta = 0.025$ to study the effect of a reduced pressure support. This can occur as part of a large-scale pressure bump, which could form around the snow line \citep{Kretke_2007}.


\subsection{Epstein vs Stokes regimes}\label{sect:lambda}

Particles with $R < 9\lambda/4$ experience Epstein drag, where the mean free path $\lambda$ is given by

\[
    \lambda \, = \, \frac{1}{n \, \sigma},
\]
where $\sigma = 2\times 10^{-15} \cm^2$ is the collision cross section of an H$_2$ molecule, and $n$ is the particle number density, which is

\[
    n \, = \, \frac{\rho}{\mu \, m_{\rm p}}.
\]
Using Eq.\ \ref{eqn:rho}, along with $H = \cs/\Omegak$ and Eq.\ \ref{eqn:cs}, one can compute the mean free path. For the MMSN, the mean free path is $\lambda \sim 43.6\cm$ at 2.5 AU. That means that particles smaller than $R \sim 98\cm$ lie in the Epstein regime. From Eq.\ \ref{eqn:particle-size} we find that all our particles with $\tauf \le 1$ lie in the Epstein regime.


\subsection{Simulation setup}

In our simulations we use $\cs \equiv \Omegak \equiv 1$ as a natural scaling unit for the problem. The free parameters for the problem are $\Delta$ and $\tauf$ given by Eqs.\ \ref{eqn:stokes-number} and \ref{eqn:Delta}. We use the Pencil Code\footnote{https://code.google.com/p/pencil-code/} to simulate the dynamics of gas and solid particles in a 2-dimensional vertical slice of the protoplanetary disk \citep{Youdin_2007}. We model a box of dimensions $0.2 H \, \times \, 0.2 H$ in the radial-vertical plane centered on the midplane of the disk. The box has periodic boundary conditions. Our baseline simulations
have a $128 \times 128$ grid resolution with 16,384 ``super-particles'' representing the solids. Each super-particle represents a large number of physical particles. We run a total of 102 baseline simulations as we vary the key simulation parameters:

\begin{itemize}
\item We test seventeen particle sizes spaced equally in log scale from $\tauf = 0.001$ to $\tauf = 10$.

\item We test both the canonical pressure support $\Delta = 0.05$, and a reduced pressure support $\Delta = 0.025$.

\item We run each ($\tauf$, $\Delta$) simulation three times, with different initial particle positions. The particle positions are chosen from a random distribution, uniform within the box.
\end{itemize}

Every simulation begins with a solid fraction of $Z = 0.005$ (Eq.\ \ref{eqn:Z}). First, we allow the solids to sediment to the midplane. In most cases, the sedimentation is complete within 50 orbits. For the $\tauf < 0.003$ simulations we allowed 250 orbits for the sedimentation phase. Once the particles reach an equilibrum scale height, we begin the numerical experiment: We reduce the gas density $\rho$ exponentially, so that it halves every 50 orbits. We keep $\tauf$ constant, which is equivalent to reducing the particle size $R$ at the same rate as $\rho$ (see Eq.\ \ref{eqn:tauf}). This procedure is also equivalent to increasing the particle density $\rhop$ while keeping $\rho$ and $R$ fixed. The objective of the experiment is to determine the value $Z$ needed to produce particle clumps for a given $\tauf$.

\subsection{System response time}
\label{sec:response-time}
It is desirable that the rate of gas removal be slower than the response time of the system. Otherwise our procedure will tend to over-estimate the $Z$ value needed to produce particle clumps. The response time of the system is in the order of the particle crossing time $t_{\rm cross}$, which is 

\begin{equation}\label{eqn:t-cross}
        t_{\rm cross} = \frac{0.2 H}{|\vr|},
\end{equation}
where $0.2 H$ is the width of the box and $\vr$ is the particle radial velocity, which is given by

\begin{equation}\label{eqn:radial-speed}
    \vr = \frac{- 2 \eta \vk}{\tauf + \tauf^{-1}}
        = \frac{- 2 \cs \Delta}{\tauf + \tauf^{-1}},
\end{equation}
\begin{equation}\label{eqn:t-cross-2}
    \Rightarrow
    t_{\rm cross} = \frac{0.1 H}{\cs \Delta}(\tauf + \tauf^{-1}).
\end{equation}
With this one can show that $t_{\rm cross} < 50$ orbits $\Leftrightarrow \tauf + \tauf^{-1} < 50\pi$, for $\Delta = 0.05$. This roughly corresponds to the interval $0.0064 < \tauf < 157$. This means that for most of our simulations the system response time is less than 50 orbits, but for $\tauf < 0.0064$ we will overestimate the $Z$ needed for particle clumps. For $\tauf < 0.0064$, our results should be considered robust upper limits. This is a necessary limitation because these simulations also suffer from small time steps which force long computation times.


\section{Results}\label{sect:results}

\begin{figure}[hb!]
    \centering
    \includegraphics[width=0.5\textwidth]{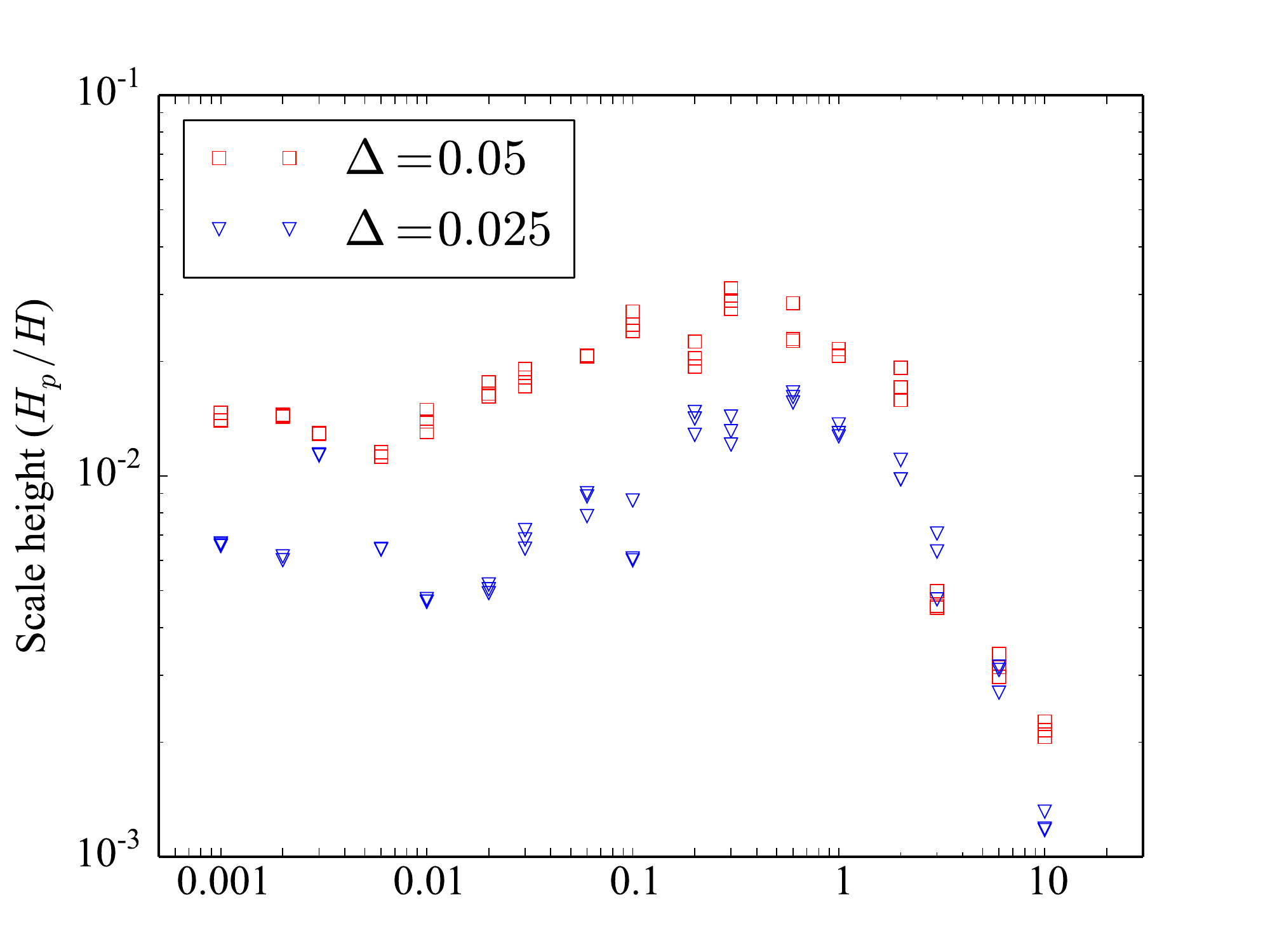} \\
    \includegraphics[width=0.5\textwidth]{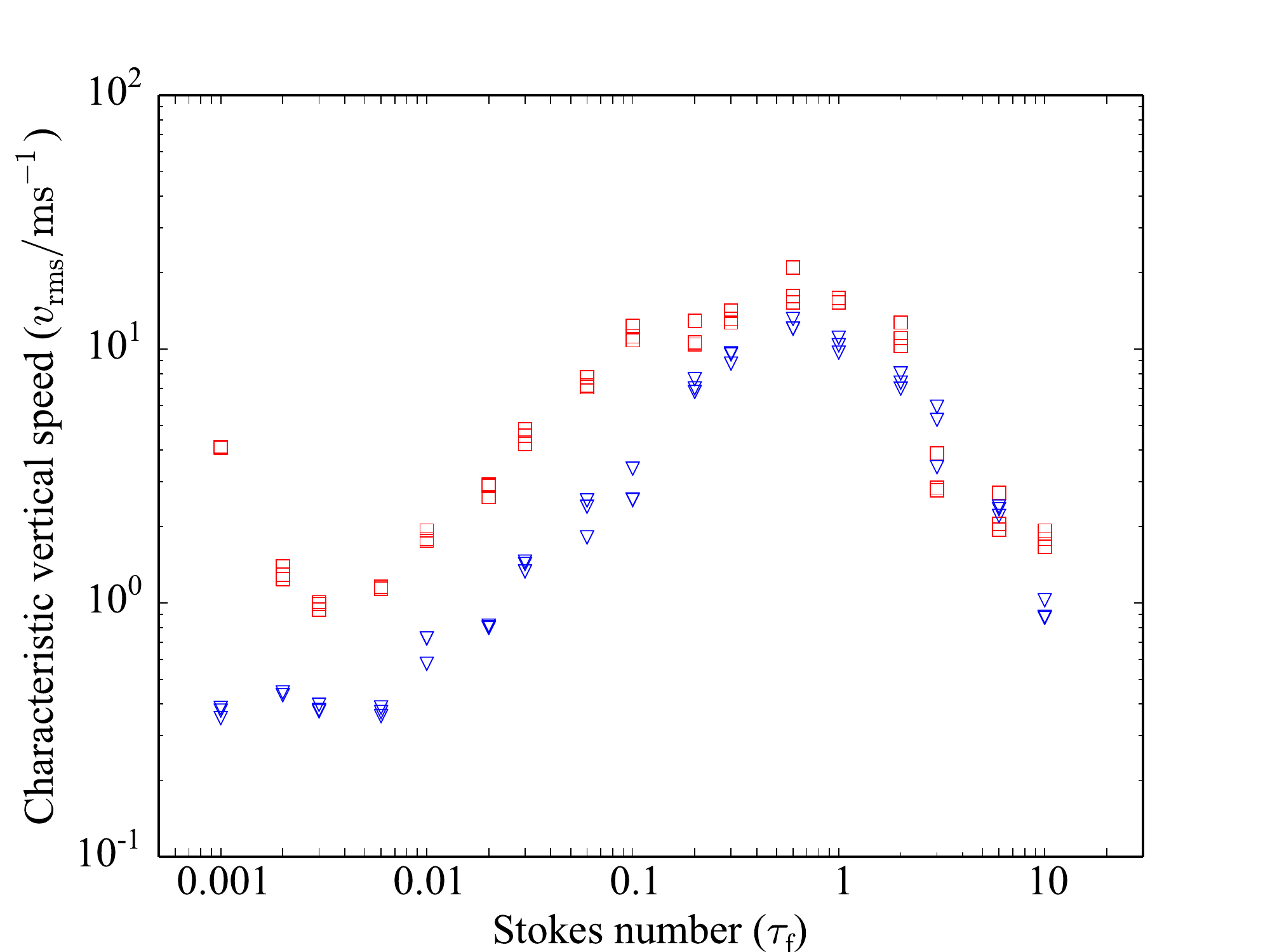}
    
    \caption{Particle scale height $\Hp$ (top; see Eq.\ \ref{eqn:Hp}), and root-mean-square of the vertical component of the particle speed $v_{\rm rms}$ (bottom; see Eq.\ \ref{eqn:vrms}) at the end of the sedimentation phase. The sound speed is set to $\cs = 800 \m \sinv$. Half of the simulations (red squares) have a standard pressure support ($\Delta = 0.05$) and the other half (blue triangles) have a reduced pressure support ($\Delta = 0.025$). For particles smaller than $\tauf \sim 0.3$, lower $\Delta$ results in stronger sedimentation (lower $\Hp$).}
    \label{fig:scale-height}
\end{figure}

A reduced pressure support ($\Delta = 0.025$) is associated with slower radial drift, lower gas turbulence, and a correspondingly higher concentration of solids at the midplane. We can quantitatively measure the scale height of particles ($\Hp$) and the degree of turbulence using the root-mean-square of the particle positions ($z$ coordinate) and vertical speed,

\begin{eqnarray}
    \label{eqn:Hp}
    \Hp &=& \sqrt{\langle  z^2 \rangle}
         =  \sqrt{\frac{1}{n} \sum_i z_i^2} ,\\
    \label{eqn:vrms}
    v_{\rm rms}
        &=& \sqrt{\langle  v_z^2 \rangle}
         =  \sqrt{\frac{1}{n} \sum_i v_{z,i}^2} ,
\end{eqnarray}
where $z_i$ is the $z$ coordinate of the $i^{\rm th}$ particle, and $v_{z,i}$ is the vertical component of the particle's velocity. Figure \ref{fig:scale-height} shows $\Hp$ and $v_{\rm rms}$ across the entire range of simulations. For particles smaller than $\tauf \sim 0.3$, the lower $\Delta$ is associated with both a lower scale height and lower vertical motion. Particles larger than $\tauf \sim 0.3$ are poorly coupled with the gas, and are largely unaffected by $\Delta$. The effect of particle size and concentration is more complex.
\vspace{0.3cm}

\newcolumntype{E}{ >{\centering\arraybackslash} b{8.5cm} }
\newcolumntype{F}{ >{\centering\arraybackslash} b{9.0cm} }
\begin{figure*}[hb!]
    \centering
    \includegraphics[width=0.49\textwidth]{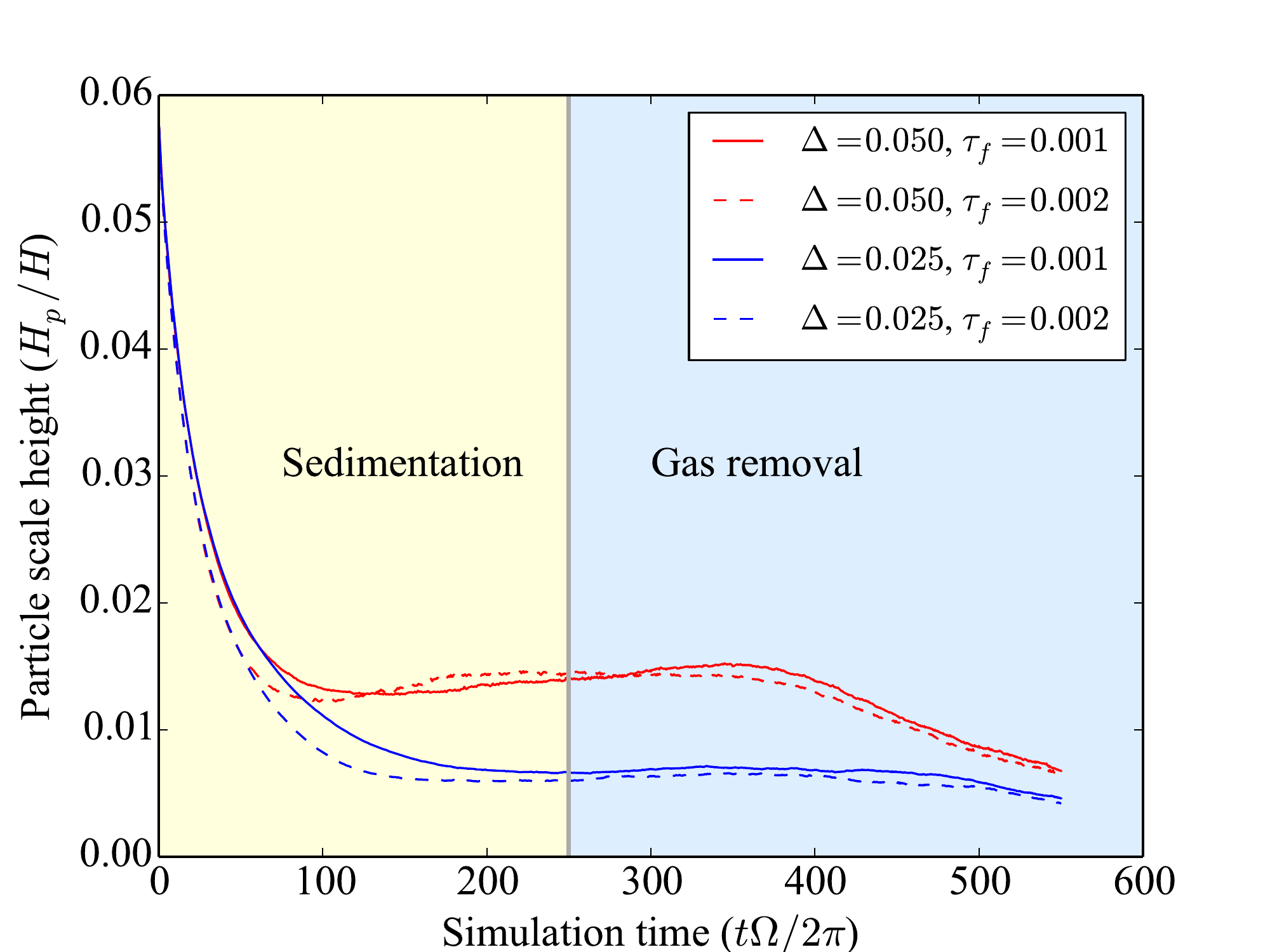}    \includegraphics[width=0.49\textwidth]{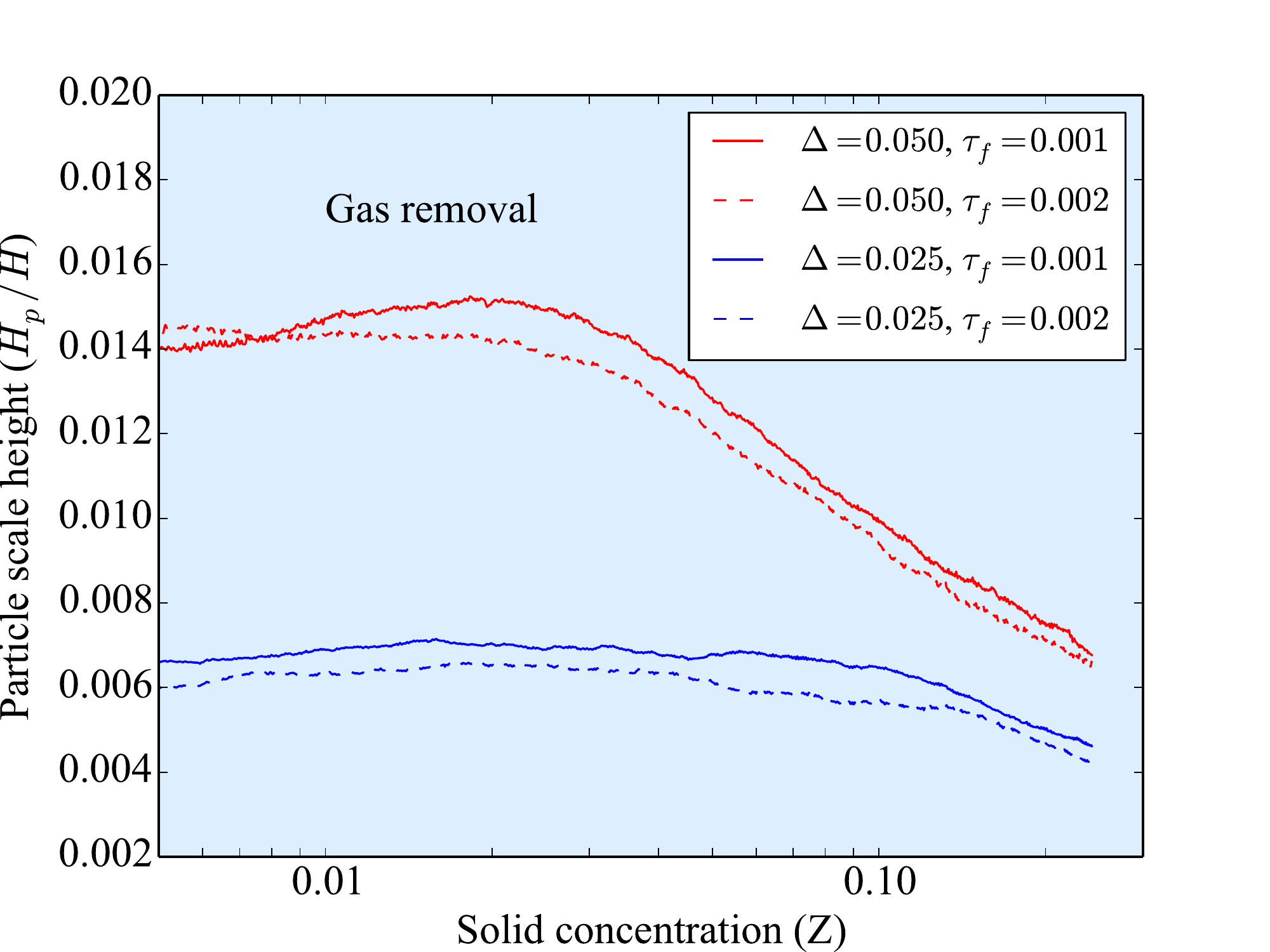}
    
    \caption{Particle scale height $\Hp$ for small particles (size $\tauf \le 0.002$, $R \le 1.5$mm) over the course of the simulation. The left figure shows the entire simulation, including the sedimentation phase. After 250 orbits, $\Hp$ has reached a steady state and the numerical experiment begins. The right figure shows only the gas removal phase. Gas is removed exponentially. In the right plot, the simulation time is replaced by the solid concentration $Z = \Sigma_{\rm solid} / \Sigma_{\rm total}$. For particles in this size scale, the scale height is determined primarily by $\Delta$, and secondly by $Z$.}
    \label{fig:suspension-Hp}
\end{figure*}

We found it helpful to divide the particles into three different size bins, based on their qualitative behavior: (1) Particles smaller than $\tauf \sim 0.003$ form a suspension, where particles cannot decouple from the gas even at high concentration and particle clumps do not occur. (2) Between $\tauf \sim 0.003$ and $\tauf \sim 0.3$ is the optimal range for particles to engage in the streaming instability. (3) Particles larger than $\tauf \sim 0.3$ suffer strong radial drift, and the over-densities that do form are quickly destroyed by gas erosion.


\subsection{Suspension regime ($\tauf < 0.003$)}

Our smallest particles have $\tauf = 10^{-3}$ and $\tauf = 10^{-2.75} \approx 0.002$. These particles do not sediment easily because they are strongly coupled to the gas. Therefore, we wait 250 orbits for the solids to sediment before we begin the experiment. Figure \ref{fig:suspension-Hp} shows the evolution of $\Hp$, including the sedimentation phase. The long sedimentation phase ensures that $\Hp$ has converged before we start removing gas.

\begin{figure*}[t!]
  \centering
  \includegraphics[width=0.49\textwidth]{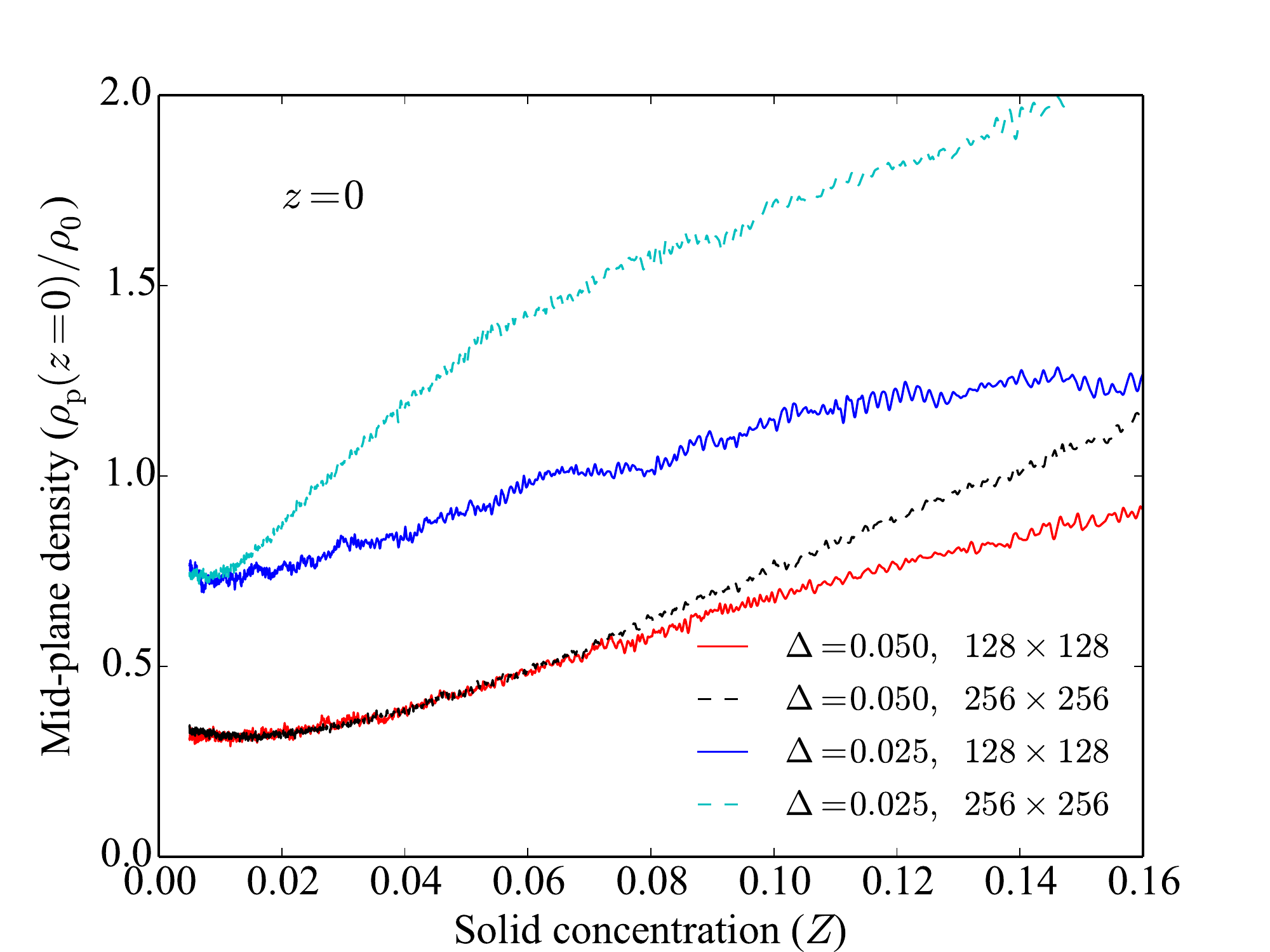} \includegraphics[width=0.49\textwidth]{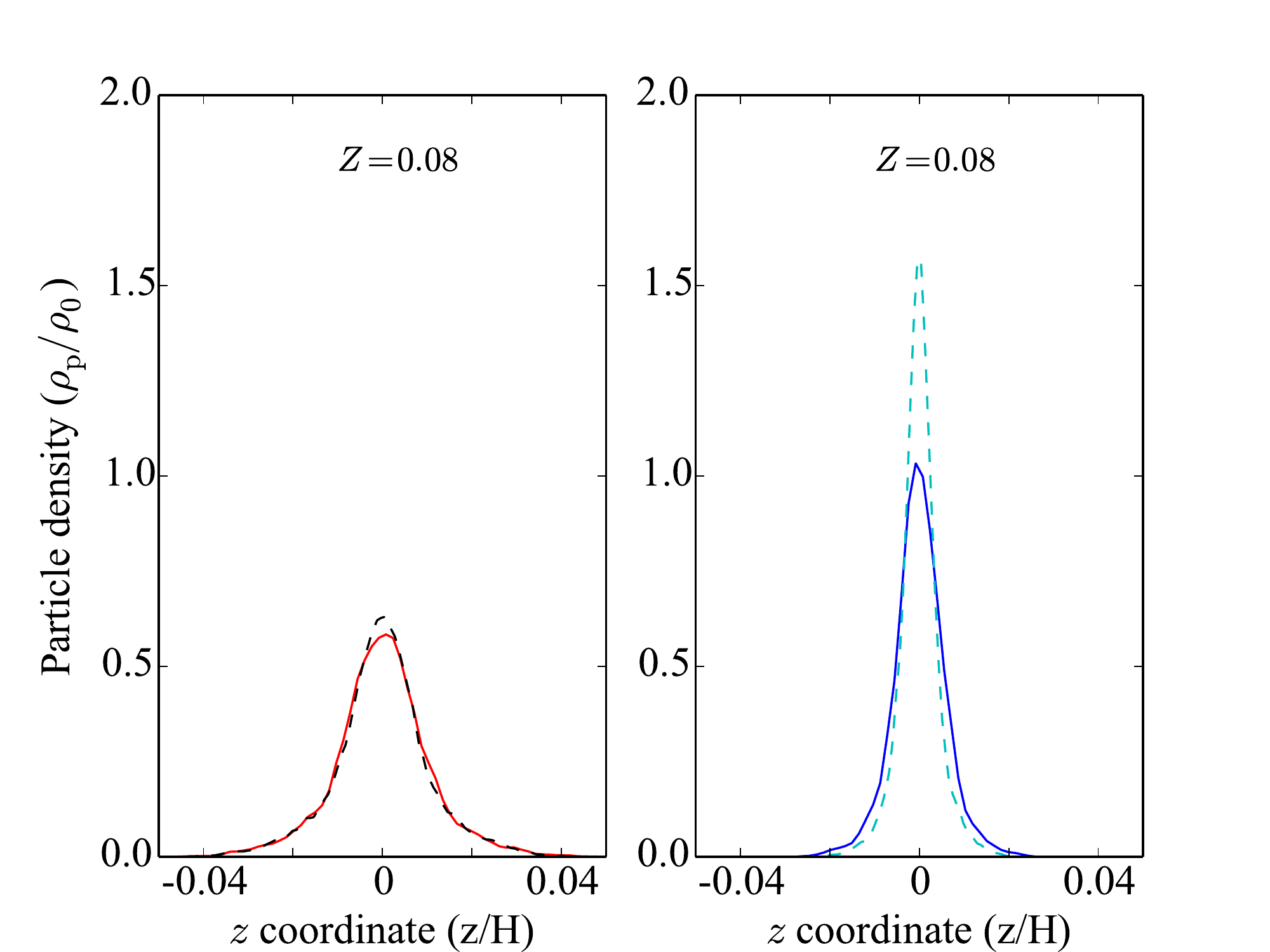}
  
  \caption{Particle density $\rhop$ as a function of solid concentration (left) and height (right). The solid concentration is $Z = \Sigma_{\rm solid} / \Sigma_{\rm total}$. The top figure shows $\rhop$ at the midplane (height $z = 0$) as a function of $Z$. The bottom figure shows $\rhop$ at $Z = 0.08$ as a function of the $z$ coordinate. The particle density is measured in terms of the initial gas density $\rho_0$. All the simulations have particle size $\tauf = 0.002$.
Two runs have a standard pressure support ($\Delta = 0.05$) and two have a reduced pressure support ($\Delta = 0.025$). Two runs have $128 \times 128$ grid resolution and two are $256 \times 256$. There is no evidence of the density cusp predicted by \citet{Sekiya_1998} and \citet{Youdin_2002}. The $\Delta = 0.05$ and $\Delta = 0.025$ runs should have clear cusps after $Z \sim 0.10$ and $Z \sim 0.02$ respectively \citep{Youdin_2002}. These simulations may have an additional source of mixing that inhibits the formation of the Sekiya-Youdin-Shu cusp. The result is the same for smaller, $\tauf = 0.001$ particles (not shown).
  }
  \label{fig:suspension-midplane}
\end{figure*}

Figure \ref{fig:suspension-Hp} shows that, for $\tauf < 0.003$, the particle size has very little effect on $\Hp$. Instead, $\Hp$ is determined by $\Delta$. For $\Delta = 0.05$, the particle concentration is also important for $Z > 0.04$. Some authors have proposed that, for well-coupled particles such as these, there is a critical $Z$ beyond which turbulent mixing becomes ineffective, resulting in a high-density ``cusp'' at the midplane \citep{Sekiya_1998,Youdin_2002}. The critical value of $Z$ depends on the disk conditions, and range from $Z \sim 0.02$ for a cold disk, and $Z \sim 0.10$ for a MMSN \citep{Youdin_2002}. To test
this idea, we ran $256 \times 256$ simulations with $\tauf = 10^{-2.75}$. The particle density can be written as

\begin{equation}\label{eqn:rhopz}
    \rhop(z) \, = \, \frac{\Sigma_{\rm solid}}{\Hp \sqrt{2\pi}}
    \exp \left( \frac{-z^2}{2\Hp^2} \right),
\end{equation}
where $z$ is the vertical coordinate. Figure \ref{fig:suspension-midplane} shows the midplane particle density $\rhop(z=0)$ as a function of $Z$, and a snapshot of $\rhop(z)$ at $Z = 0.08$. Although we find an increase in $\rhop(z=0)$ with $Z$, it is not nearly of the magnitude proposed by \citet{Youdin_2002}. We suspect that our simulations have an additional source of mixing that interferes with the formation of the Sekiya-Youdin-Shu cusp. For $r = 1 \AU$, our
$128 \times 128$ grid cells have a physical size around $5,000 - 8,000 \km$. Both the $128 \times 128$ and $256 \times 256$ runs should be able to resolve the cusp seen in Fig.\ 1 of \citet{Youdin_2002}. On the other hand, the discrepancy between the $256 \times 256$ runs and the $128 \times 128$ runs indicates that our results have not converged, and higher resolutions may reveal higher densities.

\subsection{Streaming regime ($0.003 \leq \tauf \leq 0.3$)}

\begin{figure*}
  \begin{tabular}{ABCCBA}
  \multicolumn{4}{r}{
      \includegraphics[width=10cm]{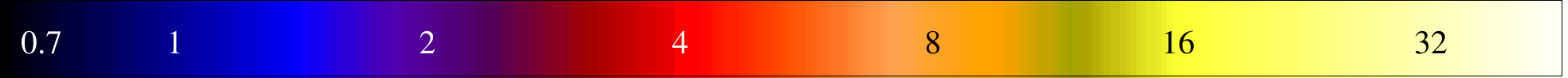}
  } &
  \multicolumn{2}{l}{
      $\Sigma_{\rm solid} / \langle \Sigma_{\rm solid} \rangle$
  } \\
  &\hspace{1cm}$\tauf = 0.001$ & $\tauf = 0.003$ & $\tauf = 0.01$ &\hspace{-1cm}$\tauf = 0.03$ & \\
  \begin{sideways} Simulation time ($t/2\pi\Omega^{-1}$) for $\Delta = 0.05$ \end{sideways}\vspace{3.0cm} &
  \includegraphics[width=4.00cm]{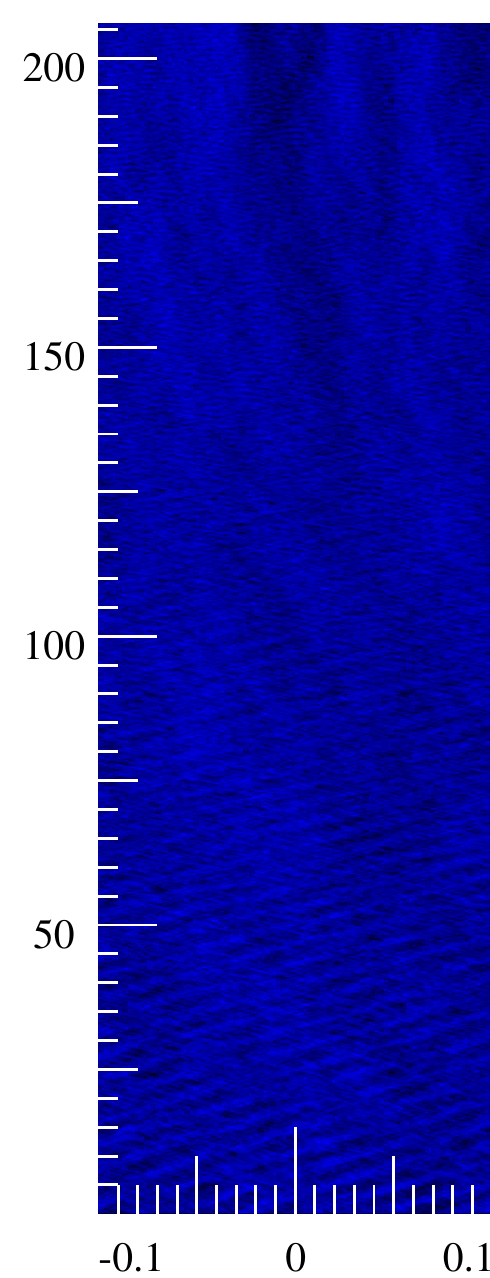} &
  \includegraphics[width=3.20cm]{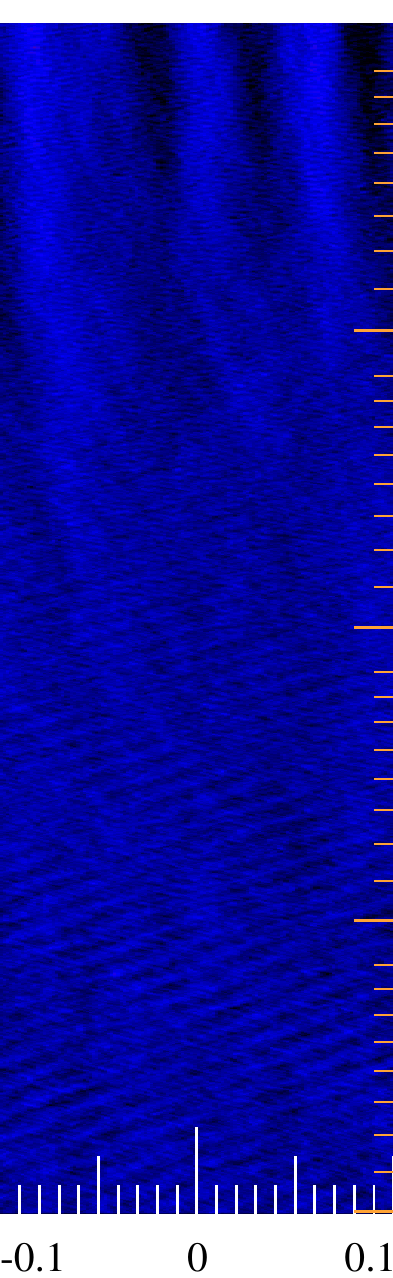} &
  \includegraphics[width=3.20cm]{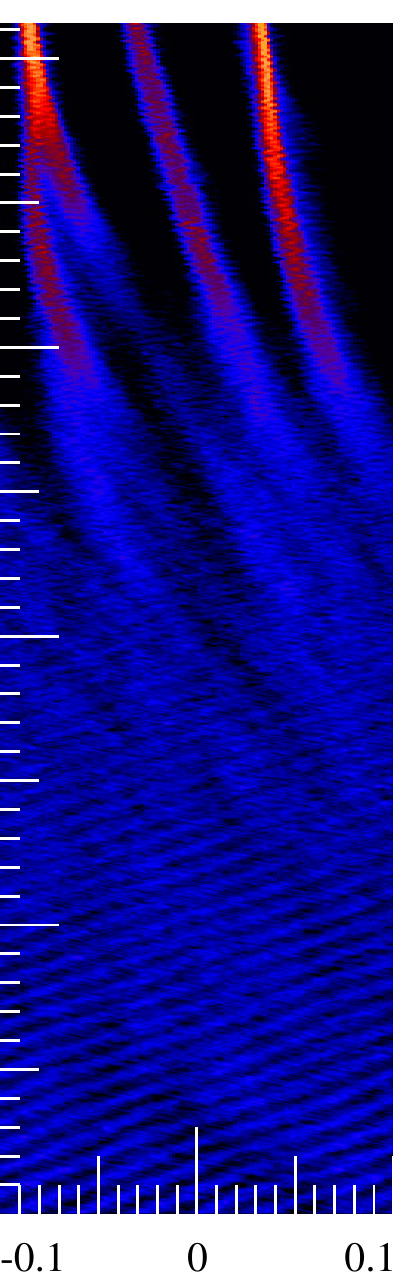} &
  \includegraphics[width=4.00cm]{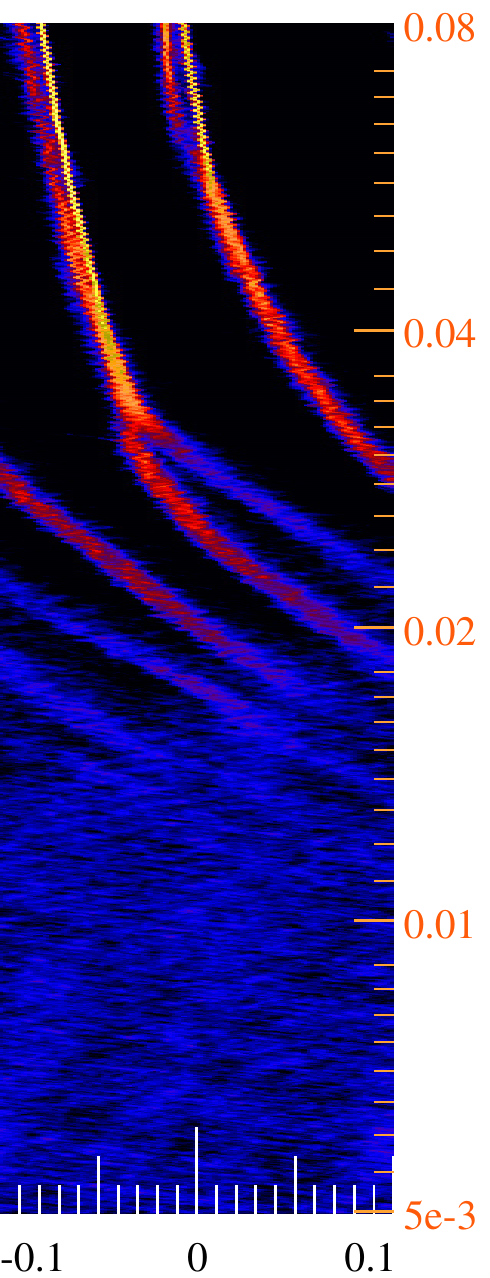} &
  \begin{sideways} Solid concentration ($Z$) \end{sideways}\vspace{4.0cm} \\
  \begin{sideways} Simulation time ($t/2\pi\Omega^{-1}$) for $\Delta = 0.025$ \end{sideways}\vspace{3.0cm} &
  \includegraphics[width=4.00cm]{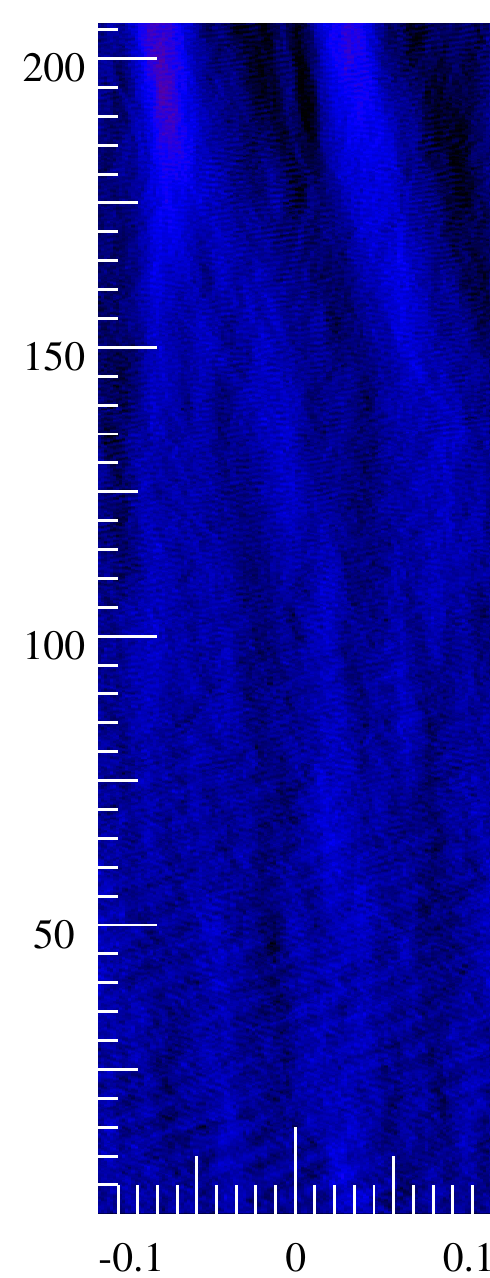} &
  \includegraphics[width=3.20cm]{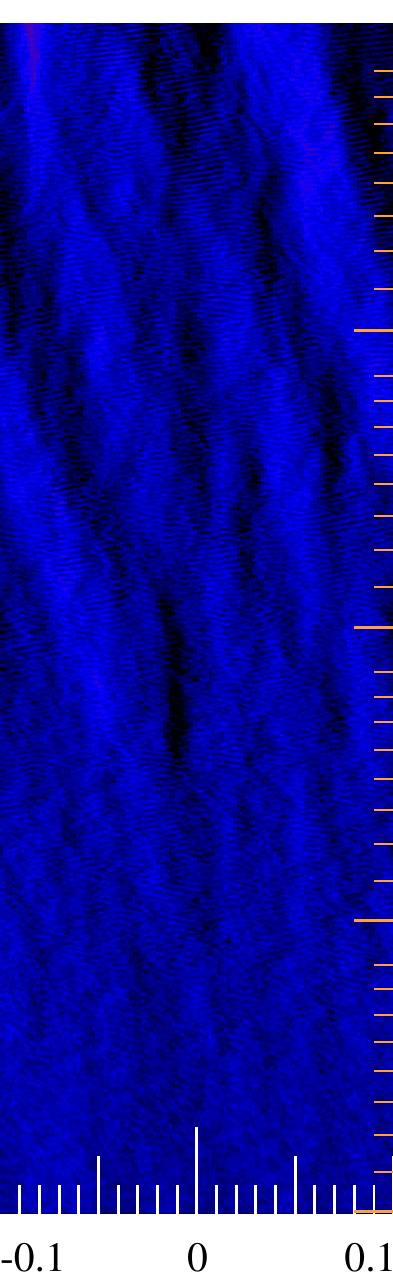} &
  \includegraphics[width=3.20cm]{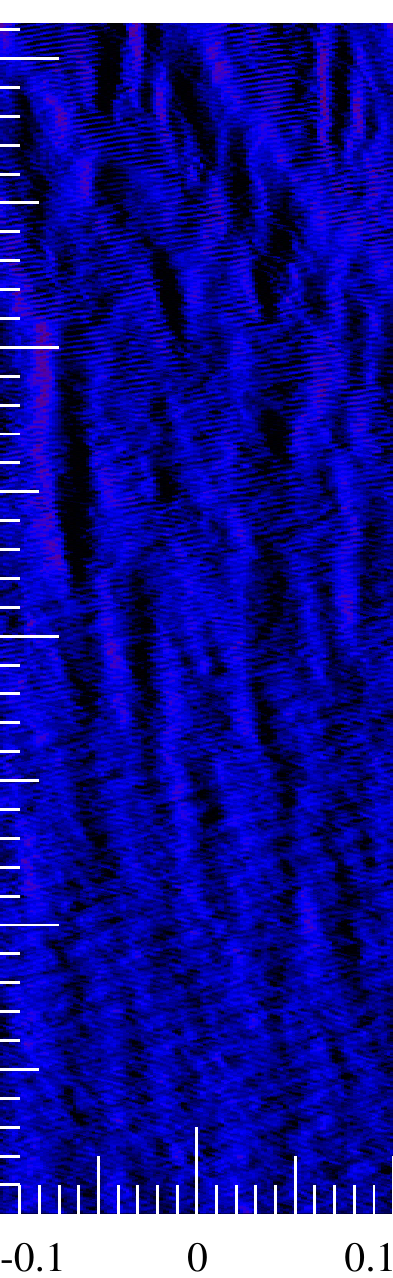} &
  \includegraphics[width=4.00cm]{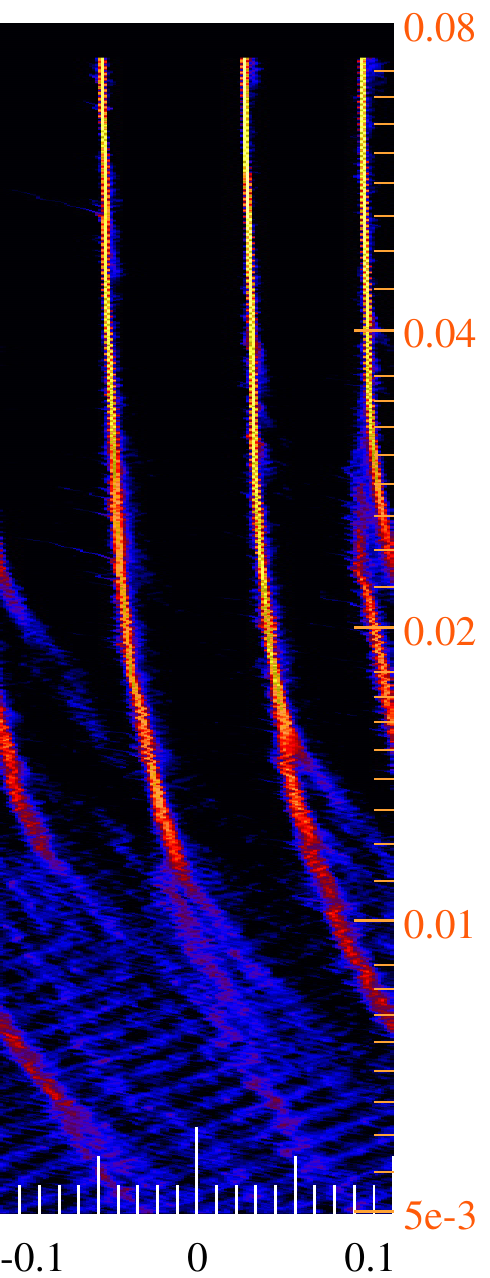} &
  \begin{sideways} Solid concentration ($Z$) \end{sideways}\vspace{4.0cm} \\
  \multicolumn{6}{c}{Radial coordinate ($x/H$)}
  \end{tabular}
  
  \caption{Spacetime diagrams showing the solid surface density $\Sigma_{\rm solid}$ as a function of the radial coordinate $x$ and simulation time. The right hand axis shows the mean solid concentration $Z = \langle \Sigma_{\rm solid} \rangle / \langle \Sigma_{\rm total} \rangle$. The figure shows selected runs with $\tauf \le 0.03$. The top row has runs with standard pressure support ($\Delta = 0.05$) and the bottom row has runs with a reduced pressure support ($\Delta = 0.025$). The surface density is shown by a color   scale. Some of the runs form visible filaments. For $\Delta = 0.05$ there is a clear trend where lower $\tauf$ requires higher $Z$ to form filaments. For
$\Delta = 0.025$ the behaviour is less predictable -- while the particle concentration is higher, the low radial drift speed interferes with the streaming instability.
  }
  \label{fig:spacetime-1}
\end{figure*}

\begin{figure*}
  \begin{tabular}{ABCCBA}
  \multicolumn{4}{r}{
      \includegraphics[width=10cm]{plots/idl/spacetime/colorbar.pdf}
  } &
  \multicolumn{2}{l}{
      $\Sigma_{\rm solid} / \langle \Sigma_{\rm solid} \rangle$
  } \\
  &\hspace{1cm}$\tauf = 0.1$ & $\tauf = 0.3$ & $\tauf = 1$ &\hspace{-1cm}$\tauf = 3$ & \\
  \begin{sideways} Simulation time ($t/2\pi\Omega^{-1}$) for $\Delta = 0.05$ \end{sideways}\vspace{3.0cm} &
  \includegraphics[width=4.00cm]{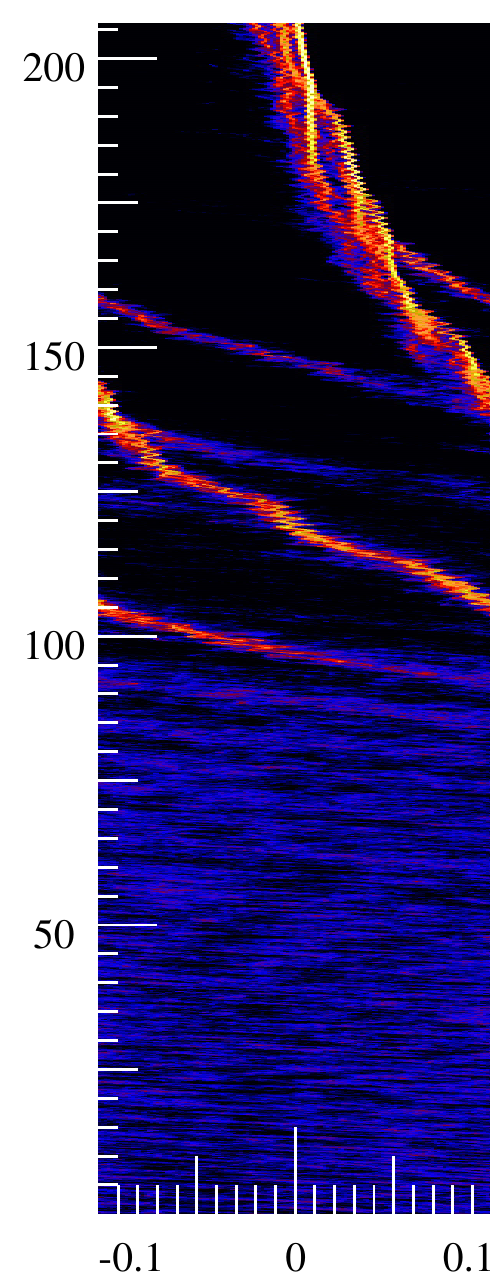} &
  \includegraphics[width=3.20cm]{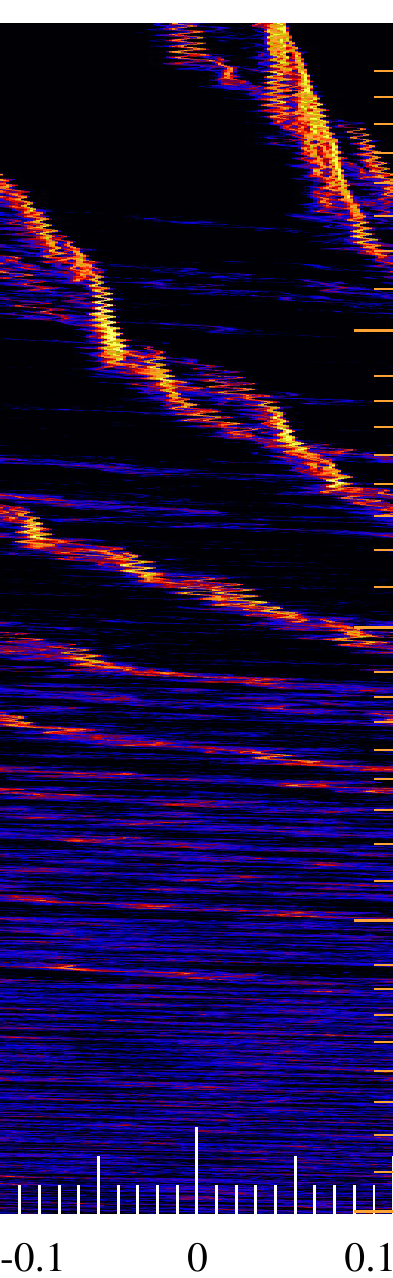} &
  \includegraphics[width=3.20cm]{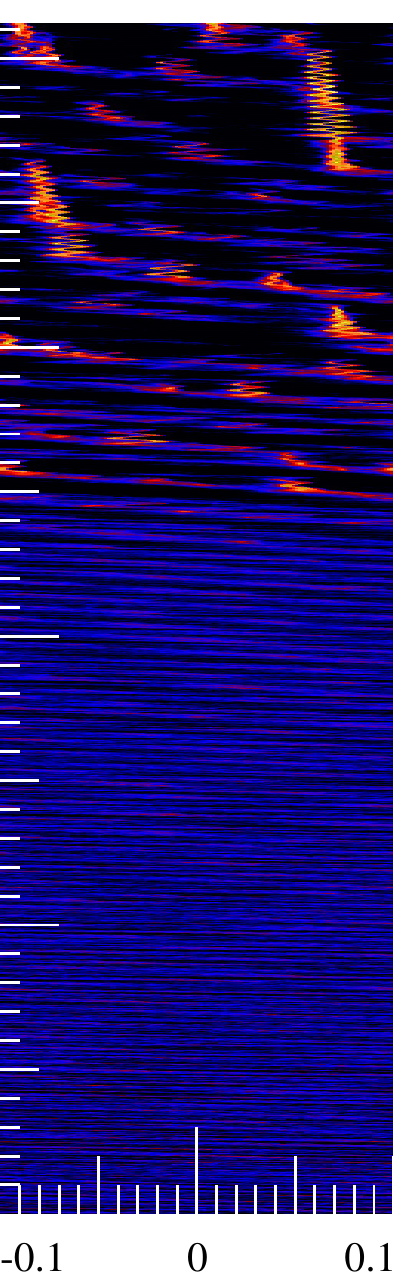} &
  \includegraphics[width=4.00cm]{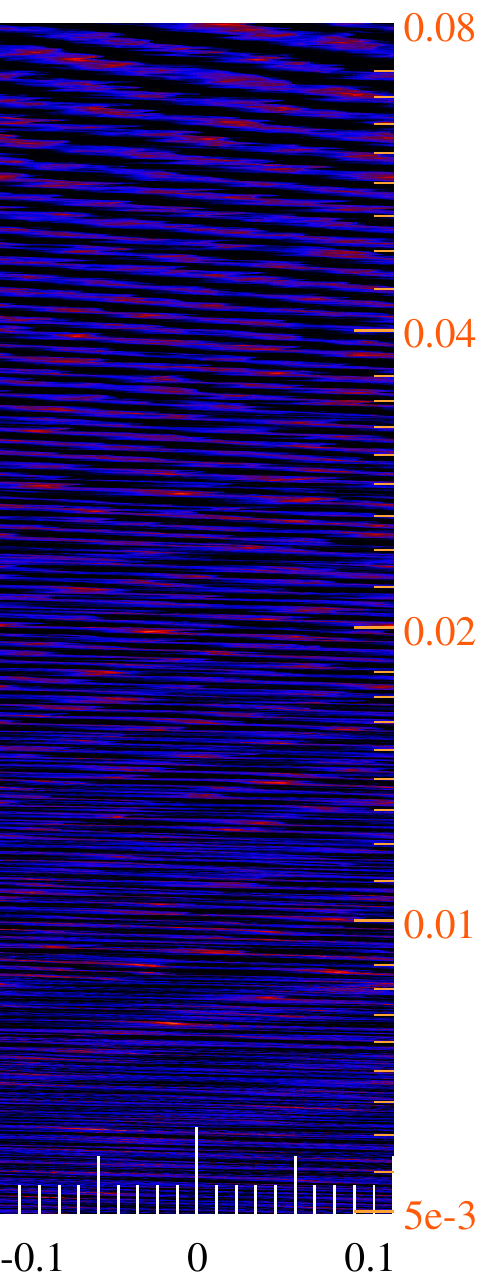} &
  \begin{sideways} Solid concentration ($Z$) \end{sideways}\vspace{4.0cm} \\
  \begin{sideways} Simulation time ($t/2\pi\Omega^{-1}$) for $\Delta = 0.025$ \end{sideways}\vspace{3.0cm} &
  \includegraphics[width=4.00cm]{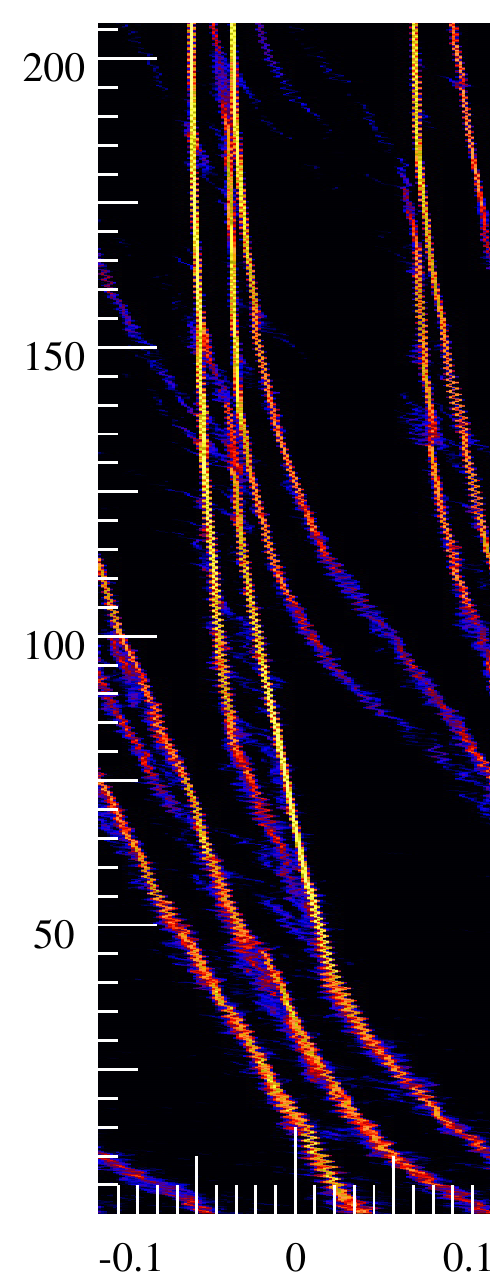} &
  \includegraphics[width=3.20cm]{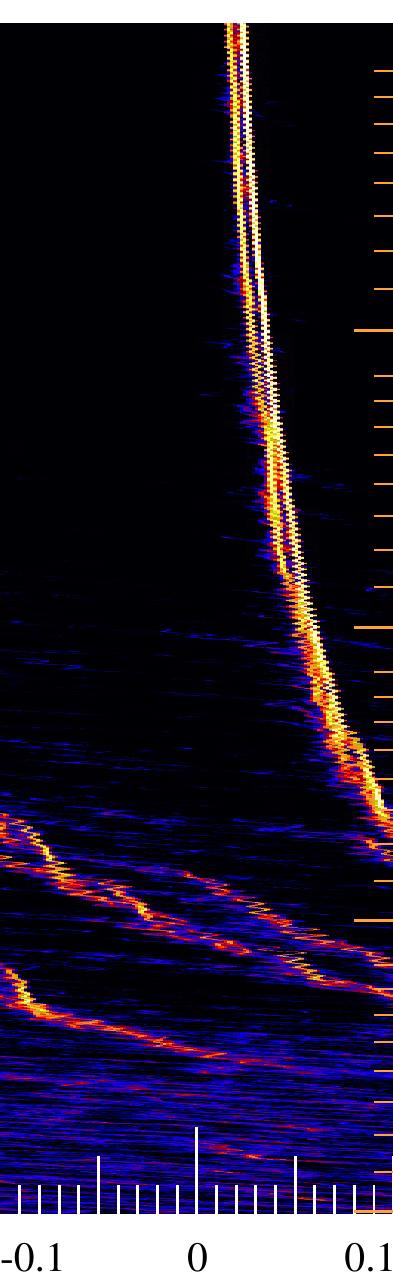} &
  \includegraphics[width=3.20cm]{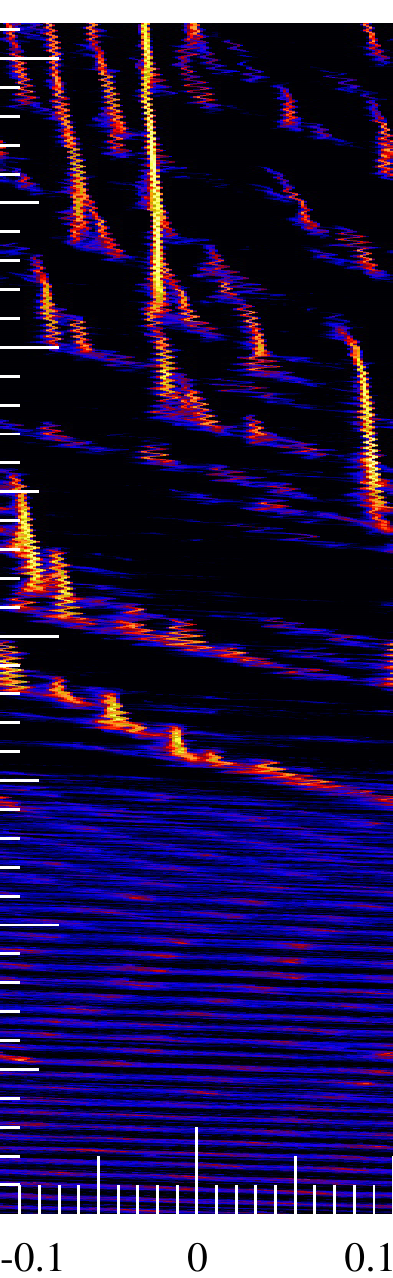} &
  \includegraphics[width=4.00cm]{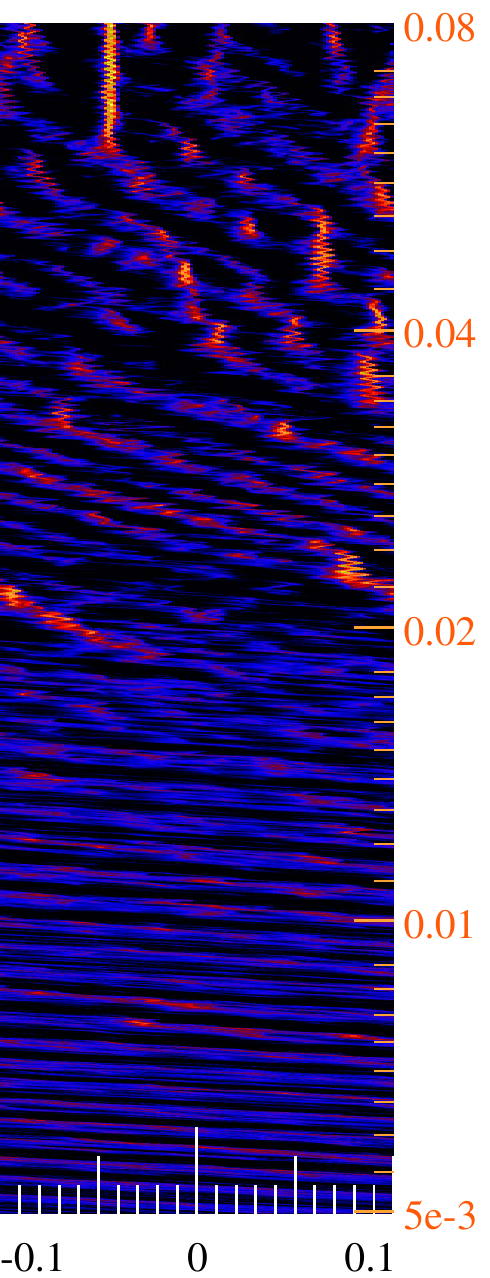} &
  \begin{sideways} Solid concentration ($Z$) \end{sideways}\vspace{4.0cm} \\
  \multicolumn{6}{c}{Radial coordinate ($x/H$)} \\
  \end{tabular}
  
  \caption{Same as Fig.\ \ref{fig:spacetime-1}, but for larger particle sizes ($\tauf = 0.1$ to 3). For $\tauf = 0.1$ and $\Delta = 0.025$ the simulation cannot resolve the minimum solid concentration ($Z = \Sigma_{\rm solid} / \Sigma_{\rm total}$) that leads to clumping. All simulations begin at $Z = 0.005$ and that is sufficient to produce visible clumps for that $\tauf$ and $\Delta$. Also, for $\tauf = 1$ and $\Delta = 0.05$, the filaments that form are short lived.}
  \label{fig:spacetime-2}
\end{figure*}

After $\tauf \sim 0.003$, solid particles begin to decouple from the gas, leading to a regime where the radial drift is sufficient to make the streaming instability effective. Figures \ref{fig:spacetime-1} and \ref{fig:spacetime-2} show spacetime diagrams for a selection of runs across all regimes, from $\tauf = 10^{-3}$ to $\tauf = 3$. For a standard pressure gradient ($\Delta = 0.05$), there are visible, long-lived filaments for $0.003 \le \tauf \le 0.3$. For $\tauf = 1$ the filaments are short lived. Somewhere between
$\tauf = 0.3$ and $\tauf = 1$ particle clumps start to become difficult again.

When $\Delta = 0.05$, the smallest particles that form visible filaments are have $\tauf = 0.003$. The pile-up process is somewhat analogous to a traffic jam: An initial overdensity of solids has a stronger back-reaction on the gas. This reduced $\Delta$, lowers the radial drift, leading to a pile-up as other solids (with faster radial drift) enter the filament and increase the density even further. Somewhere between $\tauf = 0.03$ and 0.1, clumping starts to become difficult again, as the smaller solid cross section (fewer particles, higher mass per particle) cannot lower $\Delta$ as effectively. After around $\tauf \gtrsim 1$, traffic jams are no longer easily observable.


\newcolumntype{A}{ >{\centering\arraybackslash} b{0.30cm}@{\hspace{0.2cm}} }
\newcolumntype{B}{ >{\centering\arraybackslash} p{4.00cm}@{\hspace{0.2cm}} }
\newcolumntype{C}{ >{\centering\arraybackslash} p{3.20cm}@{\hspace{0.2cm}} }

\subsection{Radial drift regime ($\tauf > 0.3$)}

\begin{figure}[h!]
    \centering
    \includegraphics[width=0.5\textwidth]{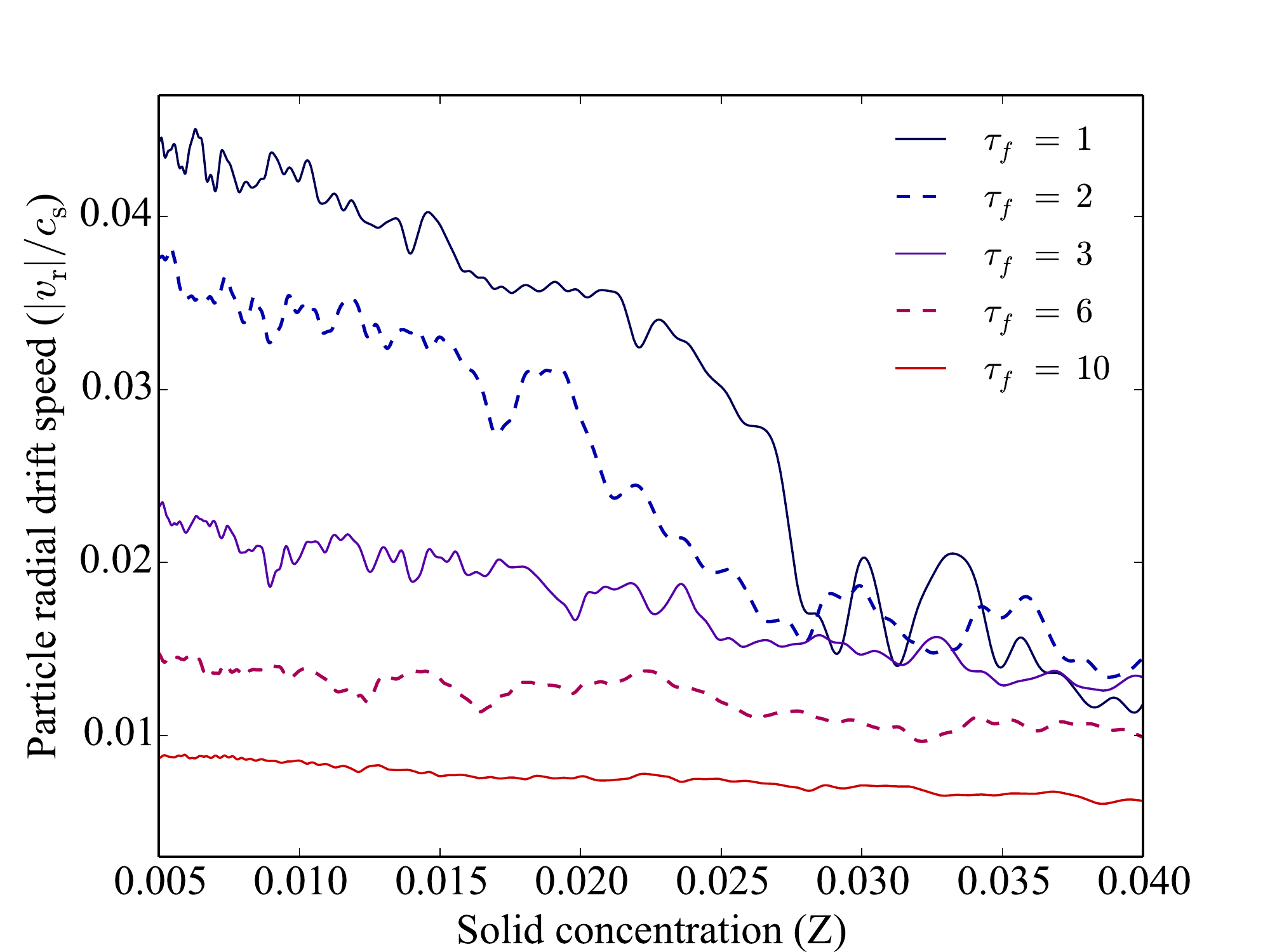}
    
    \caption{Radial drift speed $\vr$ of particles as a function of the particle size $\tauf$ and solid concentration $Z = \Sigma_{\rm solid} / \Sigma_{\rm total}$. Speed is measured in terms of the sound speed $\cs$. All solids experience gas drag. Particles of size $\tauf \sim 1$ experience very rapid drift, which decreases for larger particles. In addition, higher $Z$ forces the gas component to orbit closer to the Keplerian speed, reducing $|\vr|$. However, in all cases, the drift speed remains high. For reference, for $|\vr| = 0.005 \cs$ the particles cross the disk after a few thousand years.}
    \label{fig:radial-speed}
\end{figure}

The regime with $\tauf > 0.3$ is not a good environment for asteroid formation. Although particles in this regime do clump, they continue to rapidly drift toward the star, and the clumps that form are very transient as they are easily eroded by wind in the disk. Figure \ref{fig:radial-speed} shows the radial speed for various particle sizes and concentrations. Higher concentrations of solids are associated with a decrease in radial speed, but even at high $Z$ the radial speed remains high. In this regime, larger particles are more able to resist gas drag and exhibit lower radial drift. However, even at $\tauf \sim 10$, the radial drift is as high as $|\vr| \sim 0.01 \cs$. At that speed, these particles would cross the disk in around a thousand years.

Als note that the overdensities that do form in this regime are very short lived, as they are easily destroyed by erosion from the gas component. But in any case, it seems unlikely that this regime is at all important for realistic disks. As noted earlier, the ``particles'' in this regime are large. A $\tauf = 10$ particle is actually a 5-meter boulder. The inefficiency of sticking \citep{Guttler_2010}, and the streaming regime, probably present formidable barriers for any solids to reach the radial drift regime.


\subsection{Conditions for particle streaming}

To locate particle clumps we look for stable peaks in the solid surface density $\Sigma_{\rm solid}$. The linear stability analysis of \citet{Youdin_2005} shows that the streaming instability grows only in modes which have both a radial and a vertical variation. The non-linear simulations of \citet{Johansen_2007b} also show significant vertical variation in the particle density. However, simulations which include the vertical gravity on the particles are stratified; they have a dense midplane layer and little additional variation in the vertical direction. For this reason, the vertical average captures very well the magnitude of particle overdensities in the stratified simulations that we present here.

\begin{figure}[ht!]
  \centering
  \includegraphics[width=0.44\textwidth]{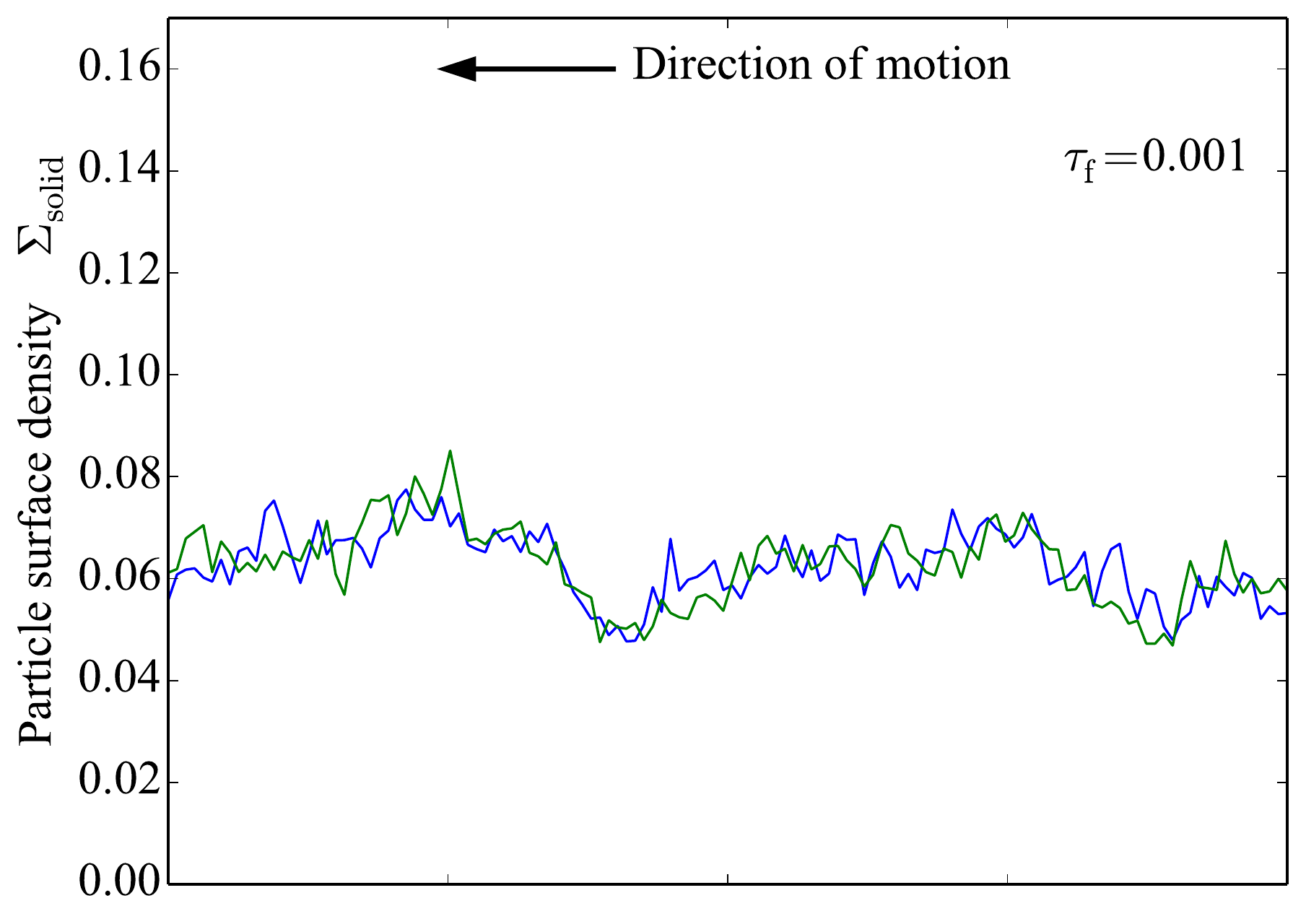}
  \includegraphics[width=0.44\textwidth]{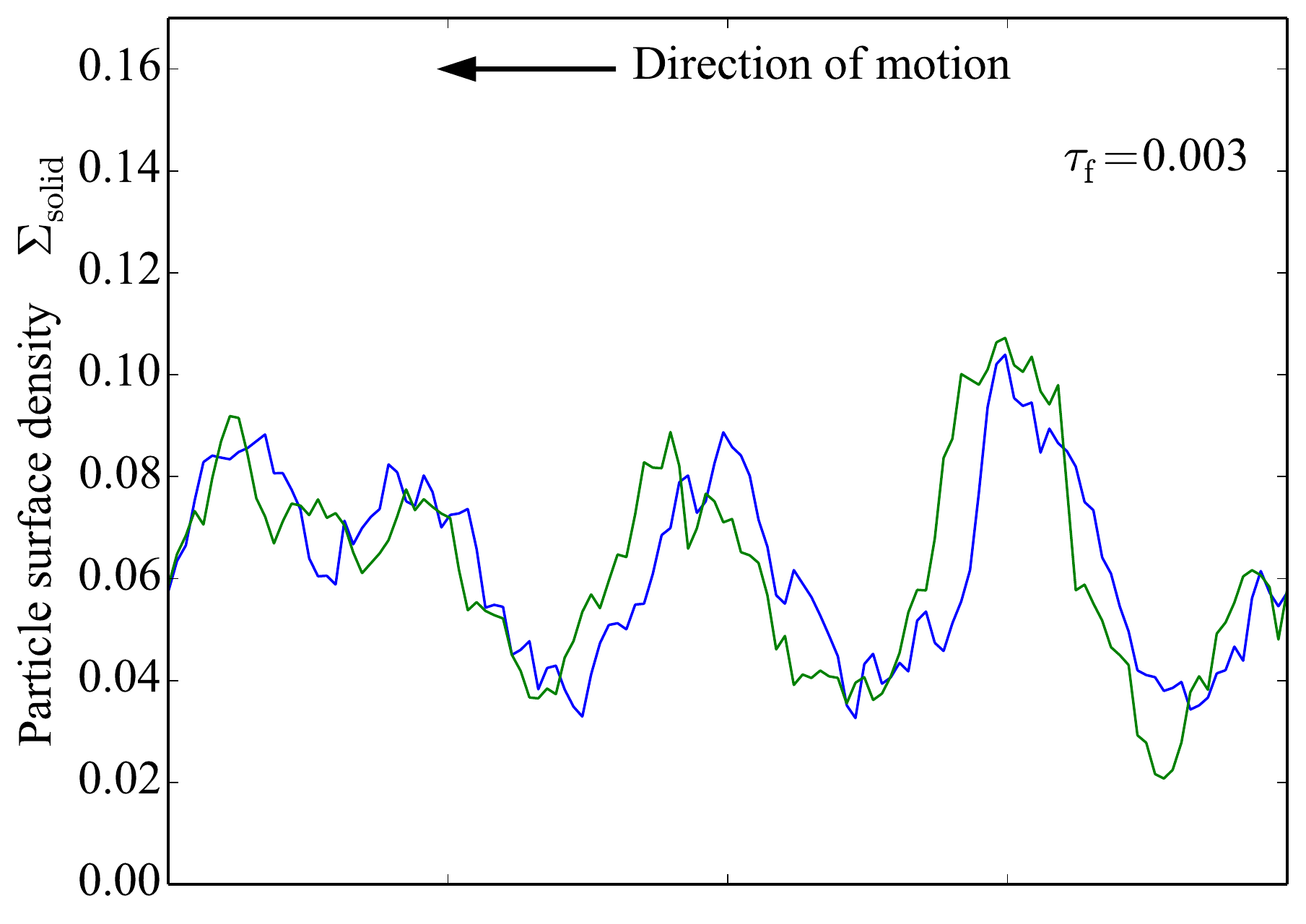}
  \includegraphics[width=0.44\textwidth]{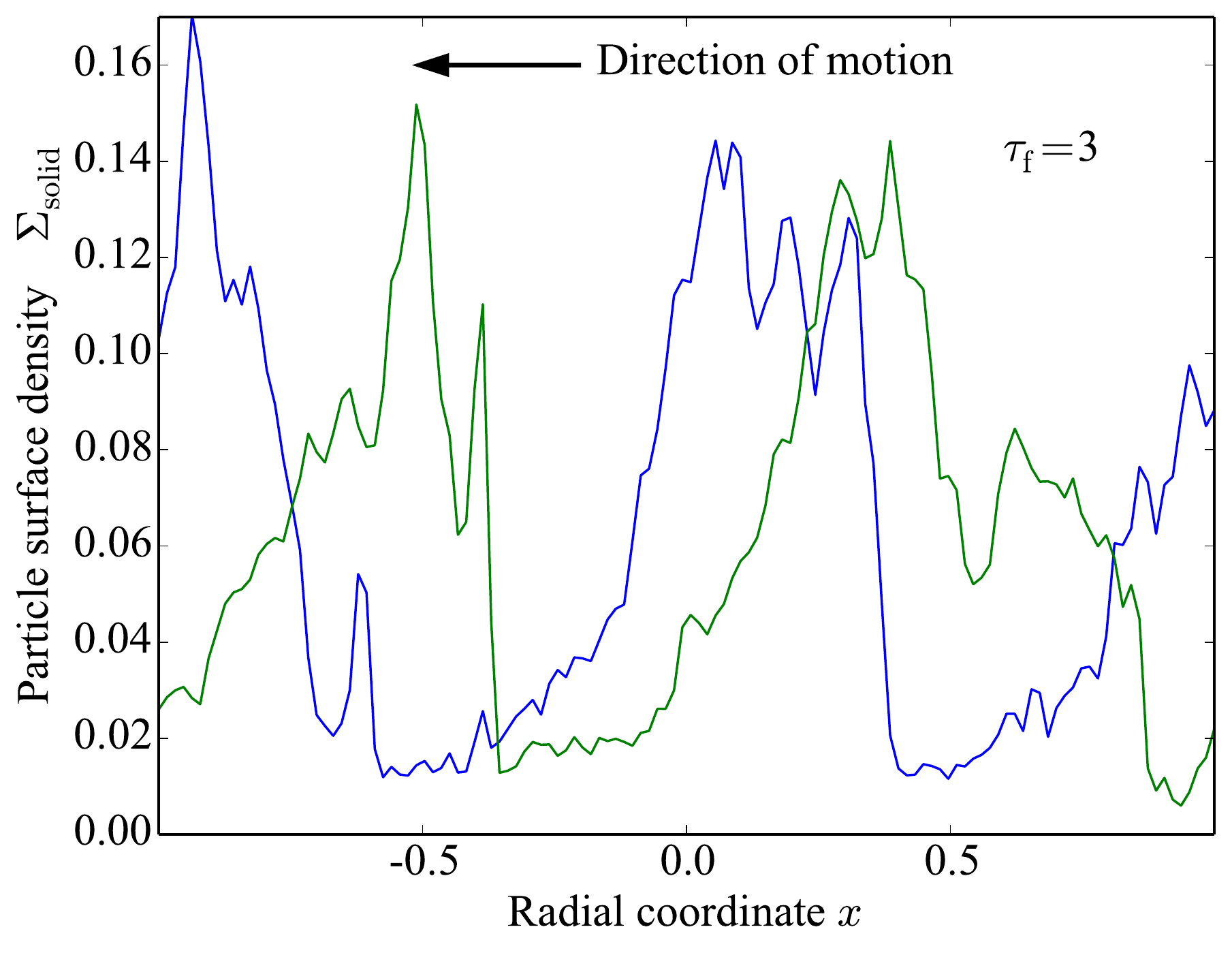}
  
  \caption{Surface density of solids $\Sigma_{\rm solid}$ as a function of the radial coordinate $x$, evaluated at two points in time 25 orbits apart (first blue, then green). The top plot is in the suspension regime (particle size $\tauf = 0.001$) and has solid concentration $Z = \langle \Sigma_{\rm solid} \rangle / \langle \Sigma_{\rm total} \rangle = 0.095$. In this regime, there is very little clumping even at high $Z$. The middle plot is the beginning of the streaming regime ($\tauf = 0.003$) and $Z = 0.095$. In this regime, clumps form and are stable. The bottom plot is in the radial drift regime ($\tauf = 3$) with $Z = 0.05$. In this regime particles form large over-densities that experience rapid radial drift.
  }
  \label{fig:rhopmx}
\end{figure}

Figure \ref{fig:rhopmx} illustrates our strategy. We are interested in identifying localized peaks in particle density that are stable over multiple orbits. That is to say, an overdensity needs to have low radial drift and high longevity. For example, for $\tauf = 0.001$ there are no strong overdensities, and for $\tauf = 3$, the overdensities are short-lived and suffer strong radial drift. Therefore, our strategy is to integrate the particle density over a 25-orbit interval and use the Kolmogorov-Smirnov test to distinguish the resulting particle distribution from a uniform distribution. We test several 25-orbit periods, spaced in steps of 10 orbits so that they overlap. The result of the KS test is a p-value, which measures the probability that two observed data sets arose from the same underlying distribution. The method is discussed in Appendix A. We divide the results of the KS test into four categories:

\vspace{0.2cm}
\begin{tabular}{ll}
   Clumping is unlikely:        & $p \geq 0.45$        \\
   Clumping is somewhat likely: & $0.25 \leq p < 0.45$ \\
   Clumping is likely:          & $0.10 \leq p < 0.25$ \\
   Clumping is very likely:     & $p < 0.10$           \\
\end{tabular}
\vspace{0.2cm}

\begin{figure*}
    \centering
    \includegraphics[width=0.9\textwidth]{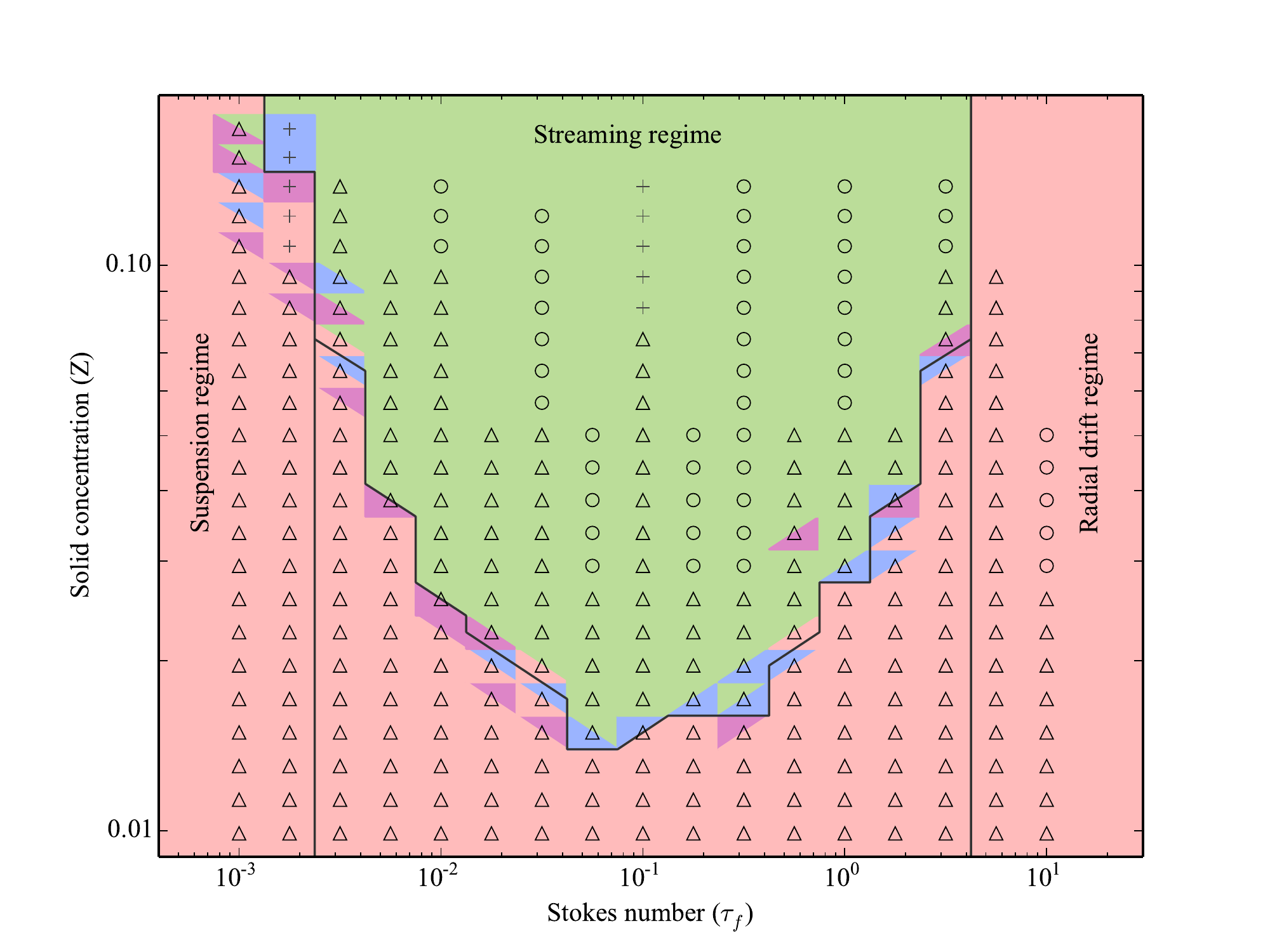}
    
    \caption{Region of the particle size vs concentration phase space where the streaming instability is active. Particle size is measured in the stopping time $\tauf$ and the particle concentration is $Z = \Sigma_{\rm solid} / \Sigma_{\rm total}$. The colors mark the probability that particle clumps can form, where red is ``unlikely'', magenta is ``somewhat likely'', blue is ``likely'', and green is ``very likely''. See appendix A for the precise definitions and methods. When different simulations give different results, the two extreme values are shown. For example, if two simulations give ``likely'' and one gives ``unlikely'', this is expressed as a square that is half red, half green. The symbols indicate the number of simulations available for the result. A triangle means three simulation, a cross means two, and a circle means one. Regions without symbols are extrapolations. The black lines indicate our (subjective) assessment of main regions where the behavior of particles is qualitatively different.}
    \label{fig:final-results}
\end{figure*}

Figure \ref{fig:final-results} summarizes the result of the KS test for all the simulations. The symbols in the figure mark the points where the test was performed. The figure is color -coded: For each point, the surrounding square has two colors (possibly the same color twice) that mark the extreme values of $p$. In this way, the figure indicates the \textit{range} of values observed across the simulations. The figure shows that, within the streaming regime, there is a distinct region in the $Z-\tauf$ phase space where particle clumps are consistently likely to form. Of special interest is the fact that particles as small as $\tauf = 0.003$ can form distinct particle clumps, at least some of the time, for solid concentrations around $Z \sim 0.065$. This value is high, but as we noted earlier, there is evidence that the ice line is associated with a particle concentration at least that high \citep{Sirono_2011}, even before we consider turbulent eddies, disk winds or photoevaporation.

Appendix B has convergence tests. Figure \ref{fig:spacetime-3d} shows the spacetime diagram of a 3D simulation with $\tauf = 0.03$; Figure \ref{fig:spacetime-res} shows the spacetime diagrams of higher resolution simulations with $\tauf = 0.003$ and $\tauf = 0.010$, and grid size $256 \times 256$. In all cases, the results are consistent with the spacetime diagrams in Figs.\ \ref{fig:spacetime-1} and \ref{fig:spacetime-2}.


\section{Constraints on planetesimal formation models}\label{sect:ppd-models}

We consider the formation of dense particle clumps, as the precursors of planetesimals. \citet{Ormel_2008} have found that $\mu$m-sized dust sticks to chondrules to form a porous rim, which absorbs some of the kinetic energy from collisions. This allows the chondrules to form small aggregates, whose size depends primarily on the speed of the turbulent eddies in the disk,

\begin{equation}
  v_{\rm eddy} \;\sim\; \cs \sqrt{\alpha}
\end{equation}
where $\cs$ is the local sound speed, and $\alpha$ is a proportionality constant that measures the strength of the disk turbulence. \citet{Ormel_2008} found that, for typical disk turbulence ($\alpha = 10^{-4}$), the chondrule aggregates have $R \sim 1 \mm$ in size. For low turbulence ($\alpha = 10^{-6}$), the chondrule aggregates have $R \sim 4 \mm$, with many aggregates as large as $R \sim 7 \mm$).

\begin{figure}[h!]
  \centering
  \includegraphics[width=0.47\textwidth]{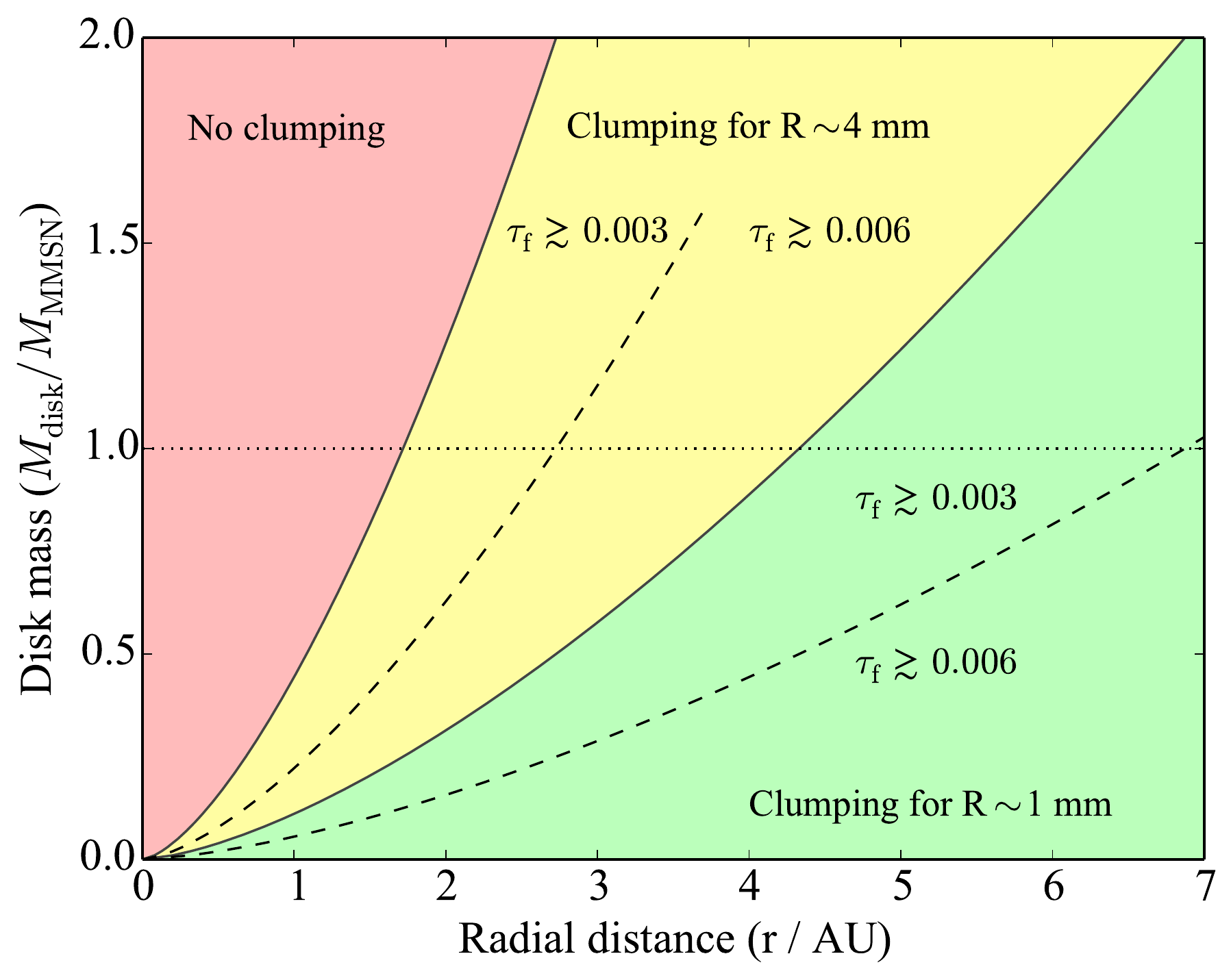}
  
  \caption{Location of the asteroid forming region in terms of disk mass $M_{\rm disk}$ and semimajor axis $r$. This can also be thought of as a ``disk age vs $r$'' phase space. The phase space can be divided into three regions. In the red region (left), particle clumps cannot form for any disk model. In the yellow region (middle), particle clumps can only form if the disk turbulence is low, so that chondrule aggregate can reach $R \sim 4-5\mm$ in size (see main text). In the green region (right), particle clumps can form for $R \sim 1\mm$ particles and low turbulence is not required. Here, $M_{\rm MMSN}$ is the mass of the minimum-mass solar nebula. The yellow and green regions can be further subdivided into an inner and outer sub-region. In the inner sub-region, particles have a stopping time of $\tauf \gtrsim 0.003$, and can only form clumps when the solid concentration is high -- i.e.
$Z \gtrsim \Sigma_{\rm solid} / \Sigma_{\rm total} = 0.065$. In the outer sub-region, the particle stopping time is $\tauf \gtrsim 0.006$ and clumps can form for $Z \gtrsim 0.04$ (see Fig.\ \ref{fig:final-results}). In all models, as the disk evolves, the disk mass drops and the planetesimal forming region moves closer to the star.
  }
  \label{fig:models}
\end{figure}

For $r > 10 \AU$, mm-sized aggregates have $\tauf > 0.01$ and the streaming instability occurs easily. For $r < 5 \AU$, Fig.\ \ref{fig:final-results} imposes some constraints on the plausible scenarios that can produce dense particle clumps. These constraints are illustrated in Fig.\ \ref{fig:models}. In the figure, the ``disk mass vs $r$'' phase space is divided into regions where different disk models can form particle clumps. Broadly speaking, there are three avenues to form planetesimals in the inner solar system:


\begin{enumerate}

\item Figure \ref{fig:final-results} shows that the streaming instability can be pushed to particles as small as $\tauf = 0.003$ if the solids can reach sufficiently high mass fractions. There are various disk processes that may enhance $Z$ sufficiently to induce clumping. They include the break-up or ice-dust aggregates \citep{Sirono_2011}, as well as large-scale pressure bumps \citep{Johansen_2009}.
\\

\item An alternative to high particle concentration is low disk turbulence. \citet{Ormel_2008} have shown that, for $\alpha \sim 10^{-6}$, chondrule aggregates can reach $R \sim 4-5$ mm. Figure \ref{fig:models} shows how low $\alpha$ can be much more effective than high $Z$ in triggering particle clumps.
\\
\item Finally, since $\tauf$ is fundamentally a measure of the particle stopping time, as the disk clears and the gas density decreases, $\tauf$ increases for all particles. Figure \ref{fig:models} shows how the disk mass ($M_{\rm disk}$) affects the location of the planetesimal forming region. The effect is most dramatic in the inner $\sim$ 1-2 AU, where particle clumping is not possible until $M_{\rm disk}$ has decreased.

\end{enumerate}

These scenarios are not exclusive. The formation of planetesimals may well rely on a combination of, say, disk evolution to decrease $M_{\rm disk}$ and pressure bumps to increase $Z$. We wish to investigate which combination of (Z, $\alpha$, $M_{\rm disk}$) are compatible with the formation of particle clumps in the disk. To do this, we generalize the Hayashi nebula to allow for different disk masses

\begin{equation}\label{eqn:hayashi-gen}
  \Sigma = 1700 \g \cminv2 
           \, \left( \frac{r}{\AU} \right)^{-3/2}
           \, \left( \frac{M_{\rm disk}}{M_{\rm MMSN}} \right),
\end{equation}
where $M_{\rm MMSN}$ is the disk mass for the Hayashi nebula. Equation \ref{eqn:stokes-number} relates $\tauf$ to $\Sigma$, the particle size $R$, and material density $\rho_\bullet$. Take a typical material density of $\rho_\bullet = 3.6 \g \cminv3$, and we obtain the minimum distance $r_{\rm pl}$ where asteroids can form by the streaming instability,

\begin{equation}\label{eqn:planetesimal-line}
    r_{\rm pl} =
       4.33 \AU
       \left( \frac{M_{\rm disk}}{M_{\rm MMSN}} \, \frac{\tau_{\rm f,min}}{0.003} \right)^{2/3}
       \left( \frac{R}{\mm} \right)^{-2/3},
\end{equation}
where $\tau_{\rm f,min}$ is the minimum Stokes number needed to trigger the streaming instability, which is determined by the particle concentration. In a disk where solids can concentrate to $Z = 0.065$, Fig.\ \ref{fig:final-results} gives a minimum Stokes number of $\tau_{\rm f,min} = 0.003$. Otherwise, take $\tau_{\rm f,min} = 0.006$ and $Z = 0.04$. Equation (\ref{eqn:planetesimal-line}) can be thought of as the location of the ``planetesimal line'', or the minimum distance where planetesimals can form. This line moves inward as the disk evolves. Figure \ref{fig:models} shows the location of the planetesimal line for different choices of $Z$, $\alpha$, and $M_{\rm disk}$.

\section{Coagulation for mm-sized particles}\label{sect:coagulation}

Particles with $R = 1 \mm$ ($\tauf = 0.003$) may be a critical stage in the formation of planetesimals. These particles lie at the upper limits of coagulation, and the lower limits of the streaming instability. The formation of planets may hinge on the ability of these particles to grow from $\tauf = 0.003$, and move deeper into the streaming regime, where gravitational collapse can occur \citep[see Eq.\ \ref{eqn:roche-density} and ][]{Bai_2010}. In this section we explore the ability of mm-sized, $\tauf = 0.003$ to grow faster than the disk accretion timescale.

Our simulations are too coarse to directly measure collision speeds between particles, but with some extrapolation, it is possible to produce robust upper limits. To do this, we note that in a small region of the disk the particle velocities should be randomized and their relative speeds should follow a Maxwell-Boltzmann distribution,

\begin{equation}
    f(v) = \sqrt{\frac{2}{\pi}}
           \left( \frac{v^2}{a^3} \right)
           \exp\left( \frac{-v^2}{2a^2} \right).
\end{equation}

The Maxwell-Boltzmann distribution is parametrized by the characteristic speed $a$. Given a series of relative speeds $\{v_{\rm i}\}_{i=1}^n$, the Maximum Likelihood Estimate for $a$ is

\begin{equation}\label{eqn:hat_a}
    \hat{a} = \sqrt{\frac{1}{3n}\sum_{i=0}^n v_{\rm i}^2}.
\end{equation}

\begin{figure}[h!]
  \centering
  \includegraphics[width=0.5\textwidth]{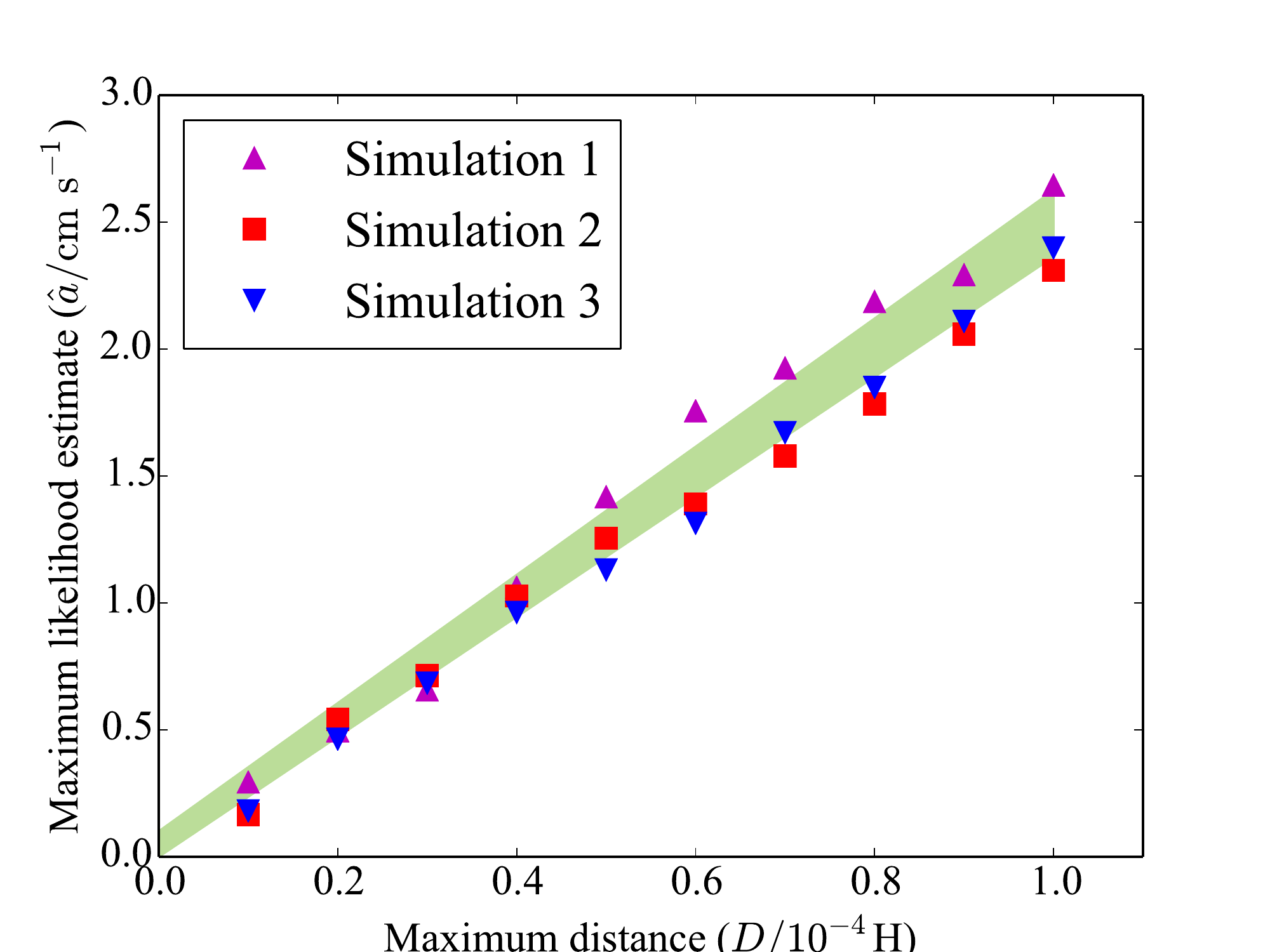}
  
  \caption{Maximum likelihood estimate $\hat{a}$ of the characteristic speed between solid particles. The particle size is $\tauf = 0.003$ and the solid concentration ($Z = \Sigma_{\rm solid} / \Sigma_{\rm total}$) is 0.074 for all the simulations. The procedure is to select particle pairs with separation less than $D$ and compute $\hat{a}$ from Eq.\ \ref{eqn:hat_a}. Extrapolating to $D \rightarrow 0$ gives a rough estimate of the characteristic collision speed $\hat{a}_0$. The region in green corresponds to
  $0 \le \hat{a}_0 \le 2\sigma_a$ and the 1-$\sigma$ confidence interval for the slope.
  }
  \label{fig:maxwellian-fit}
\end{figure}

The strategy, then, is to extract a collection of particle pairs that are closer than some distance $D$ and compute $\hat{a}$ from the relative speeds. This is shown in Fig.\ \ref{fig:maxwellian-fit} for several choices of $D$, and for three different simulations with $\tauf = 0.003$. If we extrapolate $\hat{a}$ for $D \rightarrow 0$, we obtain a rough estimate of the Maxwellian distribution of collision speeds. If we fit a linear relation, $\hat{a} = \hat{a}_0 + m \; D$, the extrapolated value of $\hat{a}$ is
$\hat{a}_0 = -0.0003 \pm 0.05$ cm/s. Since the true $\hat{a}_0$ cannot be negative, we take the range

\begin{equation}\label{eqn:maxwellian-fit}
    0 \cm \sinv \;\le\; \hat{a}_0 \;\le\; 0.1 \cm \sinv
\end{equation}

\begin{figure}[h!]
    \centering
    \includegraphics[width=0.5\textwidth]{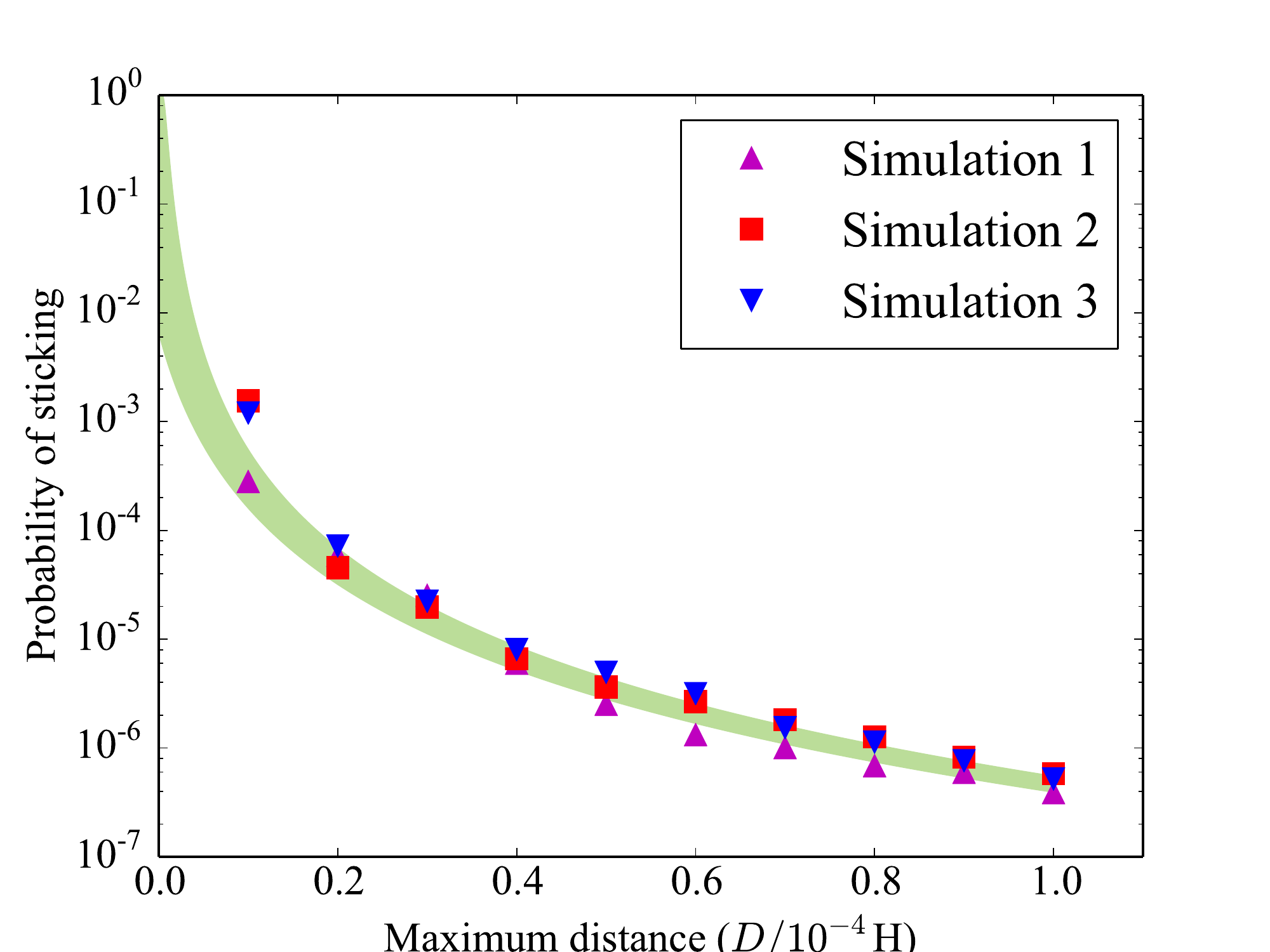}
    
    \caption{Probability that the relative speed between two solid particles lies below the sticking threshold $v\crit \sim 0.03$ cm/s \citep{Guttler_2010}. The particle size is $\tauf = 0.003$ and $Z = \Sigma_{\rm solid} / \Sigma_{\rm total} = $ 0.074 in all the simulations. Given a set of particle pairs with separations less than $D$, one can fit a Maxwellian distribution
    (Figs.\ \ref{fig:maxwellian-fit} and \ref{fig:example-maxwellian}) determine the fraction of speeds below $v\crit$. Extrapolating to $D \rightarrow 0$ gives an estimate of the fraction of particle collisions that result in sticking. The region in green corresponds to the 1-$\sigma$ confidence region marked in Fig.\ \ref{fig:maxwellian-fit}.}
    \label{fig:prob-sticking}
\end{figure}

Figure \ref{fig:maxwellian-fit} shows the extrapolation of $\hat{a}$ for $\hat{a}_0$ in this interval, and for the 1-$\sigma$ confidence region for the slope $m$. We are interested in the probability that a collision of two mm-sized particles will result in sticking. For a material density of $\rho_\bullet \sim 3.5 \g \cminv3$, the critical speed is around $v\crit \sim 0.03$ cm/s \citep{Guttler_2010}. The probability that $v < v\crit$ is given by the cumulative distribution function

\begin{equation}\label{eqn:maxwellian-cdf}
    F(v) = \int_0^{v} f(x) \, dx
    = {\rm erf}\left( \frac{v}{a \sqrt{2}} \right)
    - \frac{v}{a}
    \sqrt{\frac{2}{\pi}}
    \exp\left( \frac{-v^2}{2 \, a^2} \right).
\end{equation}
Figure \ref{fig:prob-sticking} shows the probability of a sticking collision $F(v\crit)$ for the $\hat{a}$ values from Fig.\ \ref{fig:maxwellian-fit}. As $D \rightarrow 0$, the steady decrease in particle speeds leads to a significant increase in the fraction of collisions that result in sticking.

\begin{figure*}
  \centering
  \includegraphics[width=0.75\textwidth]{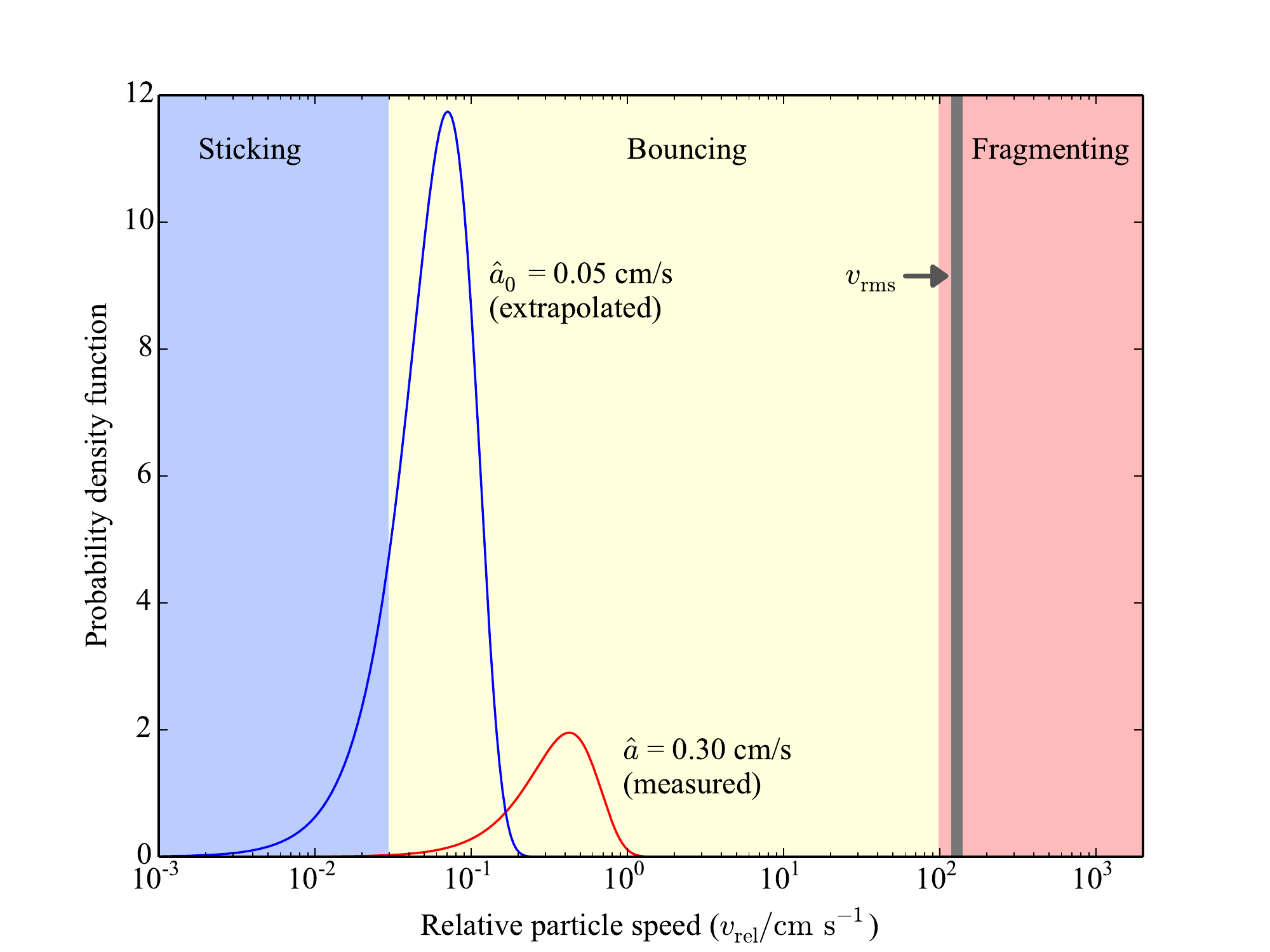}
  
  \caption{
  Maxwellian distribution of relative speeds for mm-sized particles ($\tauf = 0.003$, pressure support $\Delta = 0.05$ and solid concentration $Z = \Sigma_{\rm solid} / \Sigma_{\rm total} =$ is 0.074). For particles separated by distance $d < 10^{-5}H$ ($H$ is the disk scale height) the Maxwellian distribution has characteristic speed $\hat{a} \sim 0.3$ cm/s (red). Extrapolating as $d \rightarrow 0$ we obtain a conservative estimate of $\hat{a}_0 = 0.05$ cm/s (blue). For reference, we include the sticking,
  bouncing   and fragmenting regions from \citet{Guttler_2010}, and the root-mean-squared speed $v_{\rm rms}$ for all the particles in the simulation. Note that the shape of the curve is distorted by the log scale; in particular, the great majority of collisions are in the bouncing regime. Also note that $v_{\rm rms}$ is a poor estimate for the the collision speeds. The value $\hat{a}$ is obtained by the maximum likelihood method.
  }
  \label{fig:example-maxwellian}
\end{figure*}

Figure \ref{fig:example-maxwellian} shows the Maxwellian distributions for $\hat{a} = 0.3$ cm/s (typical value for $D = 10^{-5}$ H) and for $\hat{a} = 0.05$ cm/s (mid-range value for $\hat{a}_0$). The bulk of the collisions result in bouncing, and a small fraction result in sticking. For comparison, the figure also shows the root-mean-squared vertical speed $v_{\rm rms}$ for all the particles in the grid. The figure shows how $v_{\rm rms}$ greatly over-estimates the particle collision speeds.

\begin{figure}[h!]
  \centering
  \includegraphics[width=0.5\textwidth]{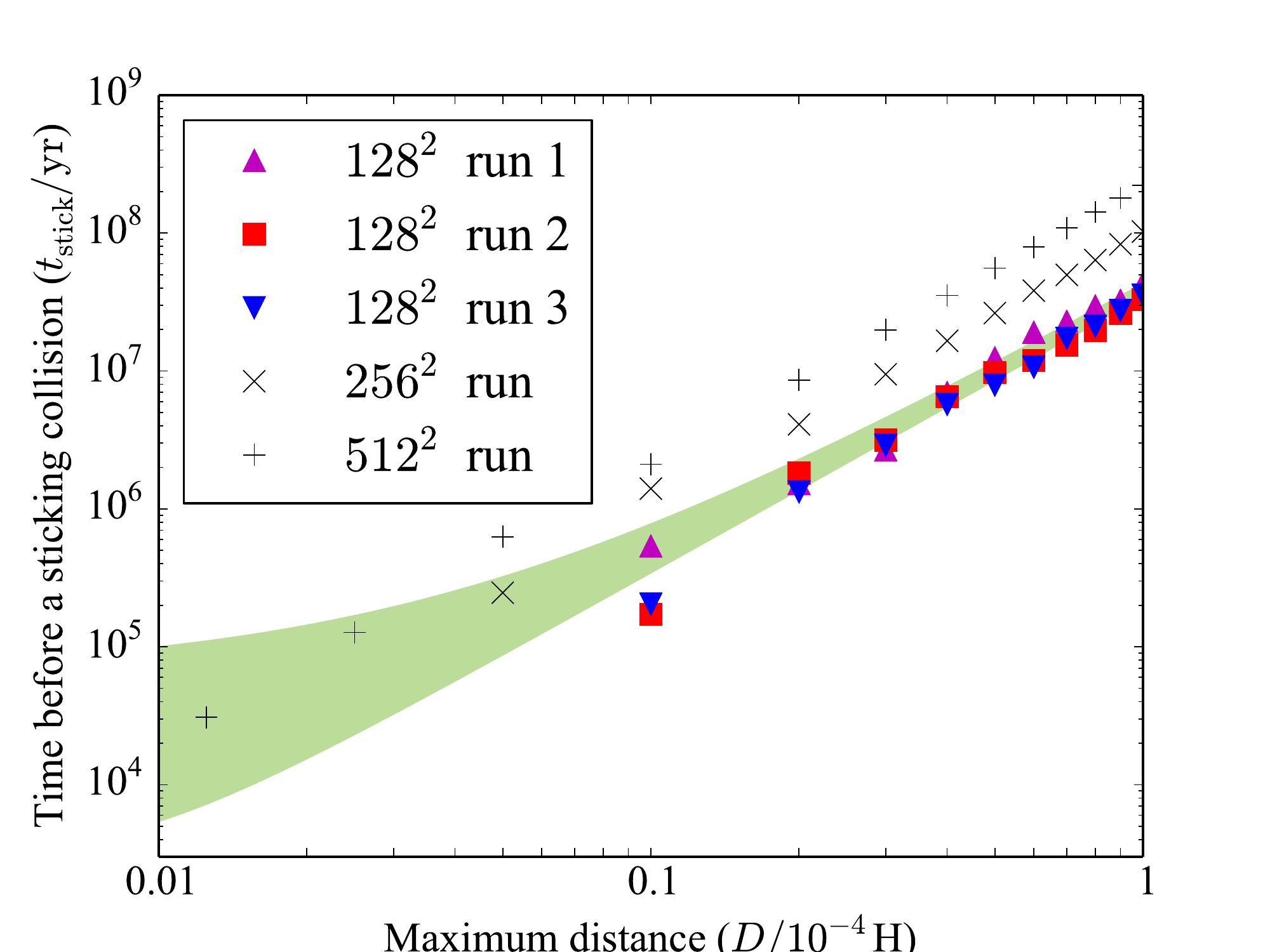}
  
  \caption{Time between sticking collisions. The figure shows the three $128^2$ runs from Fig.\ \ref{fig:prob-sticking} plus two higher-resolution runs. In all runs the particle size is $\tauf = 0.003$ and $Z = \Sigma_{\rm solid} / \Sigma_{\rm total} = $ 0.074. We fit a Maxwellian distribution to to all particle pairs with separation less than $D$. Extrapolating the fit toward small $D$ we estimate the time between sticking collisions.
  The points on the figure appear to follow a power law $t_{\rm stick} \propto D^2$. The region in green corresponds to the 1-$\sigma$ confidence region marked in Fig.\ \ref{fig:maxwellian-fit}. The figure shows that $t_{\rm stick} < 10^5$ years is a robust upper bound on $t_{\rm stick}$.
  }
  \label{fig:time-to-stick}
\end{figure}

\begin{figure}[h!]
    \vspace{0.1cm}
  \centering
  \includegraphics[width=0.49\textwidth]{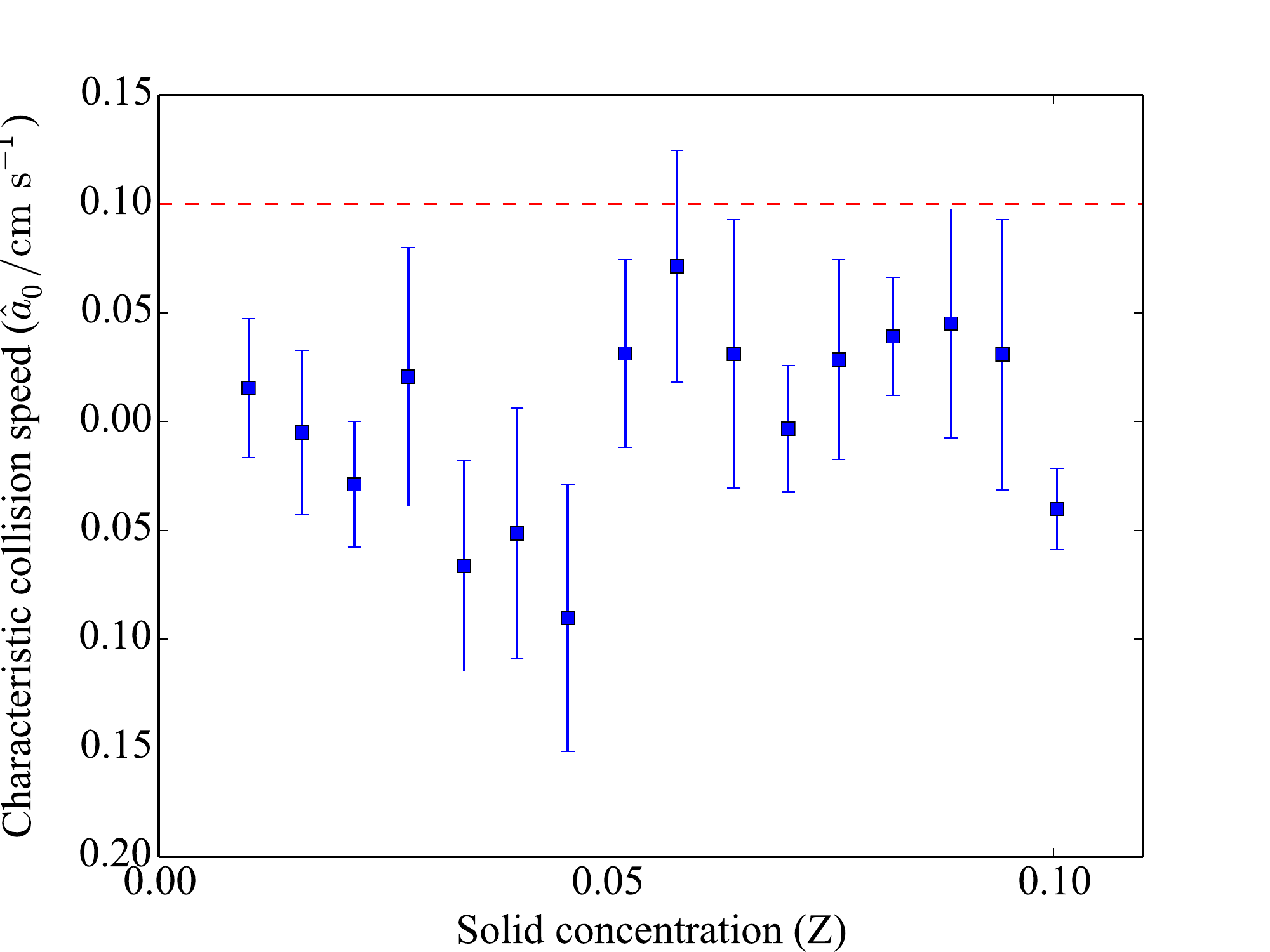}
  
  \caption{Characteristic collision speed $\hat{a}_0$ between solid particles at different points in our simulations, with 1-$\sigma$ error bars, for particle size $\tauf = 0.003$. The solid concentration ($Z = \Sigma_{\rm solid} / \Sigma_{\rm total}$) has little effect on $\hat{a}_0$. The value $\hat{a}_0 < 0.1$ cm/s (red dashed line) is a robust upper bound on $\hat{a}_0$.
  This corresponds to coagulation timescale of $t_{\rm stick} < 10^5$ years, which is significantly less than the disk lifetime ($10^6 - 10^7$ years).  Note that the fitting method can give negative values because it is a simple linear extrapolation (see main text).
}
  \label{fig:a-vs-Z}
\end{figure}

Finally, we use $F(v\crit)$ to determine the typical time needed before a 1 mm grain has a sticking collision. The key question is whether the particles can grow faster than the disk evolution timescale. The time needed for a sticking collision is given by

\begin{equation}\label{eqn:t-stick}
    t_{\rm stick} = \left(
               \pi (2r)^2 \, \bar{v} \, n_{\rm p} \, F(v\crit)
               \right)^{-1},
\end{equation}
where $n_{\rm p}$ is the particle number density, and $\bar{v}$ is the mean relative particle speed. In these simulations, $n_{\rm p} \sim 4 \times 10^{-3} \m^{-3}$ at the midplane, and for a Maxwellian, $\bar{v} = 2 \, a \, \sqrt{2/\pi}$. Figure \ref{fig:time-to-stick} shows the results for
$t_{\rm stick}$. As $D \rightarrow 0$, we get $t_{\rm stick} \sim 10^3 - 10^5$ years, which is a small fraction of the disk lifetime of $\sim 10^7$ years. We repeated this experiment for several particle concentrations from $Z = 0.01$ to 0.1. Figure \ref{fig:a-vs-Z} shows that $\hat{a} < 0.1 \cm \sinv$ is a robust upper bound, largely independent of $Z$. We conclude that the particle growth timescale
$t_{\rm stick} < 10^5$ years is also robust. Therefore, we conclude that $\tauf = 0.003$ particles can double in mass within a small fraction of the disk lifetime. As the particles grow, the streaming instability becomes more effective and the density of the particle clumps increases.

\section{Conclusion}\label{sect:conclusions}

The formation of planetesimals remains a difficult problem. The traditional bottom-up coagulation model faces important challenges from theory and observation. There is growing evidence that planetesimals must form top-down, from the gravitational collapse of dense clumps of solids \citep[e.g.][]{Morbidelli_2009}. In this paper we present the results from over 100 hydrodynamic simulations where we modeled the dynamics of solid particles inside a protoplanetary disk. We model particle sizes from sub-millimeter-sized chondrules to meter-sized rocks, and evaluate the effect of particle concentration and gas pressure gradient. Our key results can be summarized as:

\begin{enumerate}

\item We present the particle properties as a size vs concentration phase space, and we map the region where particle clumps can be expected to form (Fig.\ \ref{fig:final-results}). We measure the particle size as $\tauf = \Omegak \, \tf$, where $\tf$ is the friction time and $\Omegak$ is the Keplerian frequency. We measure particle concentration as $Z = \Sigma_{\rm solids} / \Sigma_{\rm total}$, where $\Sigma$ denotes surface density. We find that particles as small as $\tauf = 0.003$ can participate in the streaming instability and form dense clumps at $Z \sim 0.065$. Although this solid concentration is quite high, it is smaller than what can be produced by the break-up of ice-dust aggregates near the snow line \citep{Sirono_2011}. More generally, high $Z$ values may be achieved by the radial drift of solids from the outer disk into the inner disk \citep{Youdin_2002}, and by large scale pressure bumps or vortices \citep{Johansen_2009}.
\\
\item We find constraints on asteroid formation models and present them in Fig.\ \ref{fig:models}. We map the region in the $M_{\rm disk}$ vs $r$ phase space that is consistent with the formation of dense particle clumps. The location of the asteroid-forming region is primarily a function of disk mass and turbulent viscosity, as low $\alpha$ leads to larger chondrule aggregates \citet{Ormel_2008}. After that, the most important factor is the disk's ability to increase the solid concentration locally.
\\
\item We estimate the probability of sticking collisions between mm-sized particles ($\tauf = 0.003$). We find a robust upper limit on the timescale for these particles to stick and grow to larger sizes ($\sim 10^5$ years) that is significantly smaller than the lifetime of the disk ($\sim 10^7$ years). A high concentration of particles is not required for this result. As the particles grow, the streaming instability becomes more effective, leading to the formation of planetesimals.

\end{enumerate}

Altogether we find that particle concentration by the streaming instability provides a viable path to forming asteroids directly from mm-sized chondrules, particularly if weak turbulence facilitates the growth of chondrule aggregates with sizes of a few millimeters.


\begin{acknowledgements}
The authors would like to thank Axel Brandenburg, Andrew Youdin, Lennart Lindegren, Mike Alexandersen and the anonymous referee for their comments that helped improve the manuscript. We also acknowledge the support from the Knut and Alice Wallenberg Foundation, the Swedish Research Council (grants 2010-3710 and 2011-3991) and the European Research Council Starting Grant 278675-PEBBLE2PLANET that made this work possible. Computer simulations were performed using the Alarik cluster at Lunarc Center for Scientific and Technical Computing at Lund University. Some simulation hardware was purchased with grants from the Royal Physiographic Society of Lund.
\end{acknowledgements}

\bibliographystyle{aa}
\bibliography{../../References}


\appendix

\section{A measure of particle clumping}

Particle clumping is difficult to measure objectively. Previous authors have measured the peak particle density, but this is a very local measure that contains limited information about the particle behavior. In particular when small particles ($\tauf \le 0.003$) form filaments that are clearly visible in a spacetime diagram (Fig.\ \ref{fig:spacetime-1}), no single grid cell achieves a very high density. At the same time, large particles ($\tauf > 3$) have may have a few cells with very high density, but the density peaks are unstable and short lived.

Figure \ref{fig:KS_test} illustrates our approach. For each time step, we take the particle surface density $\Sigma_{\rm solid}$ (top row). Particle overdensities are clearly visible on this plot. We then average $\Sigma_{\rm solid}$ over 25 orbits ($\langle \Sigma_{\rm solid} \rangle_{\rm t}$, second row). This serves to to smooth out overdensities that are unstable, short-lived, or have too much radial drift. We chose 25 orbits because it is half of the 50-orbit timescale for $Z$ to increase. We considered using the radial speed to separate clumps that are drifting into the star. The appeal of radial speed is that it provides an instantaneous measure. However, radial speed does not capture the longevity of a particle clump. For example, as in a traffic jam, the pattern speed of the clump may be lower than the particle drift speed. At the same time, particles with low radial speed may produce short-lived clumps. Therefore, it is necessary to compare the location of a clump across time. Averaging $\Sigma_{\rm solid}$ over time is a simple way to achieve this goal.

Finally, we sort the grid cells of $\langle \Sigma_{\rm solid} \rangle_{\rm t}$ in order from highest to lowest density, and compute the cumulative distribution $F$ of the sorted grid cells. If the particle distribution is uniform, then $F$ forms a straight diagonal line $G$ from (0,0) to (1,1). We use the difference $F - G$ as a large-scale measure of clumping. The Kolmogorov-Smirnov \citep{Smirnov_1948,Wall_2003} test gives a standard way to compare $F$ and $G$. It computes a p-value that measures probability that the data set that produced $F$ follows the distribution $G$

\begin{eqnarray}
    Q(z) &=& 2 \sum_{j = 1}^\infty (-1)^{j-1} \exp\left(-2 z^2 j^2\right) \\
    p    &=& Q(D \sqrt{n}),
\end{eqnarray}
where $D$ is the maximum distance between $F$ and $G$, and $n$ is the number of independent observations. We caution the reader against interpreting these p-values as strict probabilities; we only use $p$ as a convenient metric to measure clumping. With that in mind, we divide the p-values into four broad categories:

\vspace{0.5cm}
\begin{tabular}{ll}
   Clumping is unlikely:        & $p \geq 0.45$        \\
   Clumping is somewhat likely: & $0.25 \leq p < 0.45$ \\
   Clumping is likely:          & $0.10 \leq p < 0.25$ \\
   Clumping is very likely:     & $p < 0.10$           \\
\end{tabular}
\vspace{0.5cm}

Figure \ref{fig:final-results} shows the result of this analysis across our entire range of simulations. The figure is broadly consistent with the spacetime diagrams in Figs.\ \ref{fig:spacetime-1} and \ref{fig:spacetime-2}.

\subsection{Alternate measures of particle clumping}

To confirm our results, we devised an alternate method to measure particle clumping, based as much as possible on very different principles. Our second method is to perform a chi-squared test on $\langle \rho_p \rangle_{\rm zt}$, without sorting the grid cells. We use the reduced chi-squared statistic,

\begin{equation}
    \chi_{\rm red}^2 \, = \, \frac{\chi^2}{k}
                     \, = \, \frac{1}{k}
                             \sum_{j = 1}^n \frac{(O_j - E_j)^2}{\sigma^2},
\end{equation}
where $k = n - 1$ is the degrees of freedom, $O_j$ is the $j^{th}$ observation (value of $\langle \rho_p \rangle_{\rm zt}$ on grid cell $j$), $E_j = \langle \rho_p \rangle$ is the $j^{th}$ expected density for a uniform distribution, and $\sigma^2$ is the expected variance between different measurements. If $\chi_{\rm red}^2 > 1$, that indicates that the difference between observations and the model is greater than can be explained by the variance $\sigma^2$. This can be because the model is incorrect, or because the true variance between measurements is greater than $\sigma^2$. We divide
$\chi_{\rm red}^2$ into four intervals:

\vspace{0.5cm}
\begin{tabular}{ll}
   Clumping is unlikely:        & $\chi_{\rm red}^2 < 0.7$          \\
   Clumping is somewhat likely: & $0.7 \leq \chi_{\rm red}^2 < 1.0$ \\
   Clumping is likely:          & $1.0 \leq \chi_{\rm red}^2 < 1.3$ \\
   Clumping is very likely:     & $\chi_{\rm red}^2 \geq 1.3$       \\
\end{tabular}
\vspace{0.5cm}

That leaves $\sigma^2$ as the only unspecified parameter. When the simulations start, $\sigma^2$ is almost zero, giving unrealistically high $\chi_{\rm red}^2$ values. As sedimentation proceeds, the value of $\sigma^2$ increases. We opted for a $\sigma^2$ that is typical for the last 25 orbits of the sedimentation phase. Because $\chi_{\rm red}^2$ is sensitive to $\sigma^2$, there is necessarily some amount of subjectivity in $\sigma^2$. In the end, we chose $\sigma^2$ so that the $\tauf = 0.003$ simulations wold give the same result as in Fig.\ \ref{fig:final-results}. Therefore, the true test is whether the $\chi_{\rm red}^2$ method remains consistent with the KS method across all the other particle sizes.

Figure \ref{fig:alt-final-results} shows the final results for the $\chi_{\rm red}^2$ method. The similarity between this figure and Fig.\ \ref{fig:final-results} is striking. Considering how different the two techniques are, the agreement between Figs.\ \ref{fig:final-results} and \ref{fig:alt-final-results} indicates that our core results are robust.

\newcolumntype{Y}{ >{\centering\arraybackslash} b{6.30cm} }
\newcolumntype{Z}{ >{\centering\arraybackslash} b{5.50cm} }
\begin{figure*}[ht!]
  \begin{tabular}{YZZ}
  $\tauf = 0.001$ & $\tauf = 0.003$ & $\tauf = 3$ \\
  \vspace{0.3cm}\includegraphics[height=4.9cm]{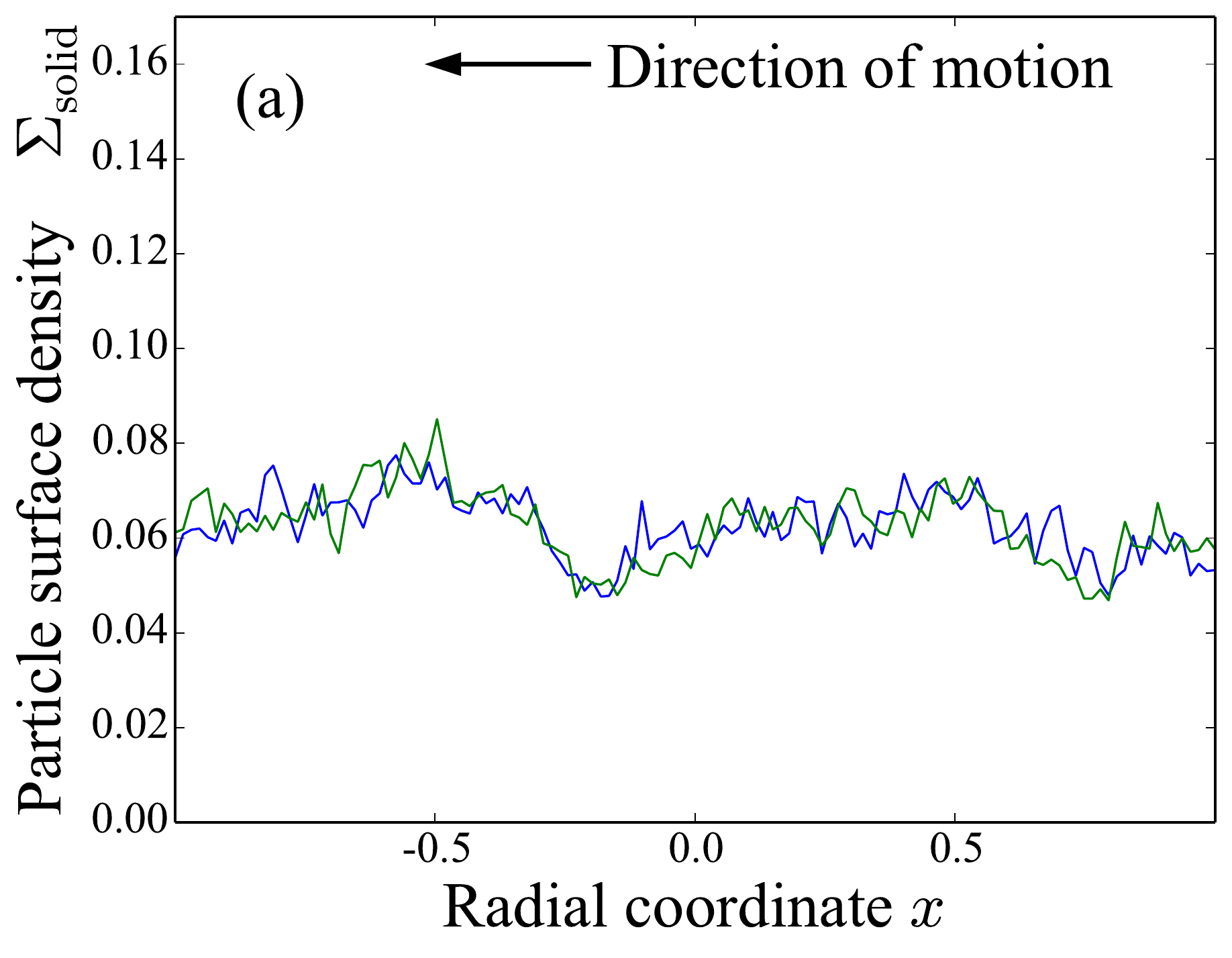} &
  \vspace{0.3cm}\includegraphics[height=4.9cm]{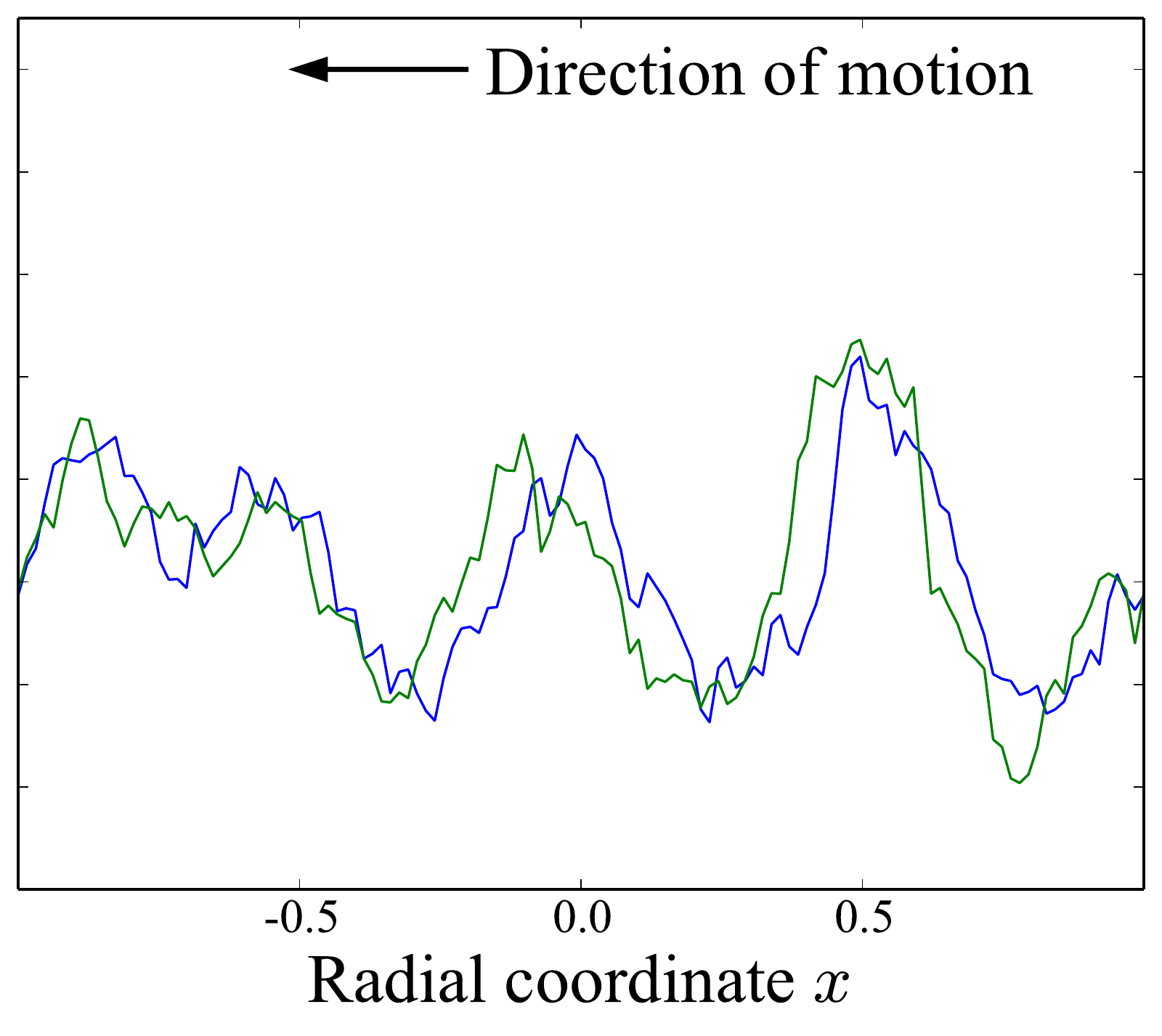} &
  \vspace{0.3cm}\includegraphics[height=4.9cm]{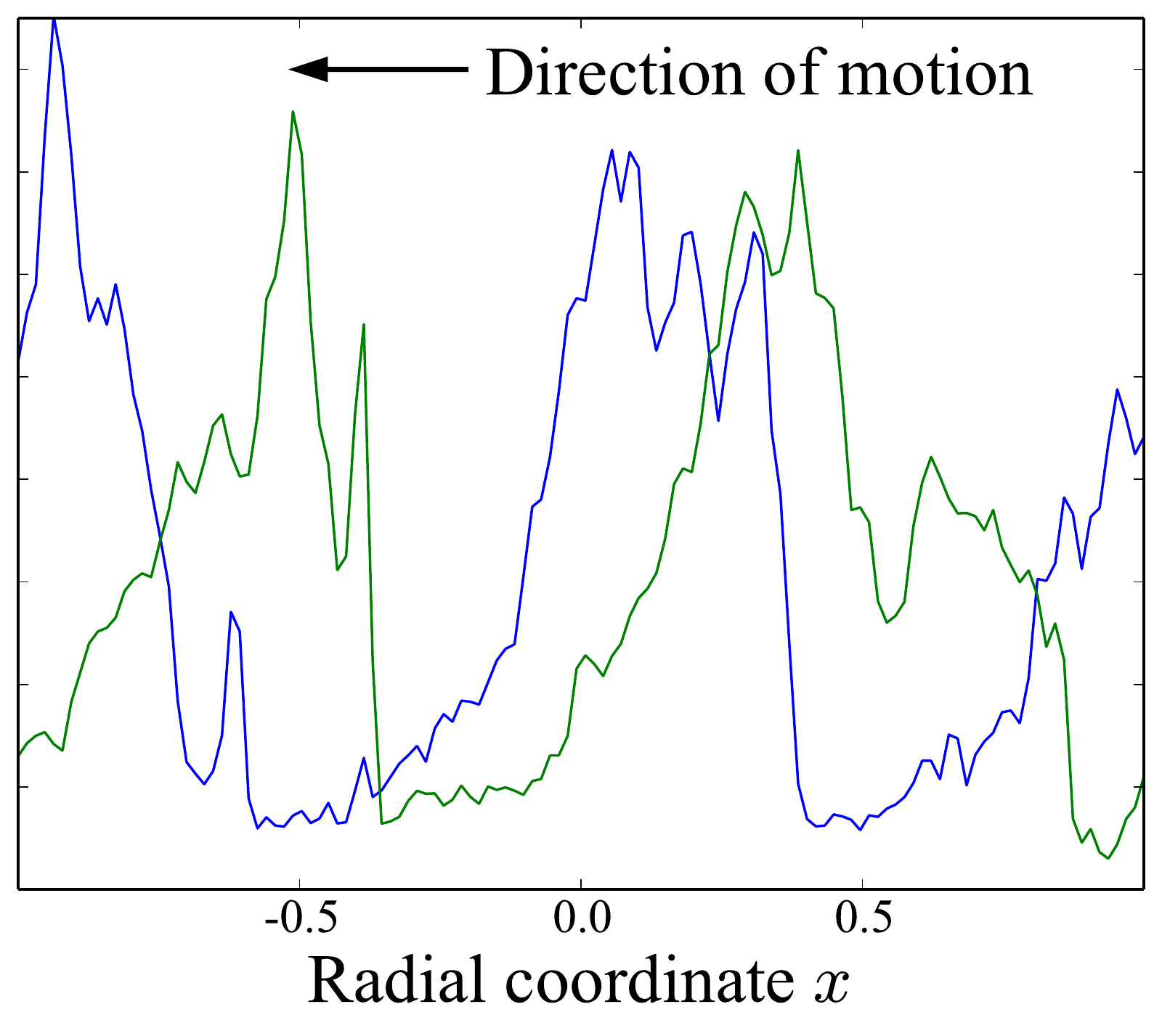} \\
  \vspace{0.3cm}\includegraphics[height=4.9cm]{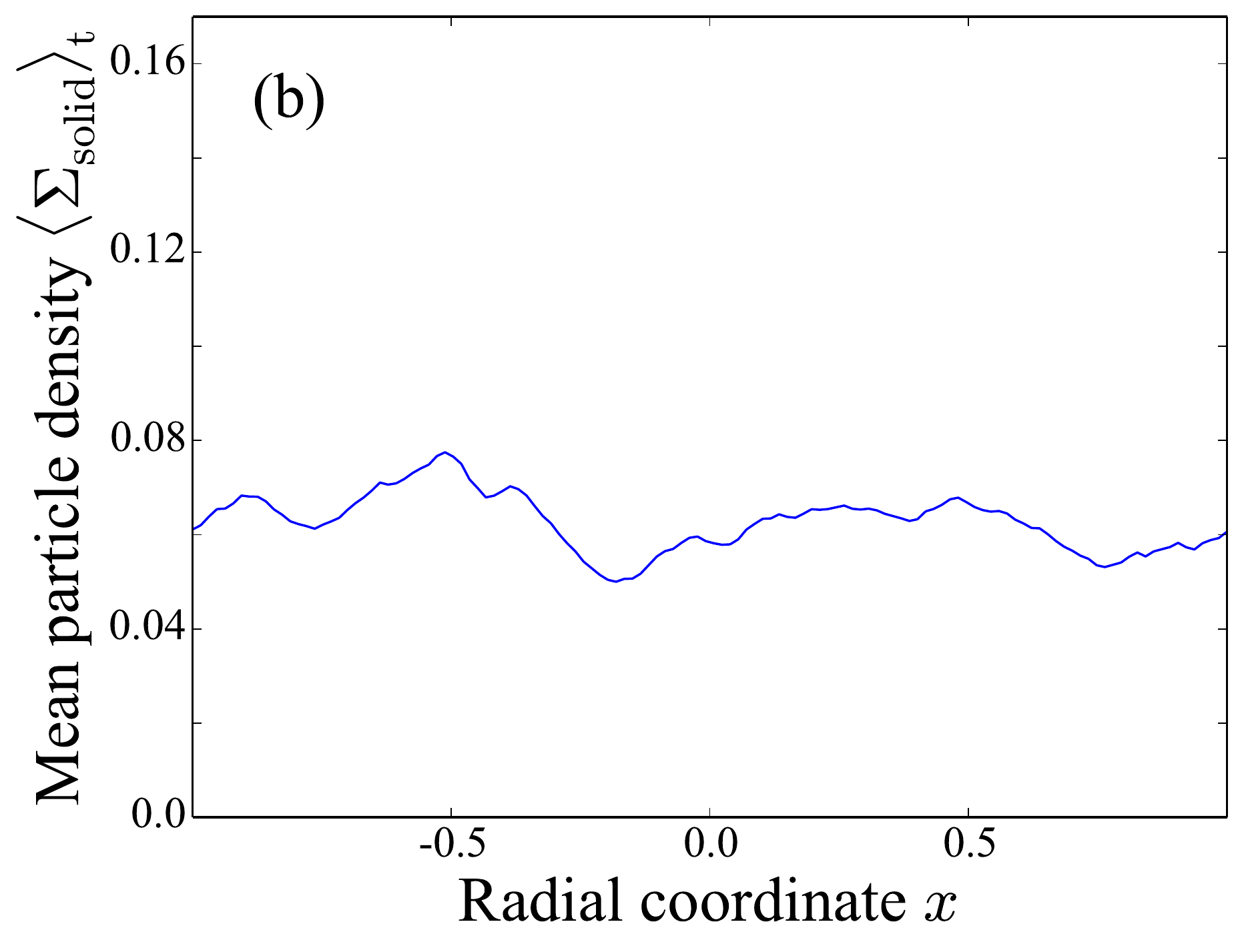} &
  \vspace{0.3cm}\includegraphics[height=4.9cm]{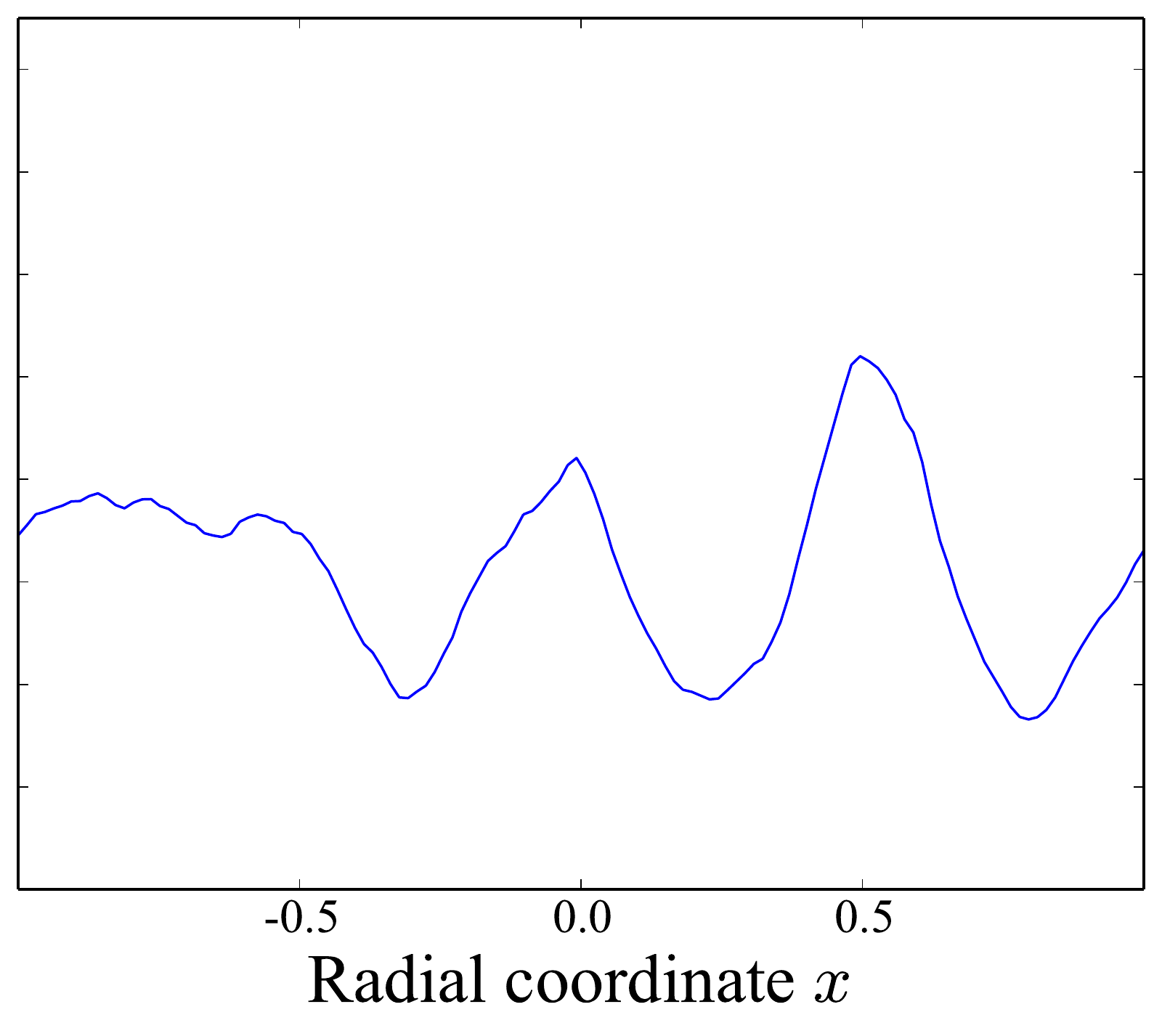} &
  \vspace{0.3cm}\includegraphics[height=4.9cm]{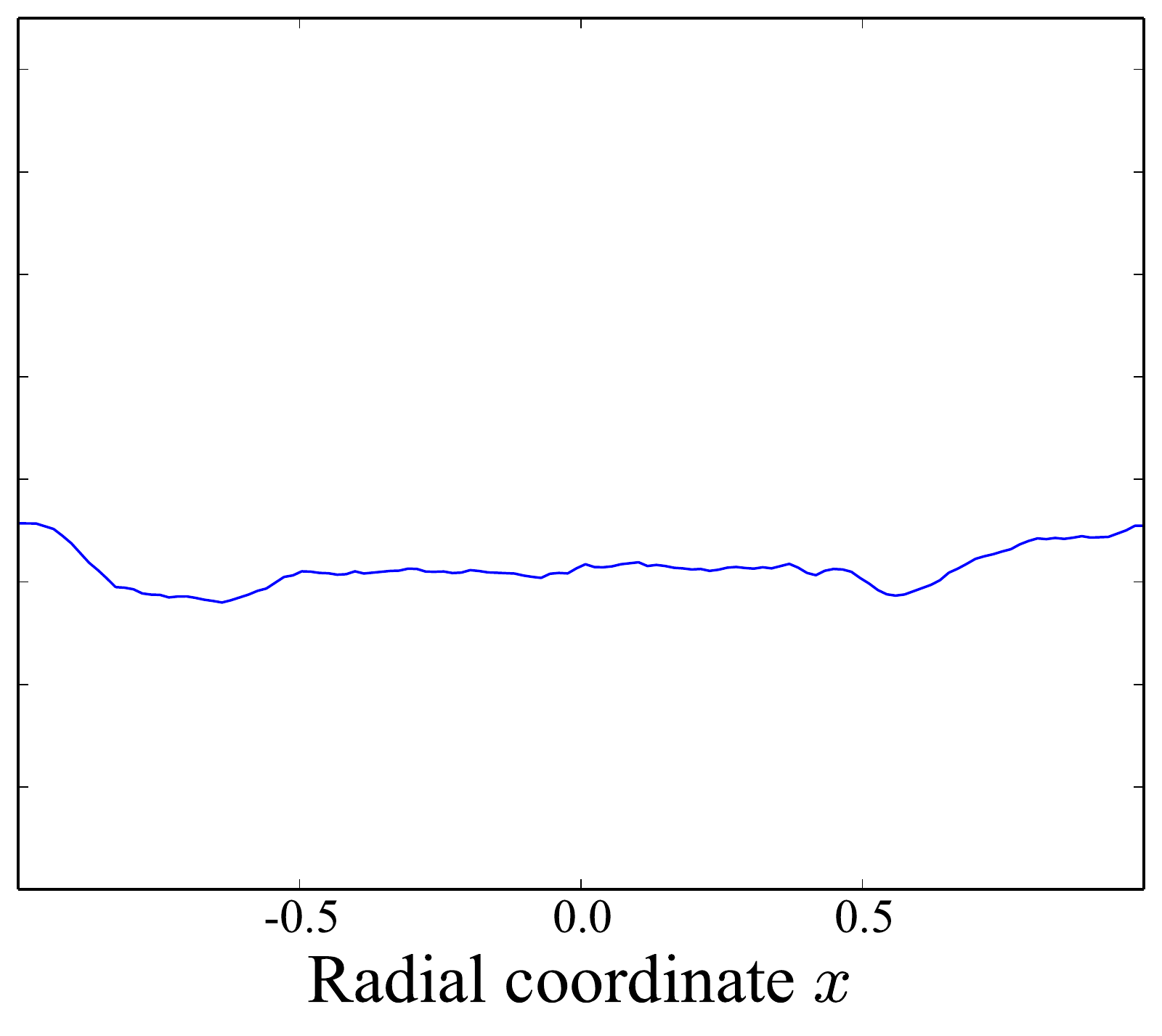} \\
  \vspace{0.3cm}\includegraphics[height=4.9cm]{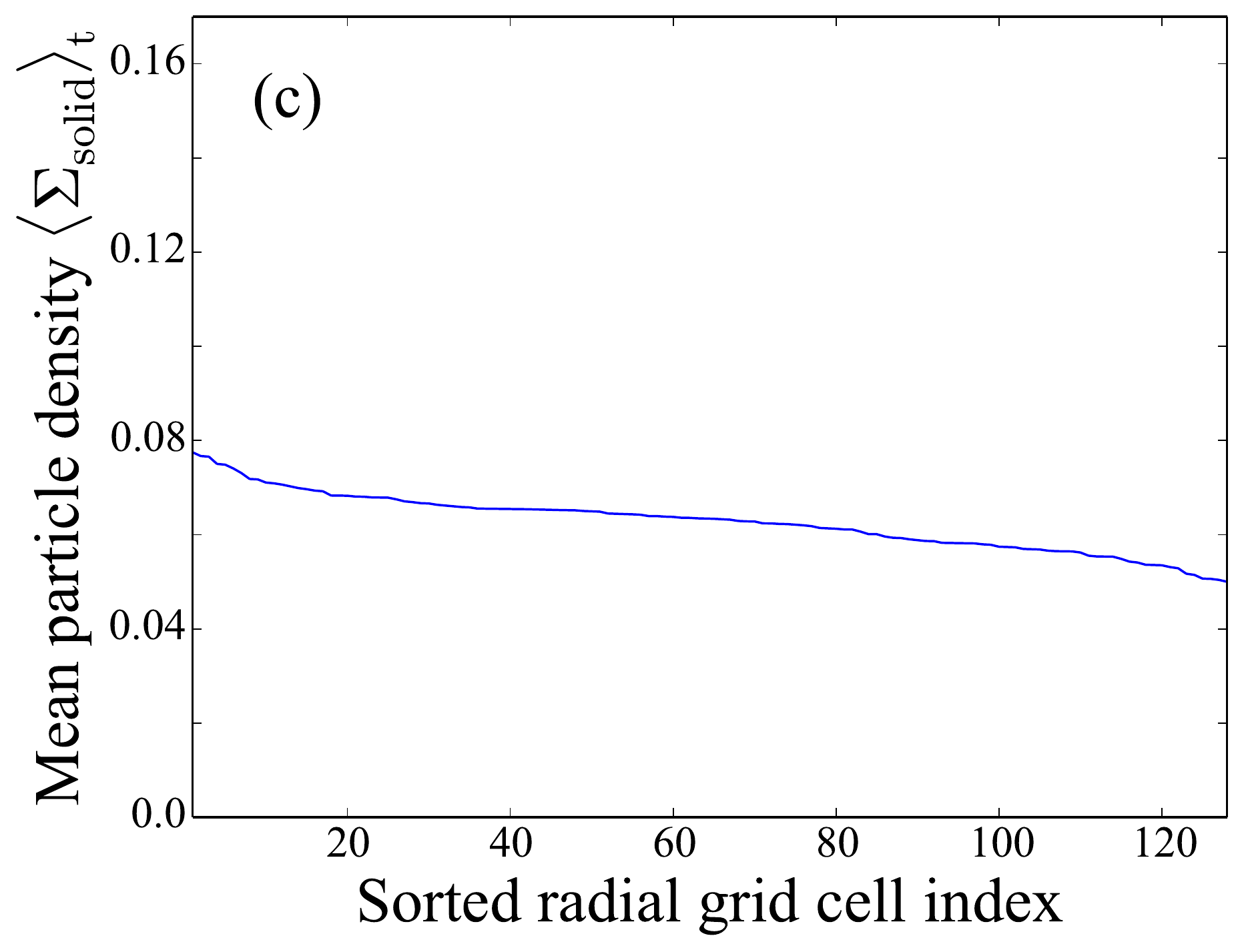} &
  \vspace{0.3cm}\includegraphics[height=4.9cm]{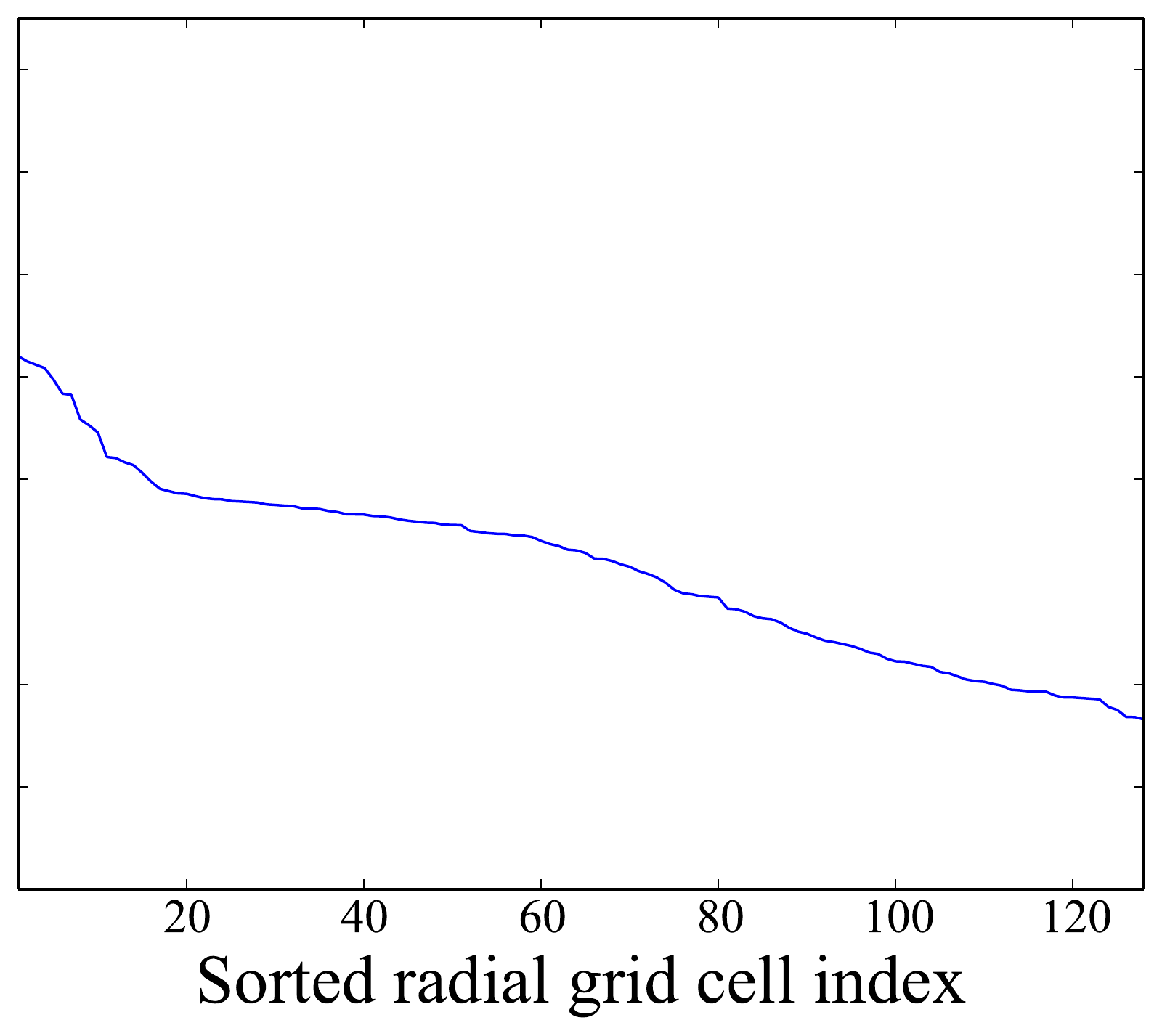} &
  \vspace{0.3cm}\includegraphics[height=4.9cm]{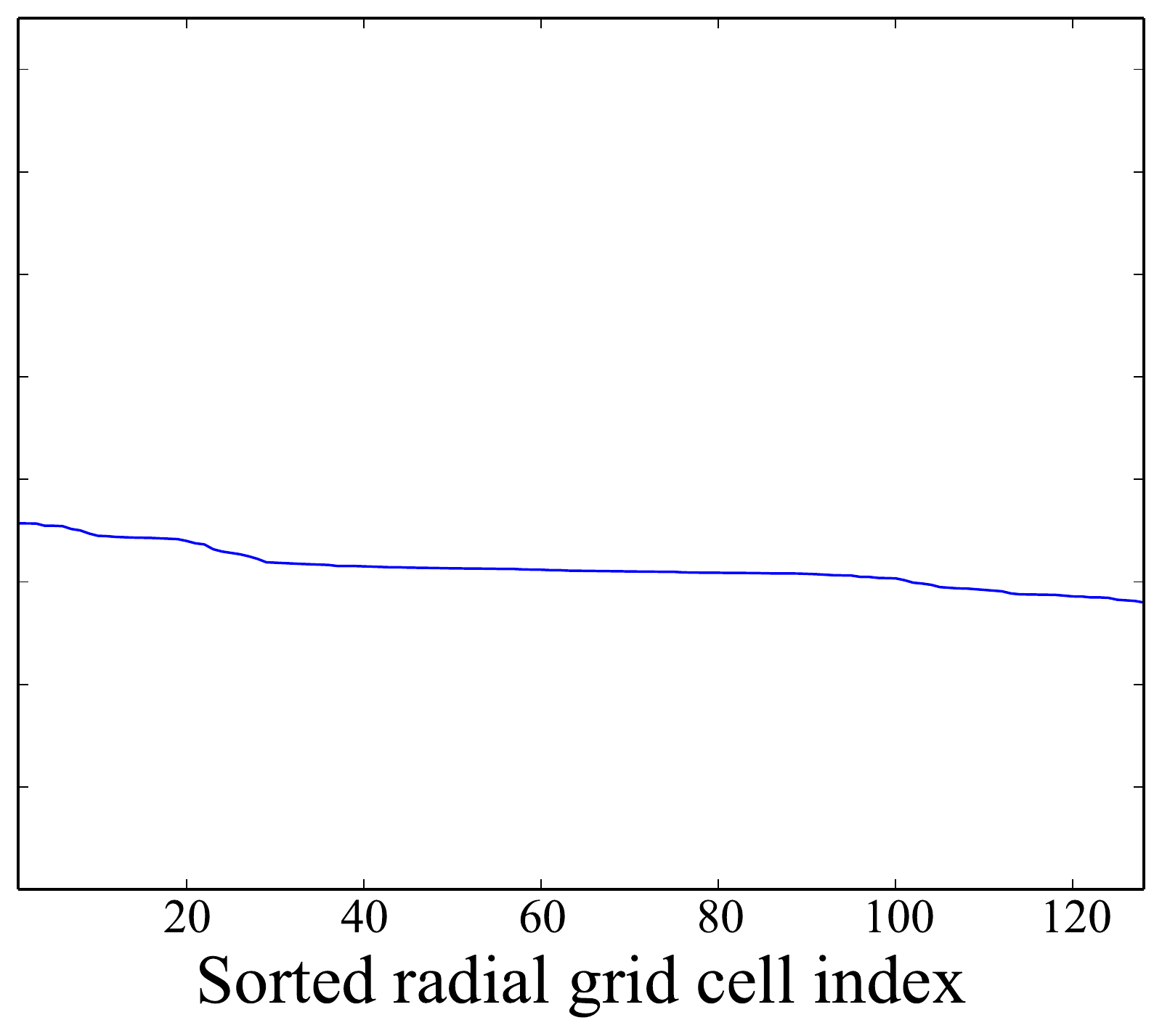} \\
  \vspace{0.3cm}\includegraphics[height=4.9cm]{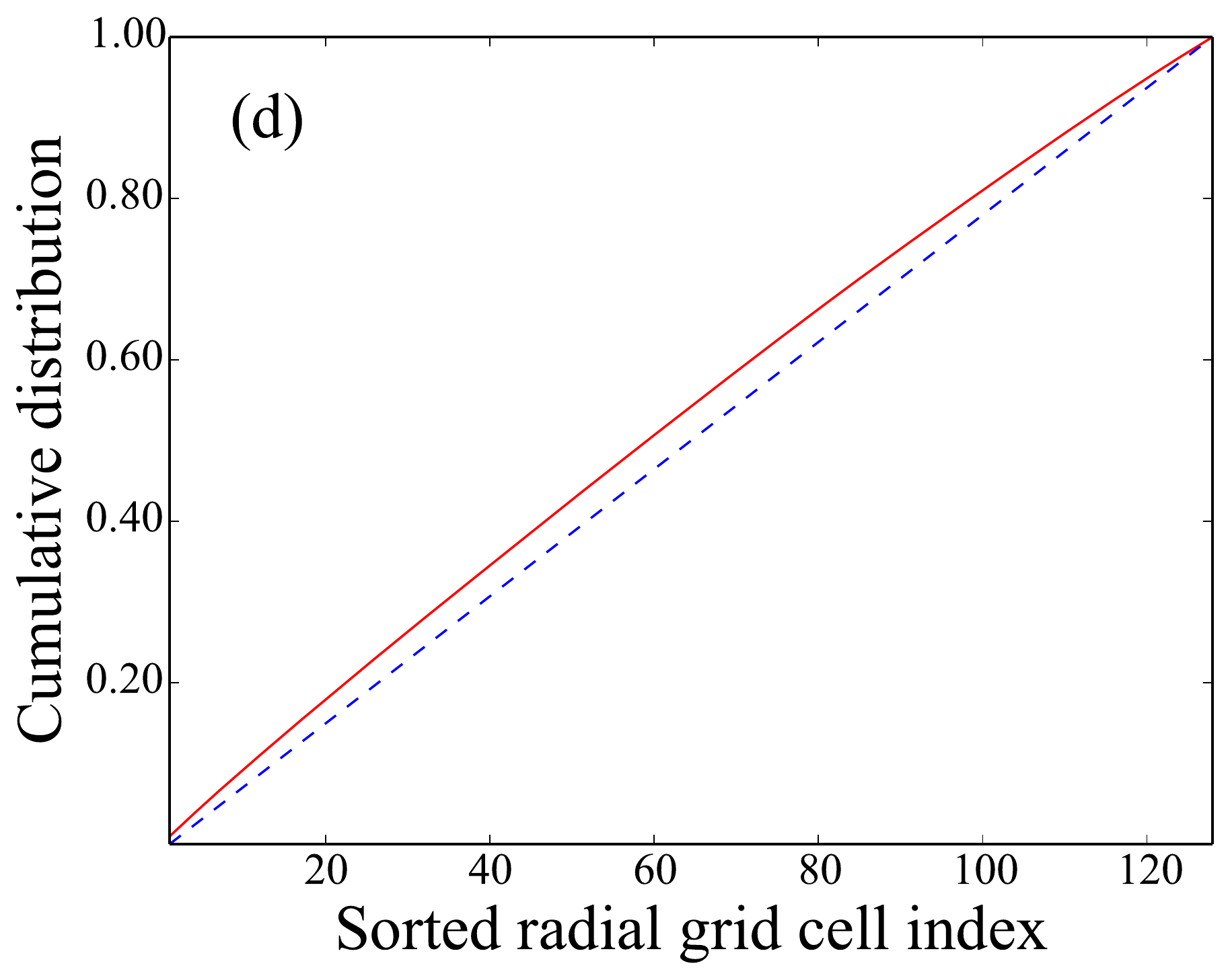} &
  \vspace{0.3cm}\includegraphics[height=4.9cm]{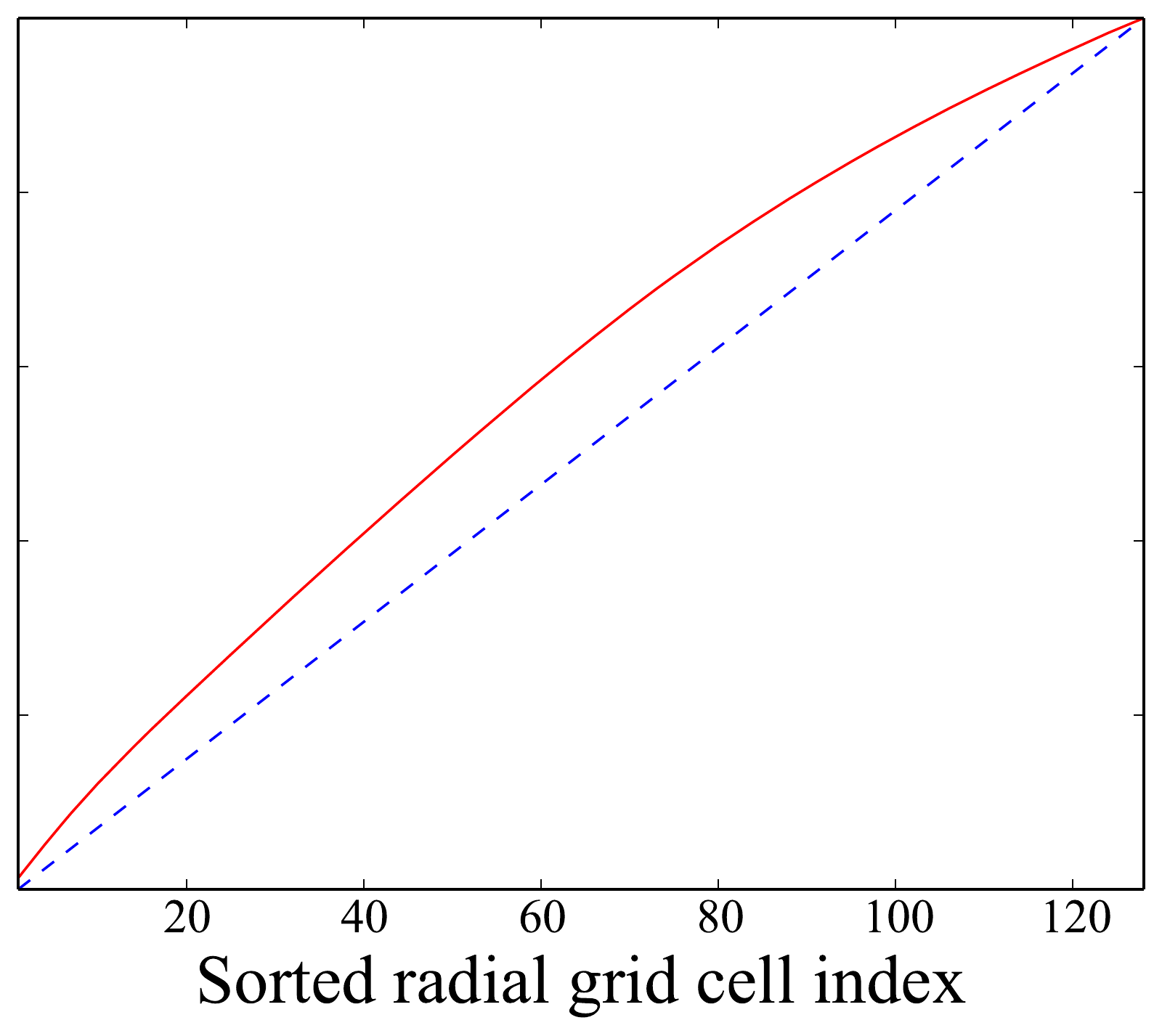} &
  \vspace{0.3cm}\includegraphics[height=4.9cm]{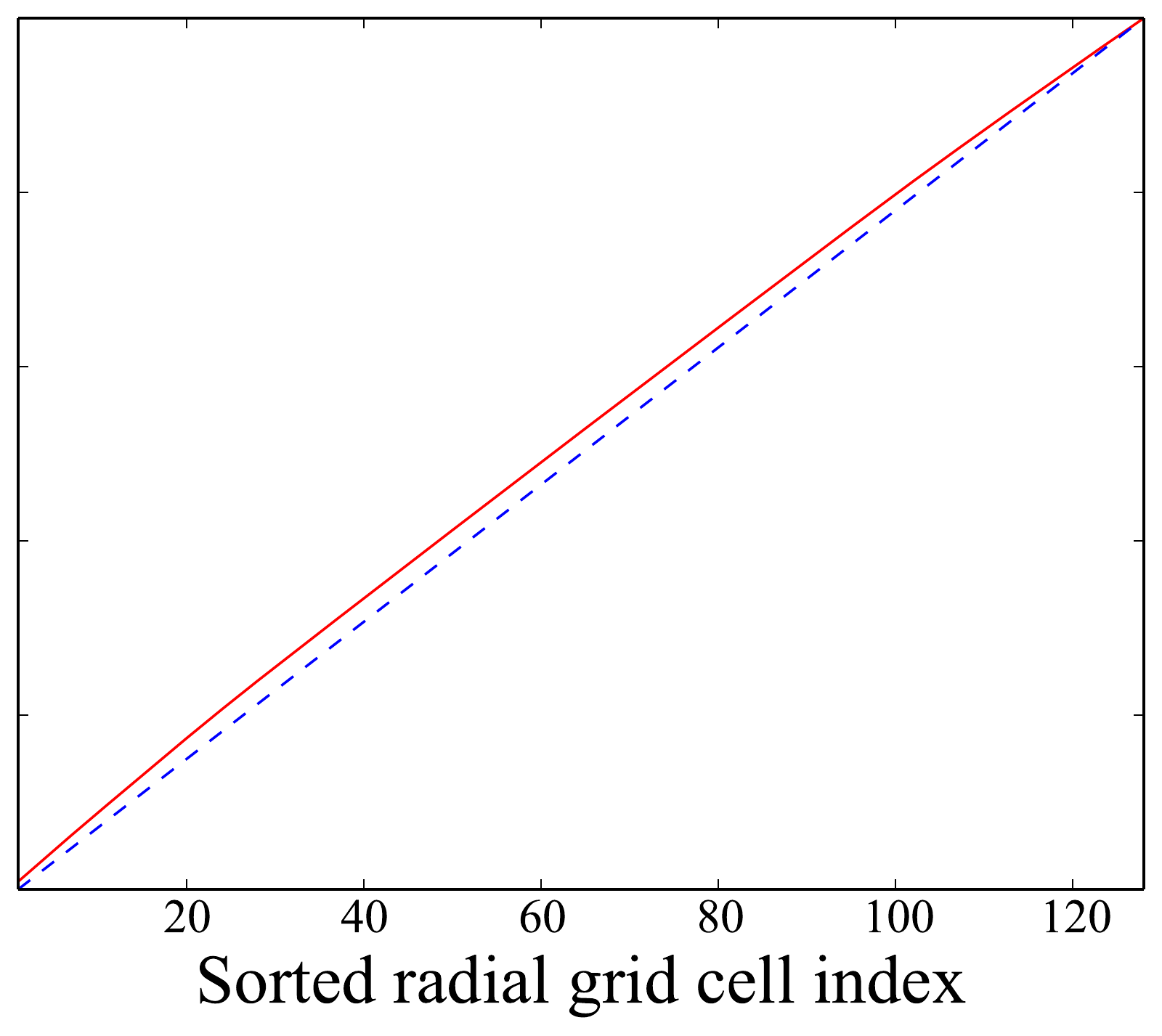}
  \\
  \end{tabular}
  
  \caption{Particle density distribution for three particle sizes. Particle size is measured in the stopping time $\tauf$, where  $\tauf = 0.001$, 0.003 and 3 correspond to the suspension regime, streaming regime and radial drift regime respectively. Row (a) is the Surface density of solids $\Sigma_{\rm solid}$ as a function of the radial coordinate $x$, evaluated at two points in time 25 orbits apart (first blue, then green). Row (b) averages $\Sigma_{\rm solid}$ over the 25-orbit interval $\langle \Sigma_{\rm solid} \rangle_{\rm t}$. Row (c) shows $\langle \Sigma_{\rm solid} \rangle_{\rm t}$  with the grid cells sorted from highest to lowest density. A steep curve indicates stable particle clumps. Row (d) shows the cumulative distribution of
$\langle \Sigma_{\rm solid} \rangle_{\rm t}$ in row (c) (solid red curve) along with the cumulative distribution of a perfectly uniform particle distribution (dashed blue line). The distance between these two curves serves as a measure of particle clumping. In these plots the particle concentrations $Z = \langle \Sigma_{\rm solid} \rangle / \langle \Sigma_{\rm total} \rangle$ are (left to right) 0.095, 0.095 and 0.05.}
  \label{fig:KS_test}
\end{figure*}

\begin{figure*}
  \centering
  \includegraphics[width=0.9\textwidth]{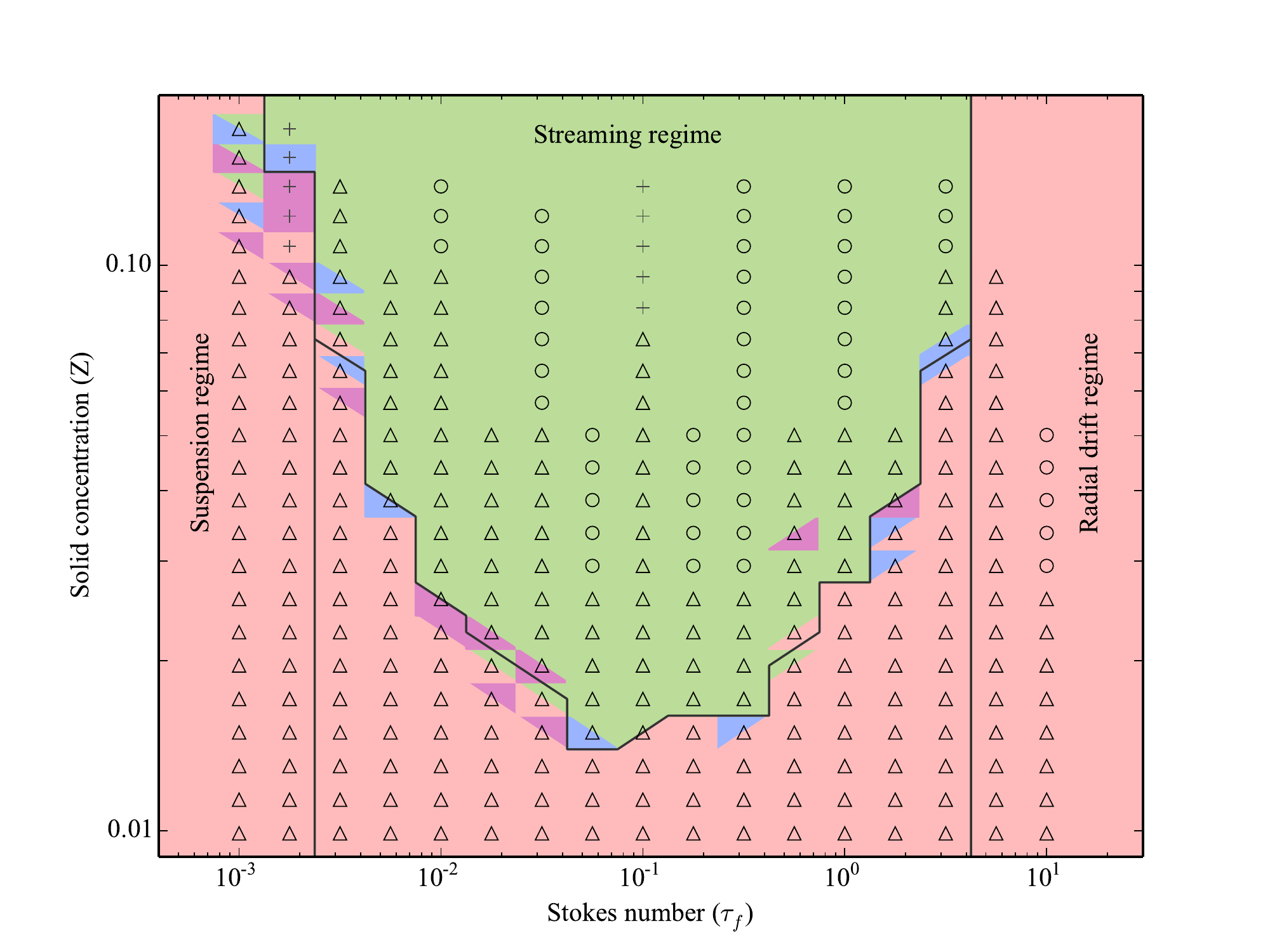}
  
  \caption{As in Fig.\ \ref{fig:final-results}, but using the $\chi_{\rm red}^2$ method to estimate the likelihood of clumping instead of the KS-derived method. The figure marks the region of the particle size vs concentration phase space where the streaming instability is active. Particle size is measured in the stopping time $\tauf$ and the particle concentration is $Z = \Sigma_{\rm solid} / \Sigma_{\rm total}$. The colors mark the probability that particle clumps can form, where red is ``unlikely'', magenta is ``somewhat likely'', blue is ``likely'', and green is ``very likely''. When different simulations give different results, the two extreme values are shown. The symbols indicate the number of simulations available. The circle, cross and triangle indicate, respectively, one, two and three simulations. Regions without symbols are extrapolations.
  To facilitate the comparison with Fig.\ \ref{fig:final-results}, the boundary marked by the black lines has been copied from Fig.\ \ref{fig:final-results} without modification.}
  \label{fig:alt-final-results}
\end{figure*}

\section{Convergence tests}
\subsection{Three-dimensional box}

Our simulations are axisymmetric, they neglect the effect of the Kelvin-Helmholtz Instability (KHI). Whether the KHI is present is dictated by the Richardson number \citep{Chandrasekhar_1961},

\[
    {\rm Ri} = \frac{g}{\rho}
               \frac{(\partial \rho / \partial z)
                   }{(\partial u/\partial z)^2},
\]
where $u$ is the gas speed, $g = \Omega^2 z$ is the vertical gravitational acceleration, and $\rho$ is the effective fluid density (treating gas and solids as a single fluid). \citet{Bai_2010b} have shown that the streaming instability is able to keep the Richardson number above the critical value needed for the KHI, so that the KHI is absent in this problem. To confirm this, we performed two 3D simulations with $\tauf = 0.03$ and $\Delta = 0.05$. The experiment performed in our 3D simulation is slightly different from that of our 2D simulations. Instead of increasing $Z$ gradually, we only test the behavior of the system at three discrete values: $Z = 0.01$, $Z = 0.02$ and $Z = 0.03$. The reason for this is that 3D simulations are very computationally expensive, especially for small particles, so that a long-term simulation is prohibitive. Our 3D runs have three phases:

\begin{itemize}
\item \textit{Phase I}: Each run starts with $Z = 0.01$, and $Z$ is held constant for 30 orbits.

\item \textit{Phase II}: The value of $Z$ is increased quickly. In run ``3D.Z2'' we increase $Z$ to $Z = 0.02$, and in run ``3D.Z3'' we increase $Z$ to $Z = 0.03$.

\item \textit{Phase III}: The value of $Z$ is once again held constant for the remainder of the run.
\end{itemize}

\begin{figure*}[hb!]
    \centering
    \begin{tabular}{ABBAA}
        \multicolumn{4}{r}{
        \includegraphics[width=10cm]{plots/idl/spacetime/colorbar.pdf}
        } &
        $\Sigma_{\rm solid} / \langle \Sigma_{\rm solid} \rangle$ \\
        & {\small 3D.Z2} & {\small 3D.Z3} & \\
        \begin{sideways}Simulation time ($t/2\pi\Omega^{-1}$)\end{sideways}\vspace{2.3cm} &
        \includegraphics[width=4cm]{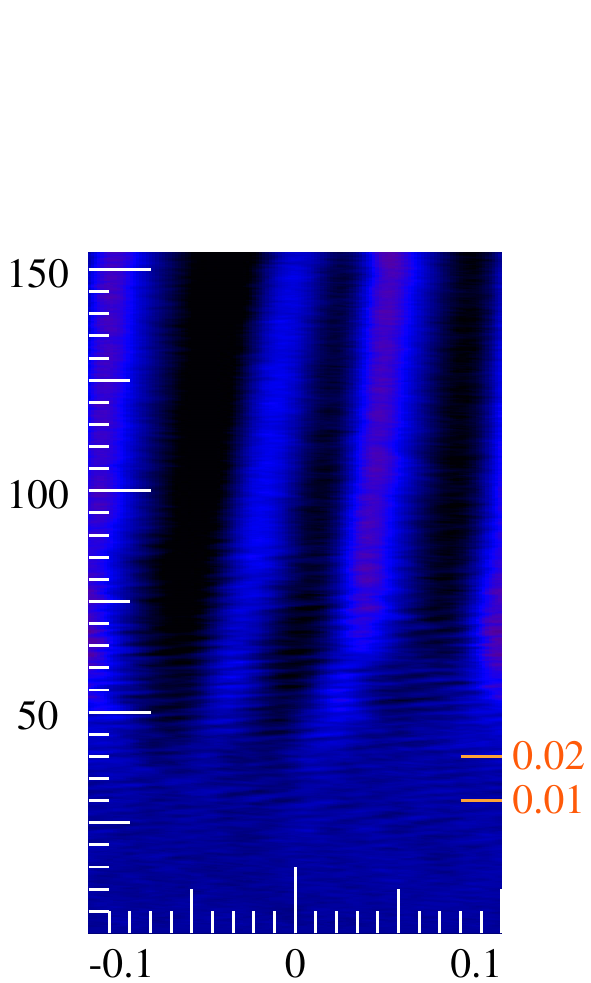} &
        \includegraphics[width=4cm]{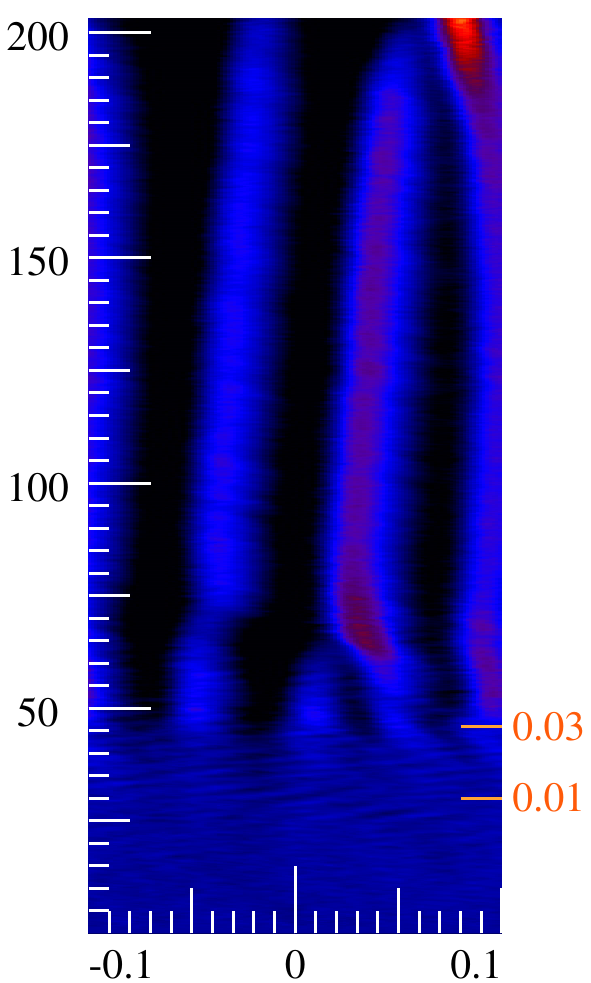} &
        \begin{sideways}Solid mass fraction (Z) \end{sideways}\vspace{2.7cm} &
    \end{tabular}
    \caption{Spacetime diagram of two 3D runs with particle size $\tauf = 0.03$ and $\Delta = 0.05$. The color indicates the column density $\Sigma_{\rm solid} / \langle \Sigma_{\rm solid} \rangle$ using the same color bar as Figs.\ \ref{fig:spacetime-1} and  \ref{fig:spacetime-2}.
    Both runs begin with $Z = 0.01$ and have $Z$ increased only for a short interval. In one run (left), $Z$ grows to 0.02 and in the other (right) $Z$ grows to 0.03. Clumps are visible for $Z \ge 0.02$, consistent with the 2D runs.}
    \label{fig:spacetime-3d}
\end{figure*}

Figure \ref{fig:spacetime-3d} shows the spacetime diagrams for both runs. The figure shows that both runs produce visible particle clumps for $Z \ge 0.02$ and no clumps for $Z = 0.01$. It is also clear that clumps form more readily with $Z = 0.03$. These results are consistent with our 2D simulations (Fig.\ \ref{fig:spacetime-1}).
For particles smaller than $\tauf = 0.03$, the small integration steps make 3D simulations prohibitive. However, as explained in section \ref{sec:response-time}, the long response time of very small particles ($\tauf \le 0.0064$) means that our simulations will overestimate $Z$; so our results can be considered robust upper bounds for this size range.

\newcolumntype{B}{ >{\centering\arraybackslash} p{4.0cm}@{\hspace{0.2cm}} }
\newcolumntype{C}{ >{\centering\arraybackslash} p{3.5cm}@{\hspace{0.2cm}} }

\begin{figure*}[hb!]
    \centering
    \includegraphics[width=11cm]{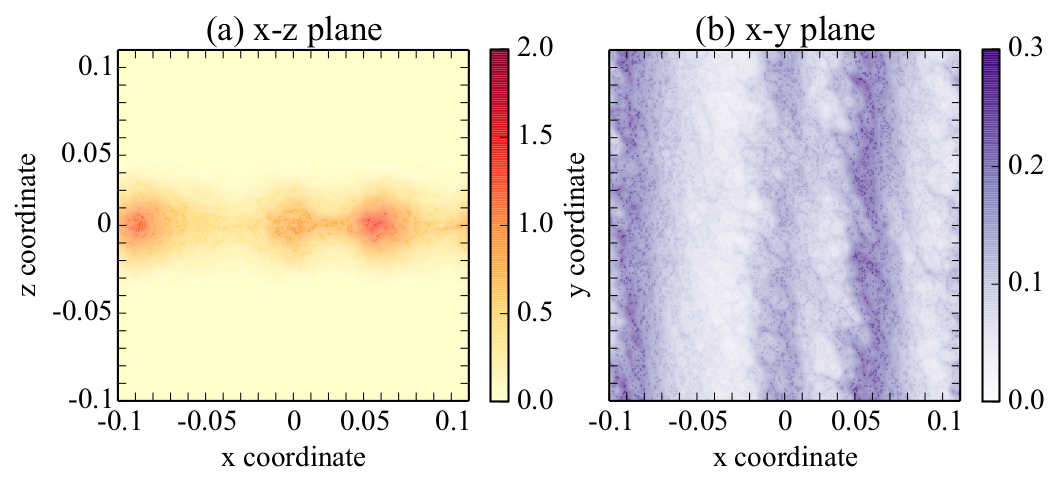}
    \caption{A snapshot of run 3D.Z2 taken at $t = 150$ orbits ($Z = 0.02$). Image (a) is the view from the side ($x-z$ plane). Image (b) is the view from above ($x-y$ plane). The color scale marks the column solid density normalized to the initial gas density ($\Sigma_{\rm solid} / \Sigma_{\rm gas,0}$). The two images have a different color scale, since the column density in the azimuthal direction is much higher. Note that the particle structure is nearly axisymmetric.}
    \label{fig:intro-3d}
\end{figure*}

\begin{figure*}[hb!]
    \centering
    \includegraphics[width=20cm]{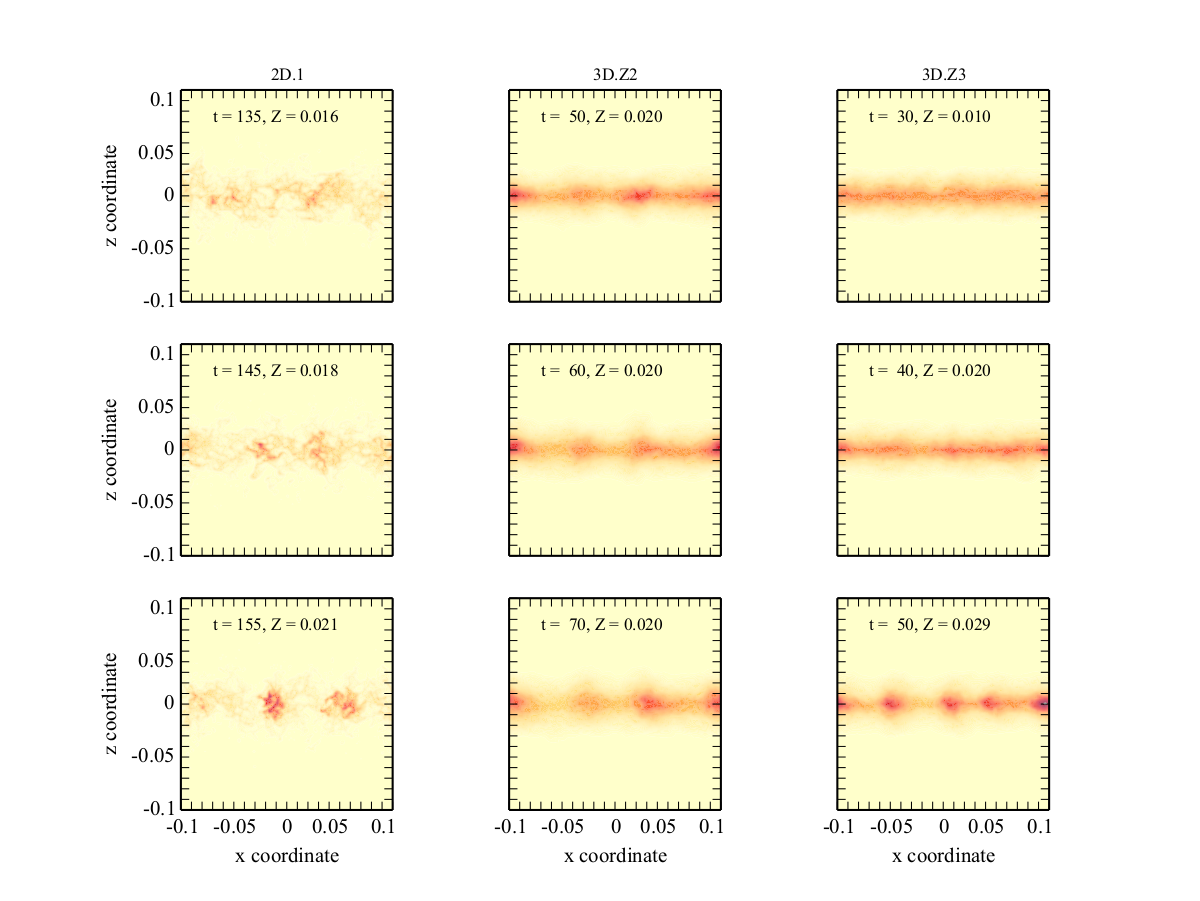}
    \caption{A sequence of vertical snapshots (x-z plane) showing the formation of distinct particle clumps in one of our 2D simulations (left) and our two 3D simulations (middle, right). For each run, the images are taken 10 orbits apart. The time and $Z$ values are shown in each plot. The times were chosen to show the formation of the first distinct clumps, and to give a similar peak density on the final image ($\rho_{p,max} = 1.8, 1.5, 2.1$ left to right). The color indicates the column density using the same color scale as in Fig.\ \ref{fig:intro-3d}(a). The other 2D runs give similar results to the one shown. The 2D run is consistent with the 3D runs. In particular, the 3D runs do not show any evidence of additional turbulence stirring compared to the 2D runs.}
    \label{fig:snapshots-3d}
\end{figure*}

Figure \ref{fig:intro-3d} gives a side view ($x-z$ axis) and a top view ($x-y$ axis) of the 3D.Z2 run at $t = 150$ orbits. The clumps and their filamentary structure are very prominent. An important feature of the 3D runs is that they form visible clumps more easily than the 2D runs. Figure \ref{fig:snapshots-3d} shows the formation of clumps in the two 3D runs and one of the 2D runs. This figure shows that the 3D runs are consistent with the 2D run. In particular, the clumps are readily visible at $Z = 0.02$ and the 3D runs show no evidence of additional stirring from the KHI.

\begin{figure*}[hb!]
    \centering
    \includegraphics[width=16cm]{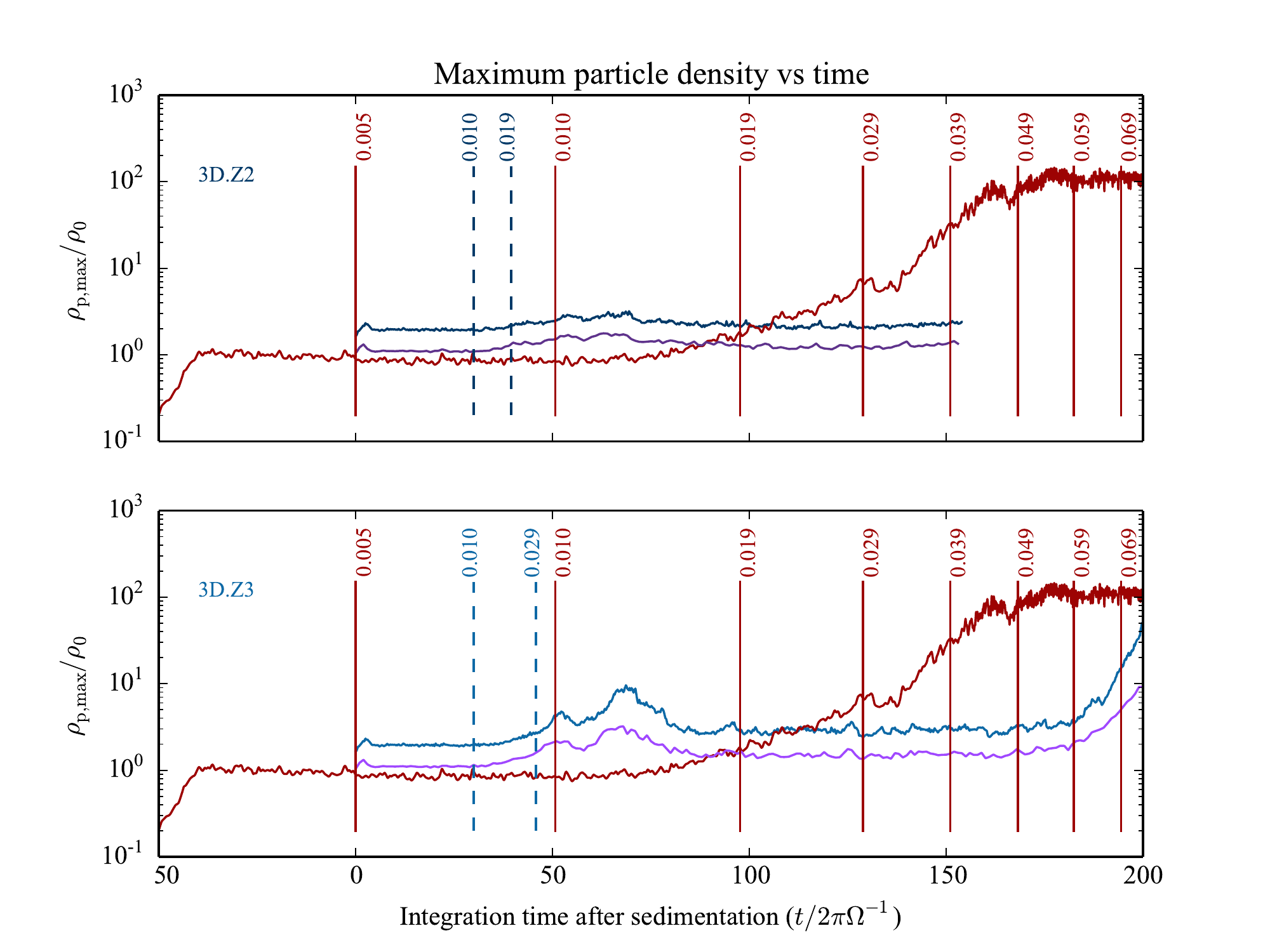}
    \caption{Maximum particle density ($\rho_{\rm p,max}$) for one of the 2D simulations and the two 3D simulations. The top plot compares run 3D.Z2 (blue) and a 2D run (red). The bottom plot compares run 3D.Z3 (blue) and the same 2D run (red). In each plot the purple line shows $\rho_{\rm p,max}$ for the 3D run after averaging along the azimuthal direction. The vertical dashed lines mark the places where the 3D runs transition from $Z = 0.01$ to $Z = 0.019$ (for 3D.Z2) or $Z = 0.029$ (for 3D.Z3). Similarly, the vertical solid lines mark the progressive increase of $Z$ during the course of the 2D runs.}
    \label{fig:density-max}
\end{figure*}

Figure \ref{fig:density-max} shows the maximum particle density ($\rho_{\rm p,max}$) for the two 3D simulations and one of the 2D simulations. One salient feature of the figure is that in the early phase of the simulations, before clumping occurs, the 3D runs consistently show a peak density 2-3 times greater than the 2D runs. This occurs because particles can also accumulate along the azimuthal direction. The figure also shows $\rho_{\rm p,max}$ for the 3D runs after averaging along the azimuthal direction (purple lines). This averaging removes the apparent discrepancy between the 2D and 3D runs. In the later stages the 2D runs reach higher peak densities because they have very high $Z$ values.

\subsection{Resolution}

\begin{figure*}[hb!]
  \begin{tabular}{ABCBC}
      \multicolumn{4}{r}{
      \includegraphics[width=10cm]{plots/idl/spacetime/colorbar.pdf}
      } &
      $\Sigma_{\rm solid} / \langle \Sigma_{\rm solid} \rangle$ \\
  &\hspace{0.9cm}$128 \times 128$; $\tauf = 0.03$ & $256 \times 256$; $\tauf = 0.03$  
  &\hspace{0.9cm}$128 \times 128$; $\tauf = 0.10$ & $256 \times 256$; $\tauf = 0.10$ \\
  \begin{sideways}
         Simulation time ($t/2\pi\Omega^{-1}$) for 
  \end{sideways}\vspace{3.0cm} &
  \includegraphics[height=11.5cm]{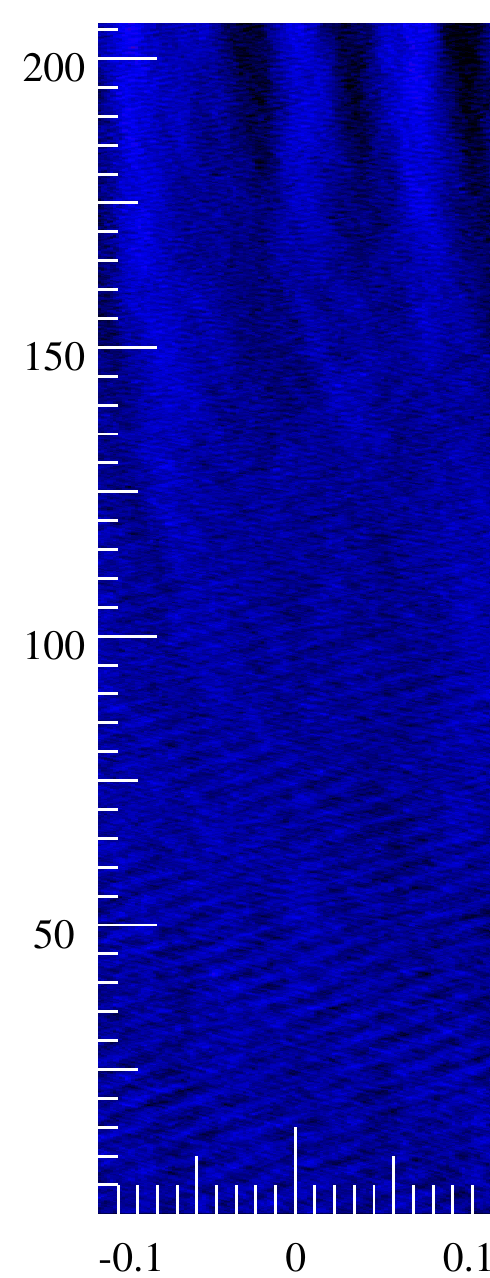} &
  \includegraphics[height=11.5cm]{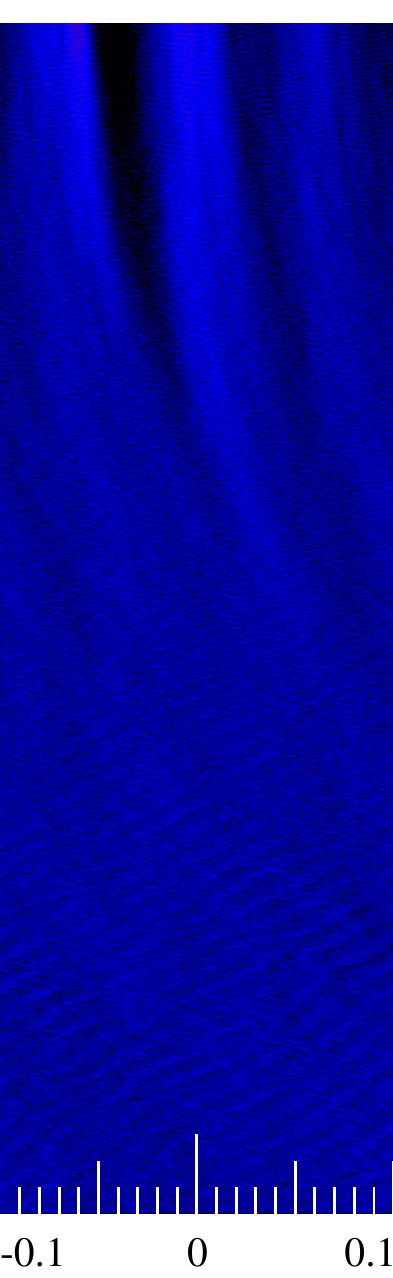} &
  \includegraphics[height=11.5cm]{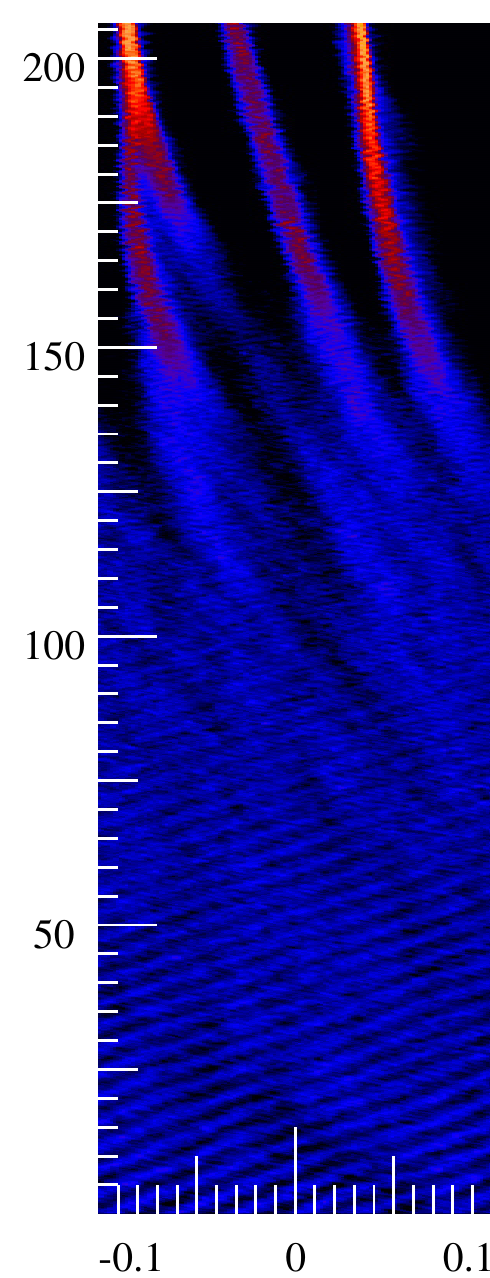} &
  \includegraphics[height=11.5cm]{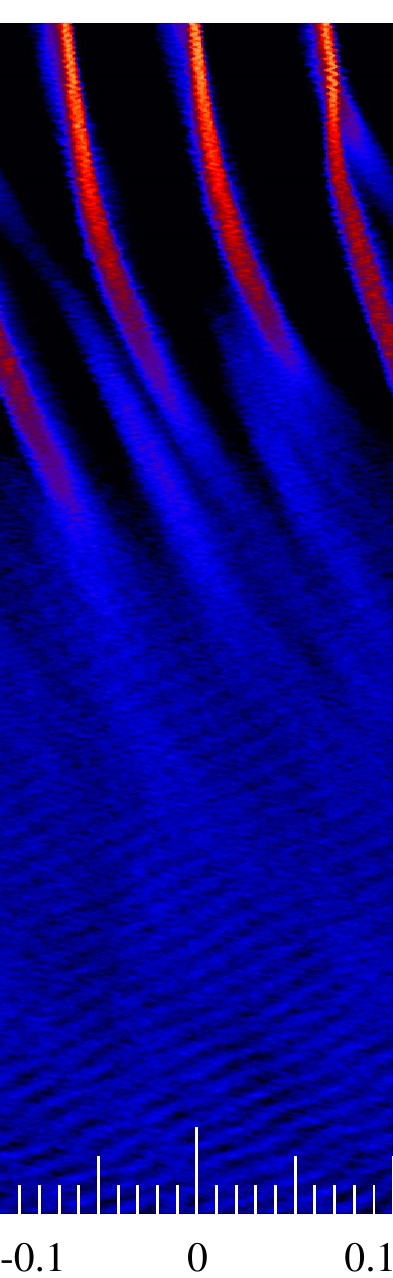} \\
  \end{tabular}

  \caption{Spacetime diagrams showing the solid surface density $\Sigma_{\rm solid}$ as a function of the radial coordinate $x$ and simulation time for $\tauf = 0.003$ and $\tauf = 0.010$ particles. The surface density is shown by color, using the same color scale as in Figs.\ \ref{fig:spacetime-1} and  \ref{fig:spacetime-2}.
  The figure shows both $128 \times 128$ simulations and the corresponding $256 \times 256$ simulations. In all simulations $\Delta = 0.05$. The higher-resolution runs produce results consistent with the $128 \times 128$ runs.}
  \label{fig:spacetime-res}
\end{figure*}

We performed two simulations at $256 \times 256$ grid resolution with $\tauf = 0.003$ particles. Figure \ref{fig:spacetime-res} shows the spacetime diagrams for both sets of simulations. Since the behavior of the $256 \times 256$ runs is almost identical to the corresponding $128 \time 128$ runs, we conclude that, in terms of resolution, our simulations are fully converged.

\newcolumntype{B}{ >{\centering\arraybackslash} p{4.40cm}@{\hspace{0.2cm}} }
\newcolumntype{C}{ >{\centering\arraybackslash} p{3.52cm}@{\hspace{0.2cm}} }

\end{document}